\documentclass{aa}
\usepackage[dvips]{graphicx}
\usepackage[dvips]{color}
\usepackage{amssymb}                       
\usepackage{wasysym}                       
\newcommand{\be}{\begin{eqnarray}}
\newcommand{\ee}{\end{eqnarray}}
\newcommand{\bequ}{\begin{equation}}
\newcommand{\eequ}{\end{equation}}
\newcommand{\doverd}[2]{\frac{\partial #1}{\partial #2}}
\newcommand{\doverdt}[1]{\frac{\partial #1}{\partial t}}
\newcommand{\doverdr}[1]{\frac{\partial #1}{\partial r}}
\newcommand{\doverdphi}[1]{\frac{\partial #1}{\partial \varphi}}
\newcommand{\subscr}[1]{_\mathrm{#1}}
\newcommand{\superscr}[1]{^\mathrm{#1}}
\newcommand{\Sect}[1]{Sect.~\ref{#1}}
\newcommand{\Fig}[1]{Fig.~\ref{#1}}
\newcommand{\Eq}[1]{Eq.~(\ref{#1})}
\newcommand{\MSun}{\mbox{$M_{\odot}$}}
\newcommand{\MStar}{\mbox{$M_\bigstar$}}
\newcommand{\MEarth}{\mbox{$M_{\earth}$}}
\newcommand{\MJup}{\mbox{$M_{\jupiter}$}}
\newcommand{\Mp}{\mbox{$M_\mathrm{p}$}}
\newcommand{\Md}{\mbox{$M_\mathrm{D}$}}
\newcommand{\Rhill}{\mbox{$R_\mathrm{H}$}}
\newcommand{\AU}{\mbox{\textrm{AU}}}
\newcommand{\Msyr}{\mbox{$\MSun\,\mathrm{yr}\superscr{-1}$}}
\newcommand{\ciro}{\textsc{Ciro}}
\newcommand{\gino}{\textsc{Gino}}
\newcommand{\pepp}{\textsc{Pepp}}
\newcommand{\wpro}{\textsc{Wpro}}
\newcommand{\elen}{\textsc{Elen}}
\begin{document}
\title{Nested-grid calculations of disk-planet interaction}
\author{Gennaro D'Angelo\inst{1,2}%
   \and Thomas Henning\inst{1}%
   \and Wilhelm Kley\inst{2}}
\offprints{G. D'Angelo,\\ \email{gennaro@astro.uni-jena.de}}
\institute{Astrophysikalisches Institut und Universit\"ats-Sternwarte,
           Schillerg\"a{\ss}chen 2-3, D-07745 Jena, Germany
      \and Computational Physics,
           Auf der Morgenstelle 10, D-72076 T\"ubingen, Germany}
\date{Received ---; accepted ---}
\abstract{%
We study the evolution of embedded protoplanets in a protostellar 
disk using very high resolution nested-grid computations.
This method allows us to perform global simulations of planets
orbiting in disks and, at the same time, to resolve in detail 
the dynamics of the flow inside the Roche lobe of the planet. 
The primary interest of this work lies in the analysis of the
gravitational torque balance acting on the planet. For this purpose 
we study planets of different masses, ranging from one Earth-mass up to 
one Jupiter-mass, assuming typical parameters of the protostellar 
disk. The high resolution of the method allows a precise determination
of the mass flow onto the planet and the resulting torques.
The obtained migration time scales are in the range from few times
$10^4$ years, for intermediate mass planets, to $10^6$ years, for
very low and high mass planets.
Typical growth time scales depend strongly on the planetary mass, 
ranging from a few hundred years, in the case of Earth-type planets,
to several ten thousand years, in the case of Jupiter-type planets.
}
\maketitle
\section{Introduction}

During the past five years radial velocity studies have allowed the detection
of planetary companions around other main-sequence stars.
Until now about sixty so-called ``extrasolar planets''
have been discovered, which orbit their stars within a distance of a few \AU.
A recent catalog of extrasolar planets, including their orbital
characteristics, is provided by Butler et al. (\cite{butler2001}) and 
up-to-date versions can be found
at \texttt{http://www.obspm.fr/encycl/encycl.html} and 
\texttt{http://exoplanets.org/}, maintained by Jean Schneider and 
the Department of Astronomy at UC Berkeley, respectively.

In contrast to the solar system, these new planets display
quite different orbital properties that challenge the accepted formation
scenario for solar planets. The major differences are 
their high minimum masses (up to $17$ Jupiter-masses), their
proximity to the central star (a fraction of the Sun-Mercury distance)
and their high eccentricities (up to $0.7$).

One of the main problems to deal with is the very close distance 
of massive planets to their parent star. 
The formation of Jupiter-type planets at these locations
is, on theoretical grounds, very unlikely. 
First of all, 
from purely geometrical arguments, the matter reservoir of 
the surrounding disk is too little so that a planet could 
hardly accrete its mass.
Second, 
the temperatures within the disk are too high 
for a rocky core to condense easily.

For these reasons it is generally believed that planets have formed 
from disk material further out, at distances of several \AU\ from the star, 
and have then migrated to their present positions.
This radial motion of the planet through the disk is primarily caused by 
gravitational torques acting on the planet. The presence of the planet in 
the disk disturbs the disk gravitationally, creating spiral density wave
perturbations, which emanate from the planet through the disk.
Hence, the disk is no longer axisymmetric which results in a net torque 
on the planet. The sign and magnitude of the vertical component of
the torque determines the direction and efficiency of the radial migration.

While initial fully non-linear hydrodynamical numerical computations of
embedded planets assumed a fixed circular orbit of the planet
(Kley \cite{kley1999}; Bryden et al. \cite{bryden1999}; 
Lubow et al. \cite{lubow1999}), more recent simulations
took into account the back reaction of the disk and allowed 
for a change in the parameters of the planetary orbit 
(Kley \cite{kley2000}; Nelson et al. \cite{rnelson2000}). 
For a Jupiter-mass planet and typical parameter values for 
the disk, the obtained orbital decay time is about $10^5$ years, 
which agrees reasonably well with previous estimates based on analytic 
linear theories (Goldreich \& Tremaine \cite{gt1980}; Ward \cite{ward1997}).

The majority of the computations, performed so far, 
have used a single grid which resolves the Roche lobe 
of a Jupiter-mass planet only with very few grid cells. 
Recently, Cieciel\c{a}g et al. (\cite{ciecielag2000a},
\cite{ciecielag2000b}) used 
an Adaptive Mesh Refinement method to resolve the immediate 
surroundings of the planet, but they didn't give any 
estimate of the mass accretion rate and magnitude of the 
gravitational torque. 
On the other hand, Armitage (\cite{armitage2001}) reduced 
the overall simulated region achieving a better resolution.  
However, also in this case the Roche lobe is only resolved 
by a few grid cells because of the low mass of the investigated
planet.

In this paper we aim at the structure and dynamics of the gas flow
in the close vicinity of the planet, while performing global disk
simulations. In order to obtain the necessary high spatial and
temporal resolution, 
we use a nested-grid formalism which allows an accurate computation 
of the mass flow onto the planet and the acting torques.

In the next section we layout the physical model followed by
a description of the numerical method (\Sect{Sect:NM}). 
We describe the setup of the various numerical models in \Sect{Sect:GMD}.
The main results are presented in \Sect{Sect:results} and our 
conclusions are given in \Sect{Sect:conclusions}. 
\section{Physical model}
\label{Sect:Phm}
For the purpose of this study, we assume that the opening angle of
the protostellar accretion disk is very small. We describe the 
disk structure by means of a two-dimensional, infinitesimal thin, 
model using vertically averaged quantities, such as the surface mass density
\[ \Sigma = \int^\infty_{-\infty} \rho dz,\]
where $\rho$ is the regular density.
We work in a
cylindrical coordinate system $(r, \varphi, z)$ whose origin is fixed at the
center of mass of the star and the planet, and where the plane of the disk 
coincides with the $z=0$ plane.

The gas in the disk is non-self-gravitating and is orbiting
a protostar having a mass $\MStar = 1$ \MSun. The total mass of the disk
\Md, within the simulated region, which extends from $2.08$ to $13$ \AU, 
is $3.5 \times 10^{-3}$ \MSun.
Embedded in this disk there is a massive protoplanet
with a mass \Mp, which ranges from one Earth-mass (\MEarth) 
to one Jupiter-mass (\MJup), depending on the considered model.
The planet is assumed to be on a fixed circular orbit throughout the
evolution. We employ a rotating coordinate system, 
corotating with the planet, whose azimuthal position is kept constant
at $\varphi\subscr{p}=\pi$. 
The angular velocity $\Omega$ of the rotating frame is then given by
\bequ
  \Omega = \Omega_p = \sqrt{\frac{G\,(\MStar + \Mp)}{a^3}},
\eequ 
where  $G$ is the gravitational constant and
$a$ is the semi-major axis of the planet's orbit.

The evolution of the disk is given by the two-dimensional
($r, \varphi$) continuity equation for $\Sigma$ and 
the Navier-Stokes equations for each of the two components 
of the velocity field $\vec{u}\equiv (u_r, u_\varphi)$.
Thus, the set of equations read
\begin {equation}
 \doverdt{\Sigma} + \nabla \cdot (\Sigma\,\vec{u})
                =  0,  \label{Sigma}
\end{equation}
\begin {equation}
 \doverdt{(\Sigma\,u_r)} + \nabla \cdot (\Sigma\,u_r\,\vec{u})
  = \Sigma \, r \, ( \omega + \Omega)^2
        - \doverd{P}{r} - \Sigma\, \doverd{\Phi}{r} + f_{r},
      \label{u_r}
\end{equation}
\begin {equation}
 \doverd{[\Sigma\, r^2\, (\omega + \Omega)]}{t}
   + \nabla \cdot [\Sigma\, r^2\, (\omega + \Omega)\, \vec{u}]
         =
        - \doverd{P}{\varphi} - \Sigma\,\doverd{\Phi}{\varphi}   
   + f_{\varphi}.
      \label{u_phi}
\end{equation}
Here $\omega=u_\varphi/\,r$ is the angular velocity and
$P$ is the vertically integrated (two-dimensional) pressure. 
The gravitational potential $\Phi$, generated by the protostar 
and the planet, is given by
\begin{equation}
    \Phi = \Phi_{\bigstar} + \Phi_\mathrm{p} =
         - \frac{G\,\MStar}{| \vec{r} - \vec{r}_{\bigstar} |}
         -  \frac{G\,\Mp}{| \vec{r} - \vec{r}_\mathrm{p} |},
         \label{potential}
\end{equation}
where $\vec{r}_{\bigstar}$ and $\vec{r}_\mathrm{p}$ 
are the radius vectors to the star and the planet, 
respectively.
The effects of viscosity are contained in the terms $f_{r}$ and
$f_{\varphi}$ which give the viscous force per unit area acting
in the radial and azimuthal ($f_{\varphi}/r$)
direction:
\[
f_{r}       = \frac{1}{r}\,\doverdr{(r\,S_{rr})}
            + \frac{1}{r}\,\doverdphi{S_{r\varphi}}
            - \frac{S_{\varphi\varphi}}{r},
\]
\[
f_{\varphi} = \frac{1}{r}\,\doverdr{(r^2\,S_{r\varphi})}
            + \doverdphi{S_{\varphi\varphi}}.
\]
Since we assume a zero bulk viscosity $\zeta$, a constant kinematic
viscosity $\nu$ and we do not include any artificial viscosity 
in our numerical models, the relevant non-zero components of the 
three-dimensional viscous stress tensor $\tens{S}$ are 
\bequ
S_{rr}      = 2\,\nu\,\Sigma\,\left(\doverdr{u_r} 
            - \frac{1}{3}\,\nabla \cdot \vec{u}\right),
            \label{S_rr}
\eequ
\bequ
S_{\varphi\varphi}= 2\,\nu\,\Sigma\,\left(\doverdphi{\omega} 
                  + \frac{u_r}{r}
                  - \frac{1}{3}\,\nabla \cdot \vec{u}\right),
            \label{S_pp}
\eequ
\bequ
S_{r\varphi}      = \nu\,\Sigma\,\left(\frac{1}{r}\,\doverdphi{u_r} 
                  + r\,\doverdr{\omega}\right),
            \label{S_rp}
\eequ
where the divergence of the velocity field can be written as
\[
\nabla \cdot \vec{u} = \frac{1}{r}\,\doverdr{(r\,u_r)} 
                     + \doverdphi{\omega}. 
\]
A more general form of the relations (\ref{S_rr}), (\ref{S_pp})
and (\ref{S_rp}), within the two-dimensional cylindrical approximation, 
is given for example in Kley (1999).

In the set of equations above we have omitted the energy equation
because in this study we will be concerned only with a relatively
simple equation of state which does not require the solution of
an energy equation.
We shall use an isothermal equation of state
where the surface pressure $P$ is related to the density $\Sigma$ 
through
\begin{equation}
        P = c\subscr{s}^2 \, \Sigma.
\end{equation}
The local isothermal sound speed $c\subscr{s}$ is given here by
\bequ
        c\subscr{s} = \frac{H}{r} \, v\subscr{Kep},
        \label{cs}
\eequ
where $v\subscr{Kep} = \sqrt{G\,\MStar/r}$ denotes the Keplerian 
orbital velocity of the unperturbed disk.
Equation (\ref{cs}) follows from vertical hydrostatic equilibrium.
The ratio $h$ of the vertical height $H$ to the
radial distance $r$ is taken as a fixed input parameter. 
With this choice, the temperature $T$ is proportional to $1/r$.
Here we use a standard value
\[      h =  \frac{H}{r} =  0.05  \label{h}, \]
which is typical for protostellar accretion disks having a mass
inflow rate of $\dot{M} \approx 10^{-7}$ \Msyr.
With this value of $h$, our kinematic viscosity coefficient
is equivalent to $\alpha=4\times10^{-3}$ at the radial
position of the planet ($\nu=\alpha\,c_s\,H$).  

Since the mass of the planet is very small in comparison to the mass
of the star, because we always use here a ratio $q = \Mp/\,\MStar$ 
smaller than $10^{-3}$, the center of mass is located
very close to the position of the star. In the following
we will often identify the radial distance
from the origin of the coordinate system with the distance 
from the central star.
\section{Numerical method}
\label{Sect:NM}
In order to study the planet-disk interaction, 
we utilize a finite difference method to solve the hydrodynamic equations 
outlined in the previous section.
As we intend to achieve a very high resolution around the planet,
we use a nested-grid technique. 
Both requirements are provided by an early
FORTRAN-Version of \textsc{Nirvana} 
(Ziegler \cite{ziegler1997}; Ziegler \cite{ziegler1998}), 
which is a 
3D, nested-grid, MHD code, based on a covariant Eulerian formalism. 
The relevant equations are solved on a mesh structure having 
a constant spacing in each direction.
For the present purposes, the code is used in a pure
hydrodynamic mode, adopting a cylindrical reference frame
where the $z$-dimension is switched off.

\textsc{Nirvana} uses a spatially, second-order accurate, explicit method.
The advection is computed by means of the second order monotonic
transport algorithm, introduced by van Leer (\cite{vanleer1977}),
which guarantees global conservation of mass and angular momentum.
It is first-order accurate in time.
The viscosity part was added and is treated explicitely.
The stress tensor has been implemented for a \textit{Newtonian fluid}
according to the \textit{Stokes hypothesis}, 
i.e.\ with a bulk viscosity $\zeta=0$.
 
This code has been already employed in similar computations, 
either in 2D (Nelson et al. \cite{rnelson2000})
and 3D (Kley et al. \cite{kley2001}), but always in a single-grid mode. 
\subsection{Nested-grid technique}
\label{Sect:NGt}
This technique is particularly useful when very high local
resolution is required at specific and predefined points
of the computational domain. In our situation, this allows us 
to simulate both the overall behavior of the disk
and the immediate surroundings of the planet.
Since this kind of numerical approach is quite new 
for the calculation of disk-planet interaction,
we describe the method to some extent, but \textit{only}
referring to our particular and specific case.

A similar numerical scheme has been adopted,
for astrophysical simulations, by a number of authors.
Ruffert (\cite{ruffert1992}) used this approach
to investigate the collision between a white dwarf and a main sequence star.
In his paper the numerical method is explained in detail.
Yorke et al. (\cite{yorke1993}) and Burkert \& Bodenheimer (\cite{burkert1993})
simulated the collapse of a protostellar cloud. 
An application to flux-limited radiation hydrodynamics can be found
in Yorke \& Kaisig (\cite{yorke1995}).

\begin{figure}
\begin{center}
\resizebox{0.9\linewidth}{!}{%
\includegraphics{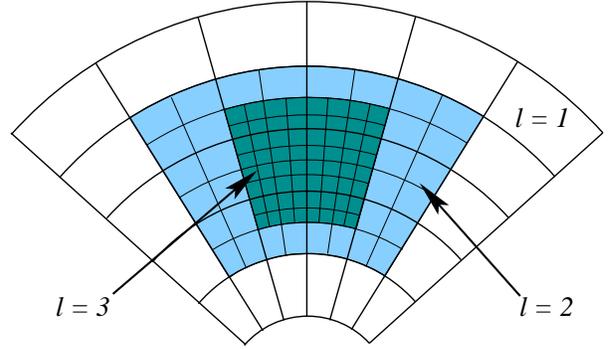}}
\end{center}
\caption{Face-up projection of a three-level grid system in 
cylindrical coordinates. On the finest subgrid ($l=3$) the
linear spatial resolution is four times as large as it is 
on the main
grid ($l=1$).}
\label{hierarchy}
\end{figure}

The method relies on the basic idea that,
whenever a greater resolution is needed in a designated region,
a finer subgrid is located inside the main grid 
(the one covering the whole computational domain).
If the resolution is not high enough yet, another subgrid 
may be placed on the underlying one.
Since any subgrid can host a finer subgrid structure, 
a grid hierarchy is generated, also called
``system of nested grids''.
In principle there is no limit to the degree of nesting.
A three-level hierarchy is shown in \Fig{hierarchy}. 

The necessary equations are then integrated,
\textit{independently}, on every grid level.
However, two neighbor subgrids must exchange the
necessary information whenever the integration
proceeds from one grid level to another.
Restrictions are imposed on the time step only
because, for stability reasons, the Courant-Friedrichs-Lewy 
(CFL) condition must be fulfilled during each integration,
on each level.
\subsubsection{Basic integration cycle}
\label{Sect:iter_cycle}
\begin{figure}
 \begin{center}
 \resizebox{0.9\linewidth}{!}{%
 \includegraphics{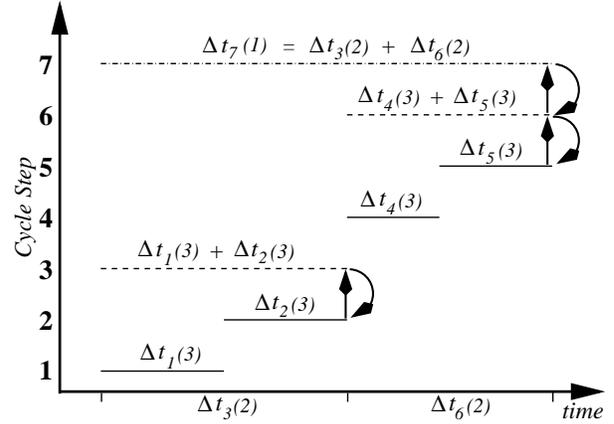}}
 \end{center}
 \caption{Scheme of a complete integration cycle for a 
          three-level grid system. Arrows indicate the direction
          of information transfer when integration proceeds
          from a level to the next lower one. Straight arrows
          stand for the solution updating process on the levels
          $2$ and $1$. Bow-arrows indicate the data transfer for setting
          the boundary quantities on the levels $3$ and $2$.}
 \label{cycle_sketch}
\end{figure}

In our calculations we use the smallest possible refinement ratio: 
\begin{equation}
\frac{\Delta r(l+1)}{\Delta r(l)}=
\frac{\Delta \varphi(l+1)}{\Delta \varphi(l)}=2,
\label{refine_ratio}
\end{equation}
where $\Delta r(l)$ and $\Delta \varphi(l)$ represent 
the mesh discretization, along each direction,
on the grid level $l$ (here $l=1$ identifies the main grid).
In order to analyze a complete integration cycle, 
let's suppose we have a three-level hierarchy at an evolutionary time $t$. 
The cycle will be completed when on each grid the system 
has evolved for the same time:
\begin{enumerate}
\item         We always start the integration from the 
              finest level.
              During the first step of the cycle, this grid is evolved
              for a time interval
\[
\Delta t_1(3) = \min\left[\Delta t\superscr{CFL}_1(3),
                     \frac{1}{2}\,\Delta t\superscr{CFL}_1(2),
                     \frac{1}{4}\,\Delta t\superscr{CFL}_1(1)\right],
\]
              where $\Delta t\superscr{CFL}_1(l)$ represents the time
              step resulting from the CFL criterion applied to the
              level $l$, after its latest integration. 
              Thus, $\Delta t_1(3)$ accounts for the CFL
              stability criterion on the whole set of grids. 
\item         The third grid has to be integrated once more because of
              \Eq{refine_ratio}.
              Since after the first step this was the only level 
              to evolve, we only have to check the new CFL time step for 
              this grid, $\Delta t\superscr{CFL}_2(3)$. 
              Then it can move further in time for an interval
\[
\Delta t_2(3) = \min\left[\Delta t\superscr{CFL}_2(3),
                     \frac{1}{2}\,\Delta t\superscr{CFL}_1(2),
                     \frac{1}{4}\,\Delta t\superscr{CFL}_1(1)\right].
\]
\item         Now level $2$ can be integrated for a time 
\[              
               \Delta t_3(2)=\Delta t_1(3) + \Delta t_2(3) 
               \leq \Delta t\superscr{CFL}_1(2),
\]
              so that numerical stability is automatically assured.
              At this point
              of the cycle the first information exchange takes place:
              the solution on the level $2$ is corrected via the more
              accurate solution of the level $3$; the boundary values 
              of the level $3$ are updated by using the solution of the 
              level $2$, which covers a larger domain. 
              These fundamental interactions will be described later.
\end{enumerate}
We have just seen that a level $l+1$ has to be visited two times as often 
as the level $l$. Then the next two cycle steps will be similar to the 
first two, provided that the appropriate CFL time steps are employed to
compute $\Delta t_4(3)$ and $\Delta t_5(3)$. During the sixth step,
the level $2$ evolves for $\Delta t_6(2)=\Delta t_4(3) + \Delta t_5(3)$.
Eventually, during the seventh step, the integration of the level $1$ 
is performed using a time step
$\Delta t_7(1)=\Delta t_3(2) + \Delta t_6(2)$ which is, by construction,
smaller than $\Delta t\superscr{CFL}_1(1)$.
The cycle is now complete and each grid level has evolved for the same
amount of time $\Delta t_7(1)$. 
The entire cycle sequence is schematically sketched in 
\Fig{cycle_sketch}.

In general, within this kind of cycle, a level $l$ is integrated 
$2^{l-1}$ times. 
\subsubsection{Downward information transfer}
\begin{figure}
 \begin{center}
 \resizebox{0.9\linewidth}{!}{%
 \includegraphics{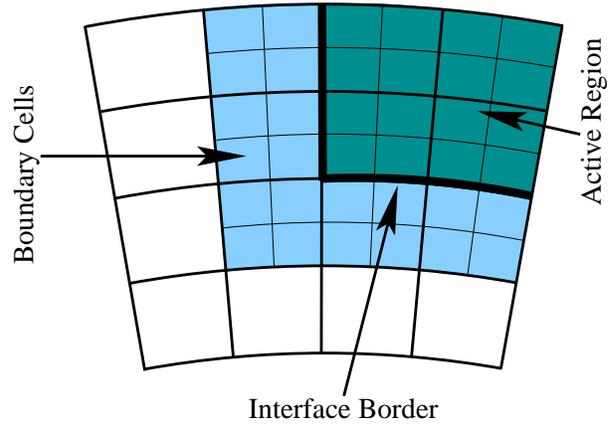}}
 \end{center}
 \caption{Interface between a subgrid and its host. The light-colored
          zone marks those cells containing boundary values needed 
          for the subgrid integration, the ``ghost cells''. 
          The darker region refers to the so-called 
          ``active zone'', where values are effectively computed 
          on the subgrid. The thick line, which
          separates the previous regions, encloses the grid
          cells on the coarse grid whose content is replaced by the more 
          accurate one coming from the subgrid.}
 \label{bc_grid}
\end{figure}

We already mentioned that after the third step of the iteration cycle,
grids $3$ and $2$ have to exchange some information. 
In general, this exchange must occur every time the grid level
$l$ evolves to the same time as the level $l+1$.
Because of the higher resolution, we assume the solution of the level $l+1$
to be more accurate than that of the level $l$. 
Therefore, the fine-grid solution replaces the coarse on the 
common computational domain. 
Whatever the level in the hierarchy is, the frame formed 
by the first and the last two grid cells are ghost cells
(see \Fig{bc_grid}).
This indicates that they contain the boundary values 
necessary to perform the algorithm integration.
Ghost cells of level $l+1$ do not contribute to the updating 
process of the solution of level $l$.

\begin{figure}
 \begin{center}
 \resizebox{0.9\linewidth}{!}{%
 \includegraphics{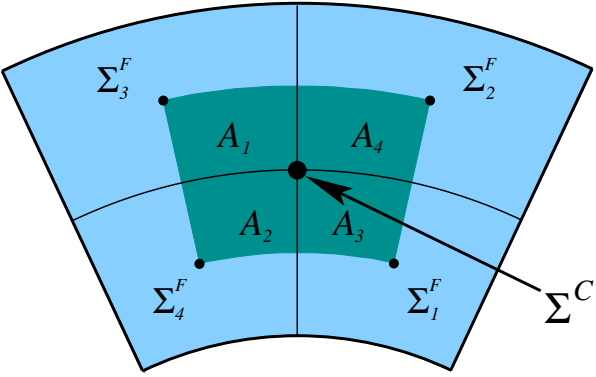}}
 \end{center}
 \caption{Surface weighted average of the surface density. 
          A \textit{coarse} cell is shown along with the four
          \textit{fine} cells it comprises.
          As a scalar, the surface density is cell-centered within
          the appropriate refinement (dots).
          $\Sigma\superscr{C}$ represents the new, interpolated, value
          of the coarse cell;
          $\Sigma\superscr{F}_i$ are the fine interpolating quantities.
          Because of the fixed value of  $\Delta \varphi(l)$, $A_1=A_4$
          and $A_2=A_3$.}
 \label{d_interp}
\end{figure}

\begin{figure}
 \begin{center}
 \resizebox{0.7\linewidth}{!}{%
 \includegraphics{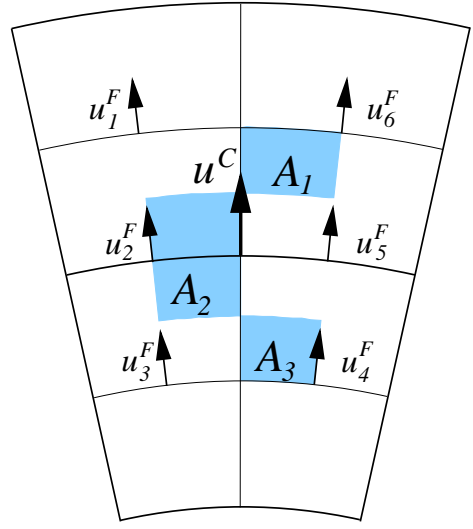}}
 \end{center}
 \caption{Surface weighted average of the radial momentum density
          $u=\Sigma\,u_r$. Two coarse cells are drawn (thick lines);
          Fine cells are delimited by thin lines.
          Because of the staggered structure of the
          grid, vectors are face-centered within the cell.}
 \label{v_interp}
\end{figure}

\begin{figure}
 \begin{center}
 \resizebox{0.9\linewidth}{!}{%
 \includegraphics{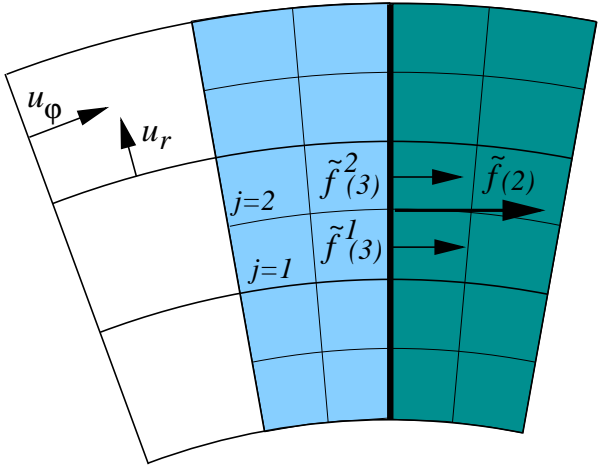}}
 \end{center}
 \caption{Momentum flux correction scheme. Momentum flux 
          components are centered as velocity field components,
          whose locations are shown in the upper left corner.  
          $\tilde{f}(2)\,\Delta r(2)=
          f_3(2)\,\Delta t_3(2)\,\Delta r(2)$ represents the
          amount of angular momentum, flowed across the 
          coarse cell border,
          during the third cycle step (see \Sect{Sect:iter_cycle}).
          $\tilde{f}^1(3)\,\Delta r(3) =
          f^1_1(3)\,\Delta t_1(3)\,\Delta r(3) + 
          f^1_2(3)\,\Delta t_2(3)\,\Delta r(3)$ represents the
          same quantity transported, during the first two cycle
          steps, across the $j=1$ fine cell border.
          The coarse quantity is replaced by
          $\left[\tilde{f}^1(3)+ \tilde{f}^2(3)\right]\,\Delta r(3)$.}
 \label{fluxcorr}
\end{figure}

The replacement procedure is straightforward: a surface weighted
average, using the nearest fine values, substitutes the coarse quantity. 
For example, referring to \Fig{d_interp}, 
the averaged coarse density ($\Sigma\superscr{C}$) is
\begin{equation}
\Sigma\superscr{C} =\frac{\sum_i \Sigma\superscr{F}_i\,A_i}{\sum_i A_i}
                 =\frac{(\Sigma\superscr{F}_1 + 
                         \Sigma\superscr{F}_4)\,A_1 +
                        (\Sigma\superscr{F}_2 + 
                         \Sigma\superscr{F}_3)\,A_2}{2\,(A_1 + A_2)}. 
                  \label{sigma_c}
\end{equation}
In our case, since the advected quantities are the
linear radial momentum density $\Sigma\,u_r$, and the angular 
momentum density $\Sigma\,r^2\,\omega$, these are the interpolated 
quantities, along with $\Sigma$.
Since velocities are centered at the sides of a cell 
(see Figs.~\ref{v_interp} and \ref{fluxcorr}), 
this average is a little more complex than
the previous one and requires six terms. 
Indicating with $u\superscr{C}$ the coarse value of the 
linear momentum density to be interpolated and
with $u\superscr{F}_i$ the surrounding fine grid values, we have:
\begin{equation}
u\superscr{C} =\frac{(u\superscr{F}_1 + u\superscr{F}_6)\,A_1 +
                     (u\superscr{F}_2 + u\superscr{F}_5)\,A_2 +
                     (u\superscr{F}_3 + u\superscr{F}_4)\,A_3}%
                     {2\,(A_1 + A_2 + A_3)}. 
               \label{u_c}
\end{equation}

In a successive step, velocities are retrieved.
No other quantity needs to be interpolated
because the potential is fixed and no energy equation is solved. 

In order to guarantee global mass and momentum conservation 
for the whole hierarchy,
we have to make sure that the momentum flux components 
$(\Delta r\,\Sigma\,u_r/\Delta t,\:
r\,\Delta \varphi\,\Sigma\,r^2\,\omega/\Delta t)$, 
across the border between a subgrid and its host grid 
(indicated in \Fig{bc_grid}), are equal in both level
solutions whenever the grid has evolved for the same time
as the subgrid.
However, each grid evolves independently and for a time interval
different from that of the lower one. 
Thus, even after the solution 
updating process described above, the amount of
momentum flowed across the borders might not 
coincide in the respective solutions.
To remove this possible discrepancy, 
at the coarse-fine grid border,
these quantities are taken from the 
fine grid integration. 
In \Fig{fluxcorr}
the situation for the azimuthal momentum flux is depicted.
Two fine cells participate in this process.
Referring to the integration cycle traced in \Sect{Sect:iter_cycle}, 
$f^j_k(l)$ represents the value of the quantity 
$r\,\Delta \varphi\,\Sigma\,r^2\,\omega/\Delta t$,
at the grid-grid interface location,
as computed during the $k$-th cycle step on level $l$.
An additional index ($j=1,2$) is needed to identify
the radial position of the two fine cells involved 
(for example, on level $3$), 
but it does not concern the coarse grid quantity to be replaced
(on level $2$).

Suppose we are at the end of the third cycle, when the first
interaction, between levels $2$ and $3$, occurs 
(first straight arrow in \Fig{cycle_sketch}).
Because of the refinement ratio established by \Eq{refine_ratio}, 
quantity $f_3(2)$ will be reset as:
\begin{equation}
f_3(2) = \frac{1}{2}\,\frac{\Delta t_1(3)\,\sum_jf^j_1(3) + 
                            \Delta t_2(3)\,\sum_jf^j_2(3)}{\Delta t_3(2)}
\end{equation}
This correction is accounted for
directly while performing the advection of radial and angular momenta.
\subsubsection{Upward information transfer}
The boundary conditions on the main grid are usually imposed
depending on the physics and geometry of the problem: symmetry,
periodicity etc. In the case of a subgrid, boundary values must
be attached, in some way, to the values of the underlying grid. 
This point turned out to be extremely delicate for our calculations.
In \textsc{Nirvana}, the boundary values of a certain level $l$ were
just set by means of a linear interpolation of the quantities
on the lower level $l-1$. 
Because of the strong variations in density and velocity,
due to the formation of shock fronts
in our simulations, this method fails and 
produces numerical inconsistencies.

Therefore, we raised the order of the interpolation. 
However, this introduces another potential trap.
In fact, a high-order interpolation (higher than the first order)
is not monotonic and can produce a new minimum.
This is not acceptable since, for example, the density is a non-negative 
quantity. Then the interpolating function should be monotonised.
In order to handle this problem we have used the same approach
as described in Ruffert (\cite{ruffert1992}), that is
by employing the \textit{monotonised harmonic mean} 
(van Leer \cite{vanleer1977}).

\begin{figure}
 \begin{center}
 \resizebox{0.9\linewidth}{!}{%
 \includegraphics{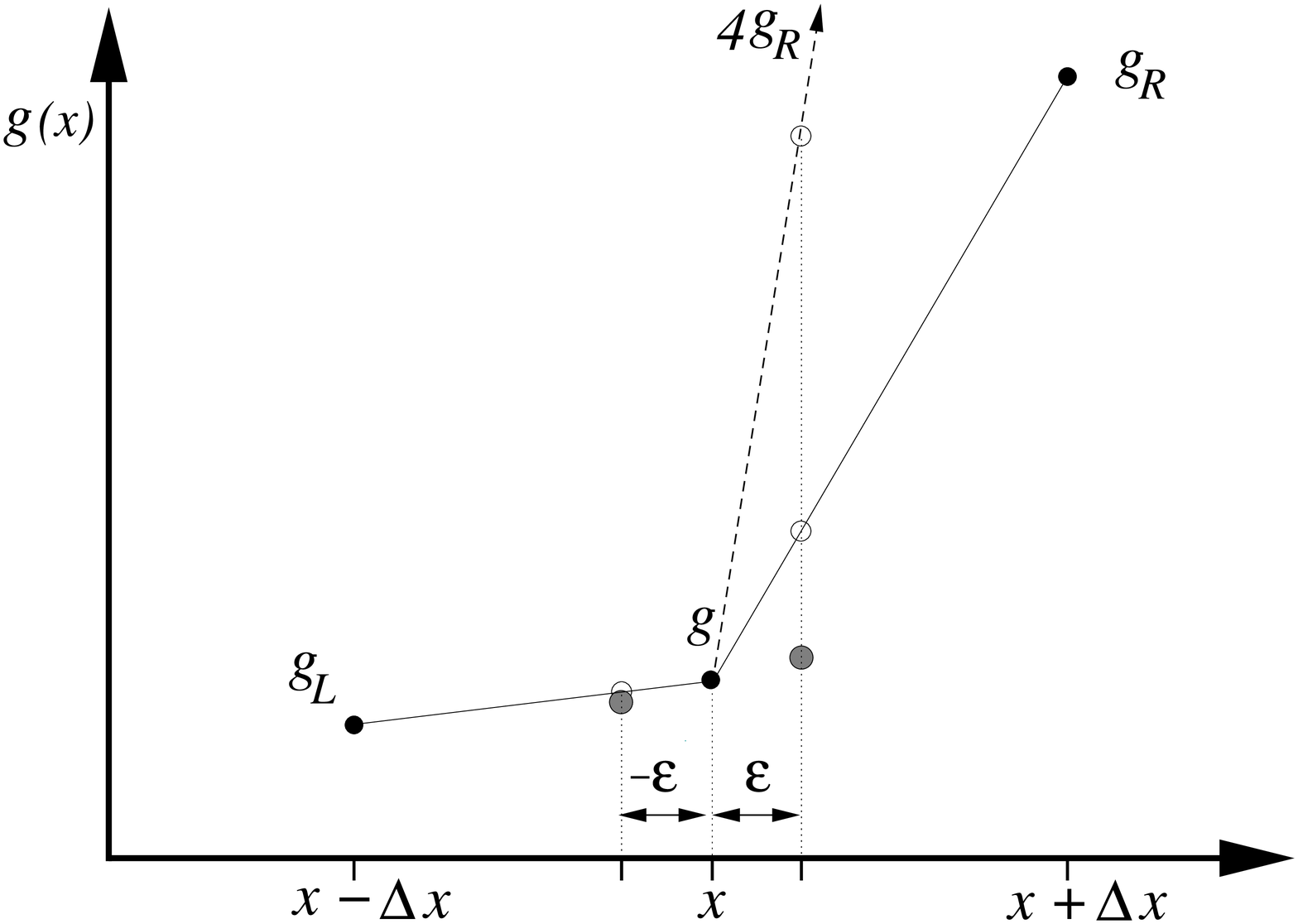}}
 \end{center}
 \caption{Behavior of an harmonic mean against a geometric mean.
          Grey filled circles indicate the \textit{harmonic} 
          values assumed by $g_\varepsilon$. 
          Open circles indicate the corresponding values 
          when the geometric mean is applied. 
          The plot intends to simulate a shock
          front. If we keep the left value $g\subscr{L}$ unchanged
          and make the right one $g\subscr{R}$ four times as large,
          the geometric mean value will increase proportionally. 
          In turn, the harmonic mean will change so 
          slightly that its variation is not visible on this plot.}
 \label{harmonic}
\end{figure}

Then, if we have a function sampled at
$x-\Delta x$, $x$ and $x+\Delta x$, with values
$g\subscr{L}$, $g$ and $g\subscr{R}$, respectively
(as shown in \Fig{harmonic}), the averaged value
at $x+\varepsilon$ is
\begin{equation}
   g_\varepsilon   = g + \frac{2\,\varepsilon}{\Delta x}\,%
                  \max\left[\frac{(g-g\subscr{L})\,(g\subscr{R}-g)}%
                  {g\subscr{R}-g\subscr{L}},0\right],
   \label{harm_mean}
\end{equation}
provided that $-\Delta x/2 \leq \varepsilon \leq \Delta x/2$.

\begin{figure}
 \begin{center}
 \resizebox{0.9\linewidth}{!}{%
 \includegraphics{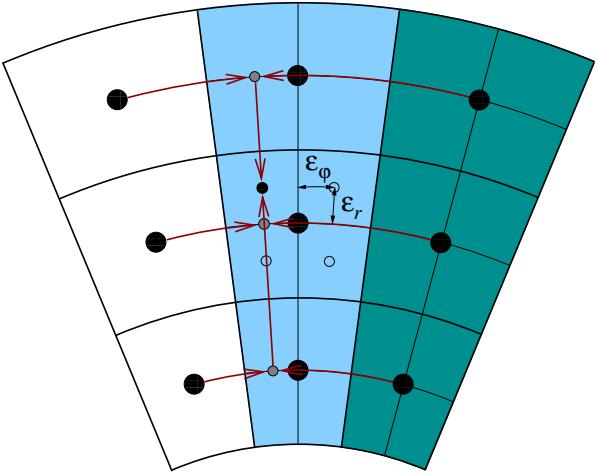}}
 \end{center}
 \caption{Harmonic average: surface density boundary values.
          The light-colored zone indicates subgrid ghost cells.
          Darker region belongs to the active subgrid zone. 
          Nine coarse values (big black circles) are engaged 
          in the average. The value to be interpolated is shown as
          a small black circle.  
          During a first step, \Eq{harm_mean} is
          applied three times along the $\varphi$-direction.
          Three new values are generated (grey small circles),
          having the correct, final azimuthal coordinate. 
          These are averaged to obtain the final value
          at the correct radial location. Small open circles refer to
          the other boundary quantities whose value depends on the
          same coarse quantities.}
 \label{d_harm}
\end{figure}

If we adopt this kind of average on a 2D-mesh, 
each interpolation generally involves $3 \times 3$ 
coarse quantities. 
It proceeds by averaging the selected coarse values along a certain 
direction, three at a time. This results in three new quantities.
A further harmonic average of these, along the other direction,
generates the subgrid boundary value at the correct position. 
\Fig{d_harm} shows how the procedure works 
in the case of the surface density.

As density is cell-centered, 
$\varepsilon_r$ and $\varepsilon_\varphi$ are always one fourth
of the coarse grid linear size.
Since the averaging process is performed for each direction 
separately, it is not affected by the metric of the mesh, 
i.e.\ it is performed as if a Cartesian grid were used.

\begin{figure}
 \begin{center}
 \resizebox{0.9\linewidth}{!}{%
 \includegraphics{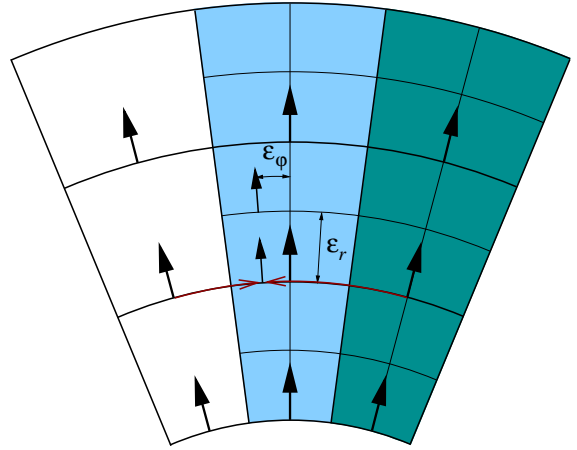}}
 \end{center}
 \caption{Harmonic average of radial velocity component.
          Two cases are shown. In one case the position
          of the velocity component lies on a coarse grid
          border (thick lines), requiring only three coarse
          quantities (one single average). 
          In the other case nine coarse quantities are needed
          as for the density interpolation. The only difference
          is that $\varepsilon_r$ is one half of the 
          radial coarse size.}
 \label{v_harm}
\end{figure}

For the velocity components we have to distinguish two different 
cases (see \Fig{v_harm}): whether the boundary value lies on
a coarse cell border or whether it does not.
In the first case, either $\varepsilon_r$ or
$\varepsilon _\varphi$ is zero. 
Then only three coarse values participate in the average, 
along either the azimuthal or the radial direction.
In the second case 
the interpolation proceeds exactly as
explained for the surface density.
\section{General model design}
\label{Sect:GMD}
The main goal of this study is the investigation of the characteristic
features of the hydrodynamic flow within the Roche lobe of the planet.
This means that we must be able to resolve a characteristic
length, the Hill radius:
\begin{equation}
\Rhill=a\,\left(\frac{\Mp}{3\,\MStar}\right)^{\frac{1}{3}}.
\label{Eq:rhill}
\end{equation}
Moreover, we intend to do that for a variety of planet-to-star
mass ratios.
In order to reach such resolutions, we build a series of 
grid systems having the planet located approximately in the middle,
at each level of the hierarchy. Smaller planets require higher
degrees of nesting.

From now on, we refer to non-dimensional units.
All the lengths are expressed in units of the semi-major
axis $a$. This is constant because the planet moves on a
fixed circular orbit, with radius
\begin{equation}
r\subscr{p}=\frac{a}{(1 + q)}.
\label{plp}
\end{equation}
Masses are in units of the central stellar mass and
time is given in units of the planet's orbital period. 
However, in order to convert them into conventional 
physical units, we assume that $a=5.2$ \AU\
and, as already mentioned, $\MStar=1$ \MSun. 
This implies that one planet orbit takes $11.8$ years.

The whole azimuthal range of the disk is taken into account
by considering a computational domain represented by
$2\,\pi \times [r\subscr{in}, r\subscr{out}]$, where
$r\subscr{in}=0.4$ and $r\subscr{out}=2.5$. 
This is
covered by a $142 \times 422$ mesh (main grid),
allowing a resolution such that
$\Delta r(1) = \Delta \varphi(1) = 0.015$ and constraining
the resolution on each other grid level, according to \Eq{refine_ratio}. 
The size of any higher grid level is $64 \times 64$.
\subsection{Smoothing of the potential}
\label{Subsec:smooth}
The perturbing action of the planet is exerted via its 
gravitational potential $\Phi\subscr{p}$. 
From a numerical point of view, it is usually smoothed
in order to prevent numerical problems near the planet.
Thus, we re-write the denominator of
$\Phi\subscr{p}$ in \Eq{potential} as 
$\sqrt{|\vec{r}-\vec{r}_\mathrm{p}|^2 + \delta^2}$.
However, the smoothing length $\delta$ cannot be the same
all over the grid system because of the different
grid size involved at each level.
Then we use the following grid-dependent length:
\begin{equation}
\delta(l) = \min\left[\frac{\Rhill}{5}, \lambda(l)\right], 
\label{delta}
\end{equation}
where
\begin{equation}
\lambda(l) = \sqrt{\Delta r^2(l) + r_p^2\,\Delta \varphi^2(l)}
          \simeq \sqrt{2}\,\Delta r(l).
\label{lambda}
\end{equation}
The value of constant part, in \Eq{delta}, has become a kind
of standard in single grid simulations (Kley \cite{kley1999}; 
Lubow et al. \cite{lubow1999}; Kley \cite{kley2000}).
This is always used on the main grid whereas the grid dependent part 
always prevails on the highest grid level.
This choice results in a very deep potential
in the immediate vicinity of the planet.  
\subsection{Accretion onto the planet}
\label{Subsec:accret}
\begin{figure}
 \begin{center}
 \resizebox{0.9\linewidth}{!}{%
 \includegraphics{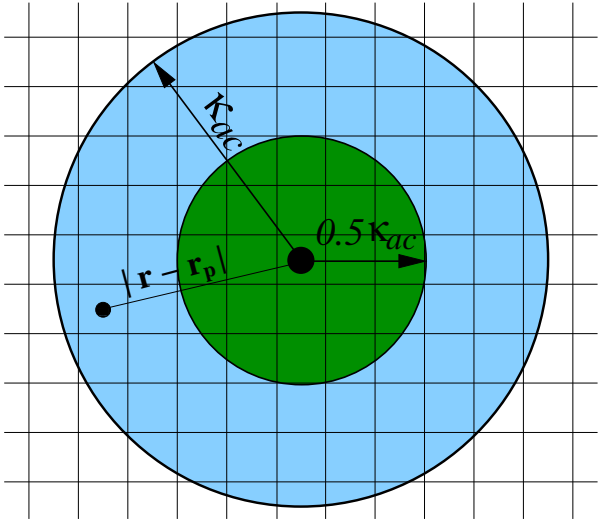}}
 \end{center}
 \caption{Planet mass accretion scheme. Wherever a cell center
          lies in the dark-colored zone, after each integration 
          its density is lowered by an amount $\Delta \Sigma$ such that
          $\Delta \Sigma/\Sigma = 2\,\pi\, \kappa\subscr{ev}\,\Delta t$,
          where $\Delta t$ is the integration time step interval.
          If the cell center falls into the light-colored zone
          $\Delta \Sigma$ is only one third of that value.
          For a typical main grid time interval and $\kappa\subscr{ev}=5$,
          we would get $\Delta \Sigma/\Sigma \approx 1.2\times10^{-2}$.
          Anyway, this evacuation process is performed only on the
          highest grid level for better accuracy.}
 \label{massacc}
\end{figure}

The presence of the planet affects the nearby disk density
also because it can accrete matter. Planet accretion
is accounted for by removing some mass from the region 
defined\footnote{Using this notation, we refer
not only to the radial extent of the accretion region but also to the
fact that the region is centered at $\vec{r}\subscr{p}$.} by 
$|\vec{r}-\vec{r}_\mathrm{p}|\leq \kappa\subscr{ac}$.
Since mass is removed from the system after each integration step,
the evacuation rate depends on an input parameter
$\kappa\subscr{ev}$ as well as on the integration time step.
The details of the accretion process are given in \Fig{massacc}. 
This removal is accomplished only on the highest (finest) hierarchy
level and the removed mass is not added to the dynamical mass of the
planet, but just monitored. A standard value $\kappa\subscr{ev}=5$ is used, 
while we set $\kappa\subscr{ac}$ between
$8.0\times10^{-2}$ and $9.4\times10^{-2}$ \Rhill.
These values are such that only few grid cells, on each side
of the planet, are involved in the gas accretion, 
making it a locally confined process. 
For the Jupiter-mass planet, $\sim 12 \times 12$ cells participate. 
The lowest number of employed cells is $\sim 8\times8$, 
which is used for the smallest planet ($\Mp =1$ \MEarth). 
The largest is $\sim 18 \times 18$ and is adopted for a planet 
with $\Mp =0.5$ \MJup\ (160 \MEarth). 
This circumstance is due to the high numerical resolution of the model
(see Table~\ref{models}).

Note that the sphere of influence of the accretion process consists
of a region typically $10^{-1}$ \Rhill\ which is smaller than the radius
of the protoplanet, which is assumed to fill its Roche lobe during
its growth phase (Bodenheimer \& Pollack \cite{bodenheimer1986};
Tajima \& Nakagawa \cite{tajima1997}). As the present study 
does not take the energy equation into account, it precludes
a detailed treatment of the internal structure of the protoplanet
(see Wuchterl et al. \cite{wuchterl2000} and references therein, and also
Fig.~\ref{core_match} below),
hence the inferred accretion rates may still be unreliable.
\subsection{Initial and boundary conditions}
\begin{figure}
 \begin{center}
 \resizebox{1.0\linewidth}{!}{%
 \includegraphics{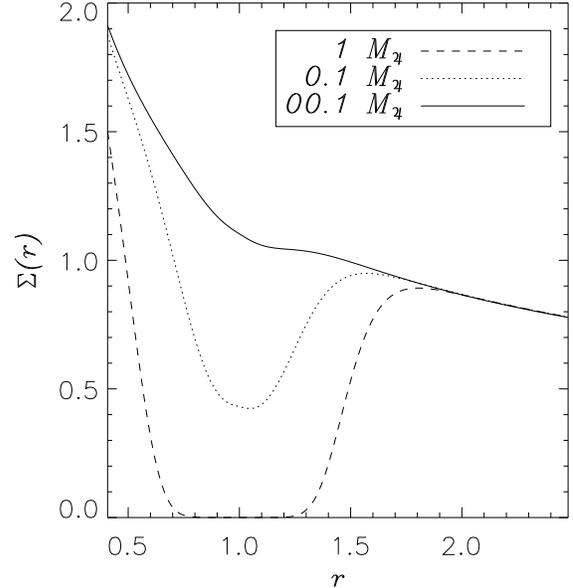}}
 \end{center}
 \caption{Initial surface density distribution for $q=10^{-3}$,
          $q=10^{-4}$ and $q=10^{-5}$. 
          $\Sigma(r,\varphi)=\Sigma(r)$ at $t=0$. 
          The gap greatly reduces when $q$ gets small enough.} 
 \label{sigma0}
\end{figure}

The initial density distribution is proportional to $r^{-1/2}$.
However, we superimpose to this an axisymmetric gap around the planet,
obtained by an approximate balance of the viscous torque and the
gravitational torque due to the planet
(P. Artymowicz, private communication). 
Figure~\ref{sigma0} shows the surface density, at $t=0$, 
for some selected planet masses.
The initial velocity field is that of a Keplerian disk.

As boundary conditions, periodicity is imposed at
$\varphi=0$ and $\varphi=2\,\pi$.
We allow matter to flow out of the computational domain at the inner
radial border ($r\subscr{in}$) whereas we set reflective boundary 
conditions at the outer radial border ($r\subscr{out}$). 
The angular velocity is set equal to the unperturbed Keplerian value
$\Omega\subscr{Kep} = \sqrt{G\,\MStar/\,r^3}$, 
both at $r\subscr{in}$ and $r\subscr{out}$.

For low-mass planets ($\Mp\lesssim 10$ \MEarth), 
boundary conditions should not affect much the system 
evolution because density waves damp before reaching 
$r=r\subscr{out}$ and are very weak when they reach
$r=r\subscr{in}$. 
For more massive planets, some reflection is seen at the 
radial outer border.
Further, the torque exerted by the planet pushes the 
inside-orbit material inwards. As a result, because of the
open boundary at  $r=r\subscr{in}$, the inner disk is partially
cleared. The higher the planet mass is the stronger 
these effects appear.     
\subsection{Model specifications} 
\label{Sect:MS}
In this paper, we are mainly interested in investigating how disk-planet
interactions change by varying $q$, the planet-to-star mass ratio.
Table~\ref{models} summarizes the values of $q$ used
in different models, along with the number of grid levels employed.
For reference, a measure of the linear resolution $\xi$, in units
of the Hill radius \Rhill,
is given for the highest level as well. 
Some models have different prescriptions than the ones
outlined above, as specified in the Table~\ref{models}.

Few models may deserve some comments.
Model \elen2 and \wpro1 aim at checking whether results from \ciro1 and 
\ciro2 (respectively) are resolution-dependent. 
This test is negative, as we show in the next sections.
Since the planet position $(r\subscr{p},\varphi\subscr{p})$
can fall anywhere within a grid cell
(according to the value of $q$ in Eq.~\ref{plp} and 
to the definition of the grid), 
some asymmetries could arise. These might have some effects on the
finest levels, due to the small value of the smoothing
factor $\delta$.
In order to achieve a complete symmetry, in the model
\gino1 the planet is placed at the corner of a main grid cell
(i.e.\ the intersecting point of four grid cells).
This property is such that, on every other gird level,
the planet always sits on the cross-point of four grid cells.
\ciro3 and its counterpart, \gino1, give almost identical
results. 
Since in the Jupiter-mass case the inner-disk is greatly depleted,
model \elen1 was run to evaluate the influence of its presence
on the gravitational torque. 
For this reason, in such model we prevent matter from draining
out of the inner radial border by setting reflective boundary 
conditions. 
\begin{table}
 \caption{Model specific parameters: 
          $q = \Mp/\,\MStar$; 
          $ng$ is the number of grid levels; 
          $\xi= \Delta r(ng)/\Rhill$ is the normalized
          grid resolution on the finest level.
          Each model is let evolve, at least, till 200 orbits.}
 \label{models}
 \begin{center}
 \begin{tabular}{lcccc}
 \hline
 \hline
 Model   &  $q$                  & $ng$ & $\xi$                 & Notes \\
 \hline
 \ciro1  &  $1.0 \times 10^{-3}$ & $5$  & $1.3 \times 10^{-2} $ &      \\ 
 \ciro2  &  $1.0 \times 10^{-4}$ & $6$  & $1.5 \times 10^{-2} $ &      \\ 
 \ciro3  &  $1.0 \times 10^{-5}$ & $7$  & $1.6 \times 10^{-2} $ &      \\ 
 \pepp1  &  $3.0 \times 10^{-6}$ & $7$  & $2.3 \times 10^{-2} $ &      \\  
 \pepp2  &  $1.5 \times 10^{-5}$ & $7$  & $1.4 \times 10^{-2} $ &      \\ 
 \pepp3  &  $3.0 \times 10^{-5}$ & $7$  & $1.1 \times 10^{-2} $ &      \\ 
 \pepp4  &  $6.0 \times 10^{-5}$ & $7$  & $8.6 \times 10^{-3} $ &      \\ 
 \wpro1  &  $1.0 \times 10^{-4}$ & $7$  & $7.3 \times 10^{-3} $ &      \\
 \wpro2  &  $1.0 \times 10^{-4}$ & $6$  & $1.5 \times 10^{-2} $ &%
         $^{(a)}$                                                      \\ 
 \gino1  &  $1.0 \times 10^{-5}$ & $7$  & $1.6 \times 10^{-2} $ &%
         $^{(b)}$                                                      \\ 
 \gino2  &  $2.0 \times 10^{-4}$ & $6$  & $1.2 \times 10^{-2} $ &      \\
 \gino3  &  $5.0 \times 10^{-4}$ & $6$  & $8.5 \times 10^{-3} $ &      \\
 \elen1  &  $1.0 \times 10^{-3}$ & $5$  & $1.3 \times 10^{-2} $ &%
         $^{(c)}$                                                      \\
 \elen2  &  $1.0 \times 10^{-3}$ & $6$  & $6.8 \times 10^{-3} $ &      \\
 \hline
 \multicolumn{5}{l}{$^{(a)}$ Same as \ciro2, but $\kappa\subscr{ev}=10$;}\\ 
 \multicolumn{5}{l}{$^{(b)}$ Same as \ciro3, but the planet is symmetrically}\\
 \multicolumn{5}{l}{$~~~$ placed with respect to the grid cells.}\\
 \multicolumn{5}{l}{$^{(c)}$ Same as \ciro1, but reflective boundary conditions}\\ 
 \multicolumn{5}{l}{$~~~$ are set at  $r=r\subscr{in}$.}
 \end{tabular}
 \end{center}
\end{table}
\section{Results}
\label{Sect:results}
Hereafter we mainly discuss three models, 
namely \ciro1, \ciro2 and \ciro3.
We concentrate on them because they cover
a mass range from $1$ \MJup\ down to $3.2$ \MEarth.
Nevertheless, whenever required by our discussion,
we mention other particular models.
Some results, concerning the whole set of models
given in Table~\ref{models}, are presented as well.
\subsection{Overall flow structure}
\label{Subsec:ofs}
\begin{figure*}
\begin{center}
\resizebox{0.85\textwidth}{!}{%
\includegraphics[bb= 0 35 480 480,clip]{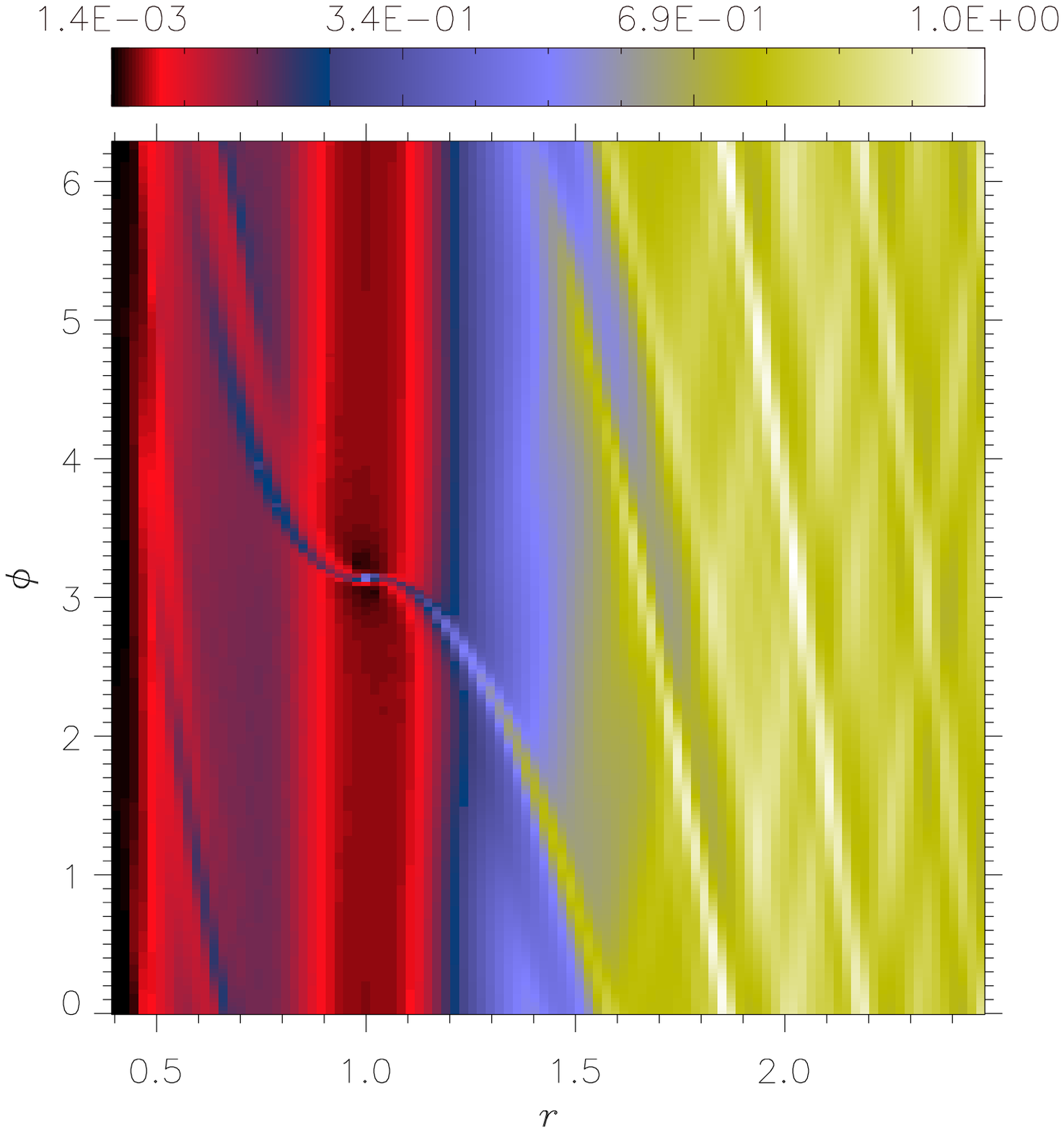}%
\includegraphics[bb=30 35 480 480,clip]{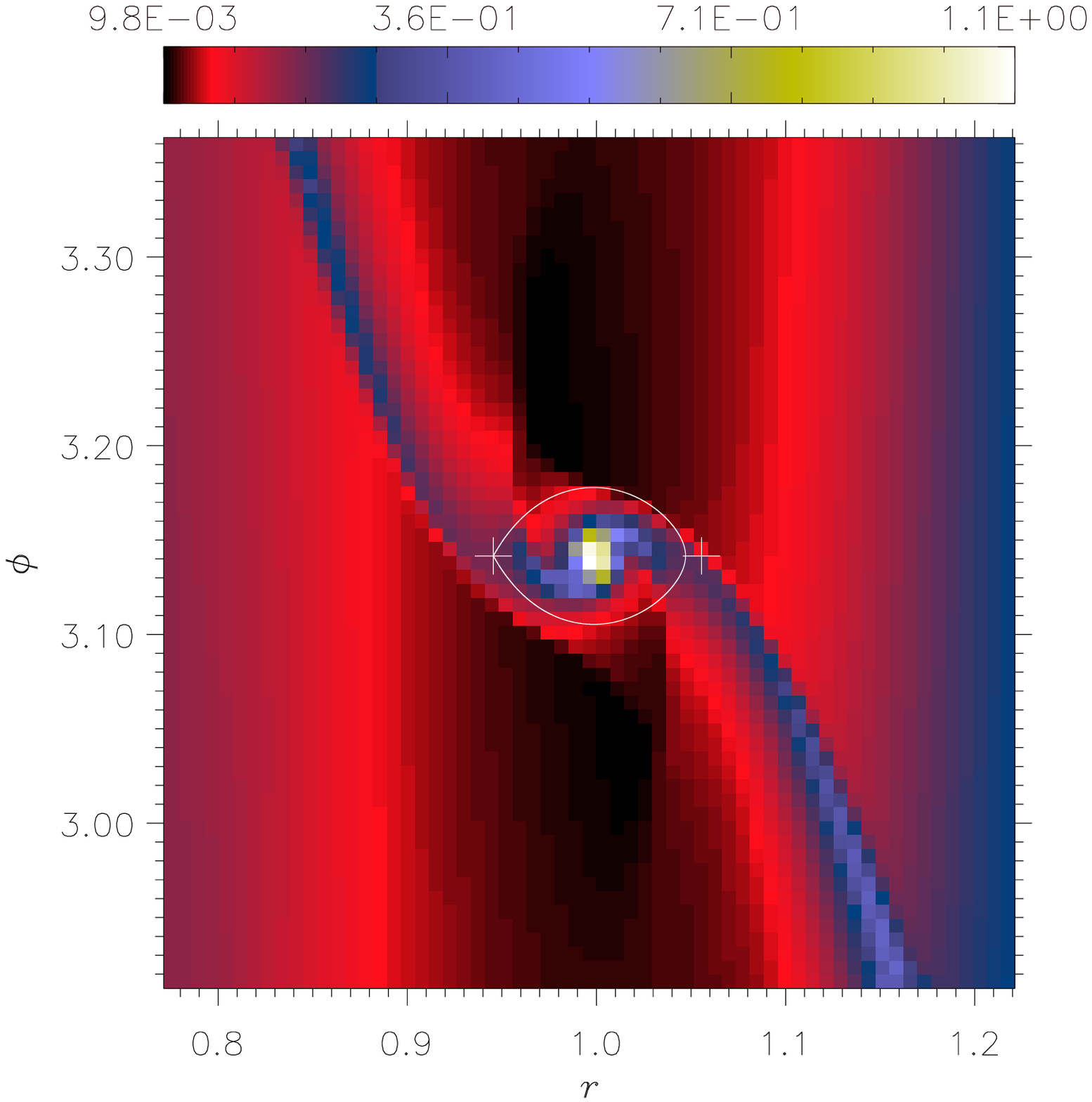}}
\resizebox{0.85\textwidth}{!}{%
\includegraphics[bb= 0 35 480 480,clip]{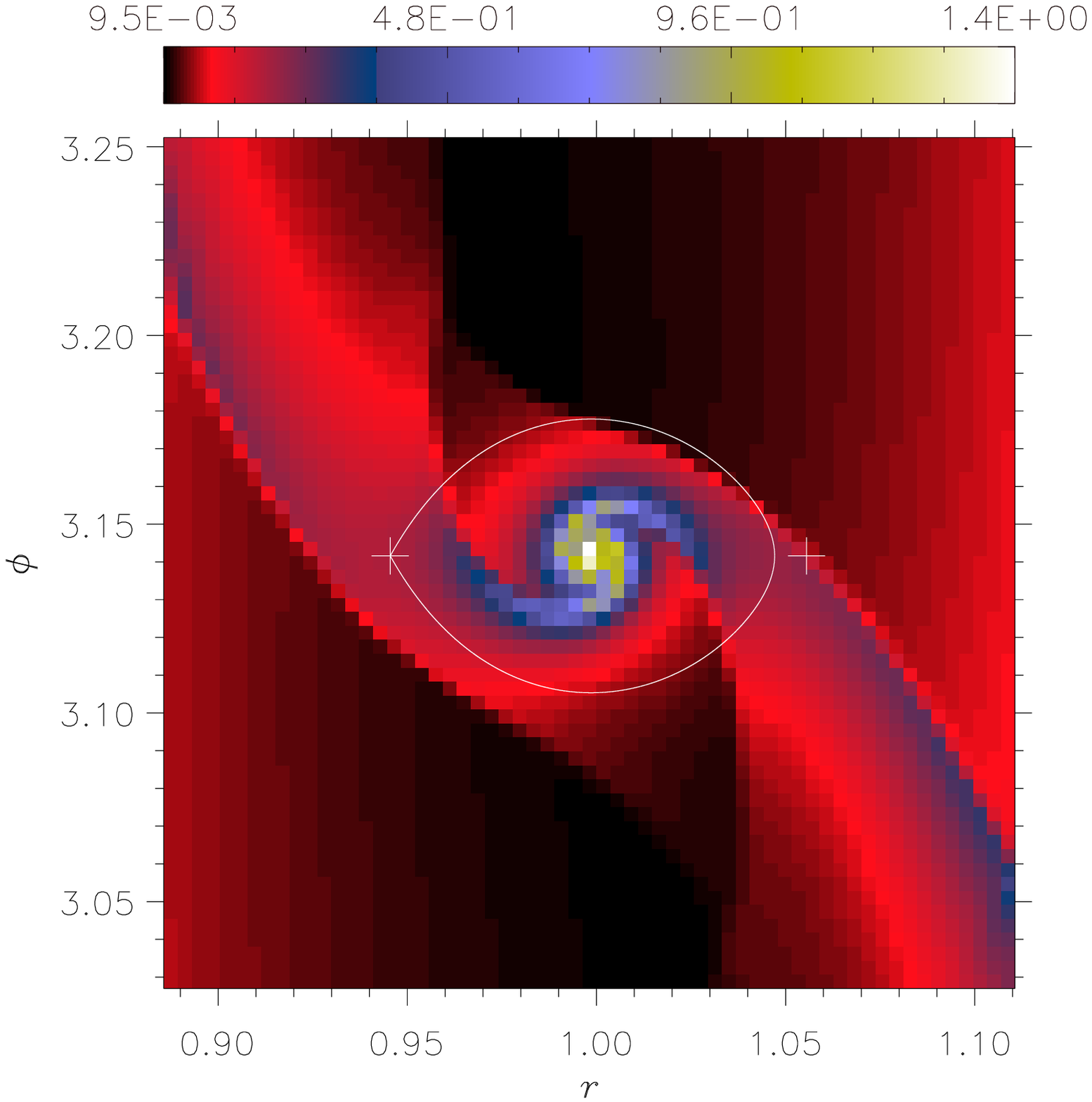}%
\includegraphics[bb=30 35 480 480,clip]{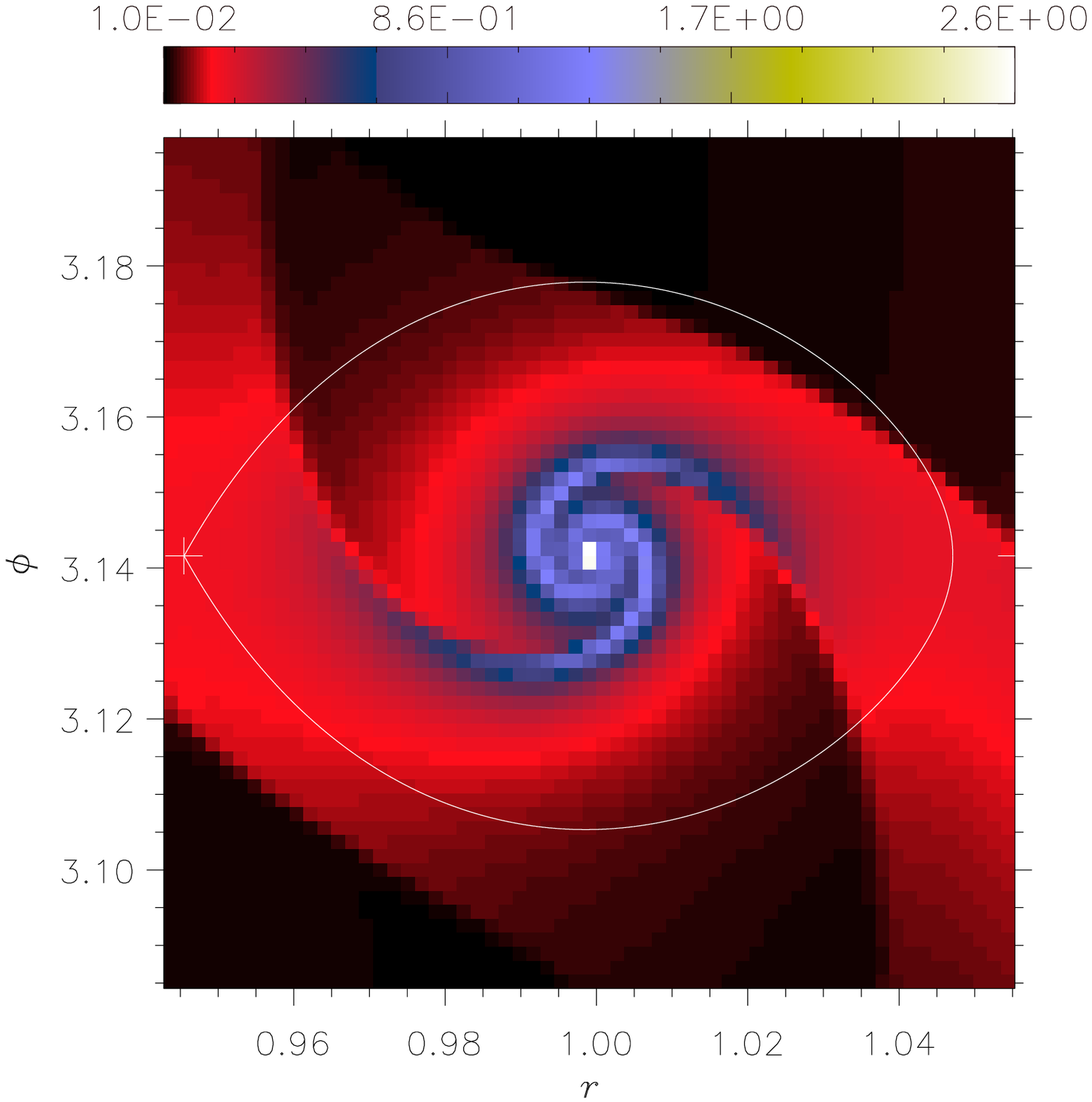}}
\resizebox{0.85\textwidth}{!}{%
\includegraphics[bb= 0  0 480 480,clip]{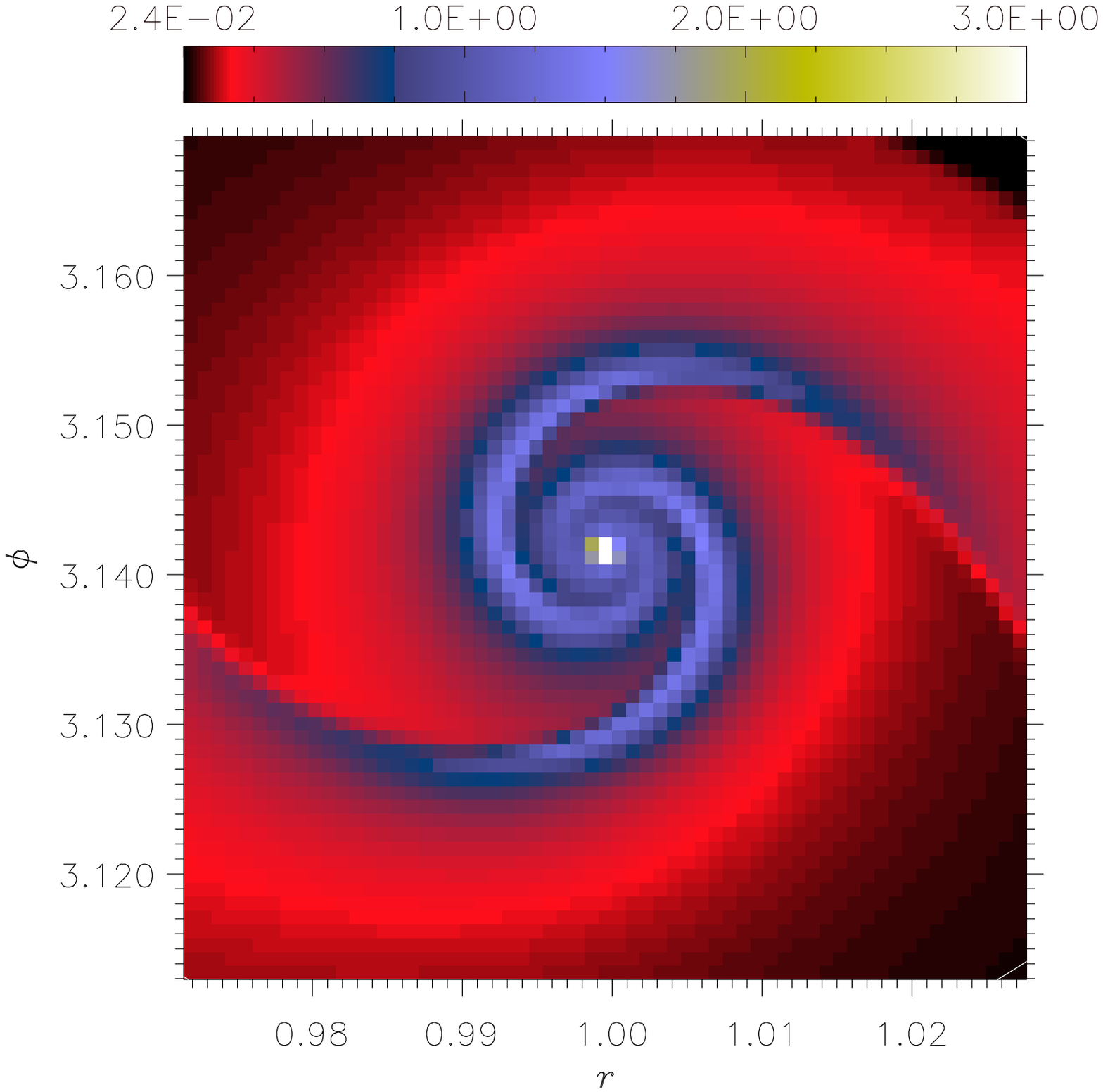}%
\includegraphics[bb=25  0 480 480,clip]{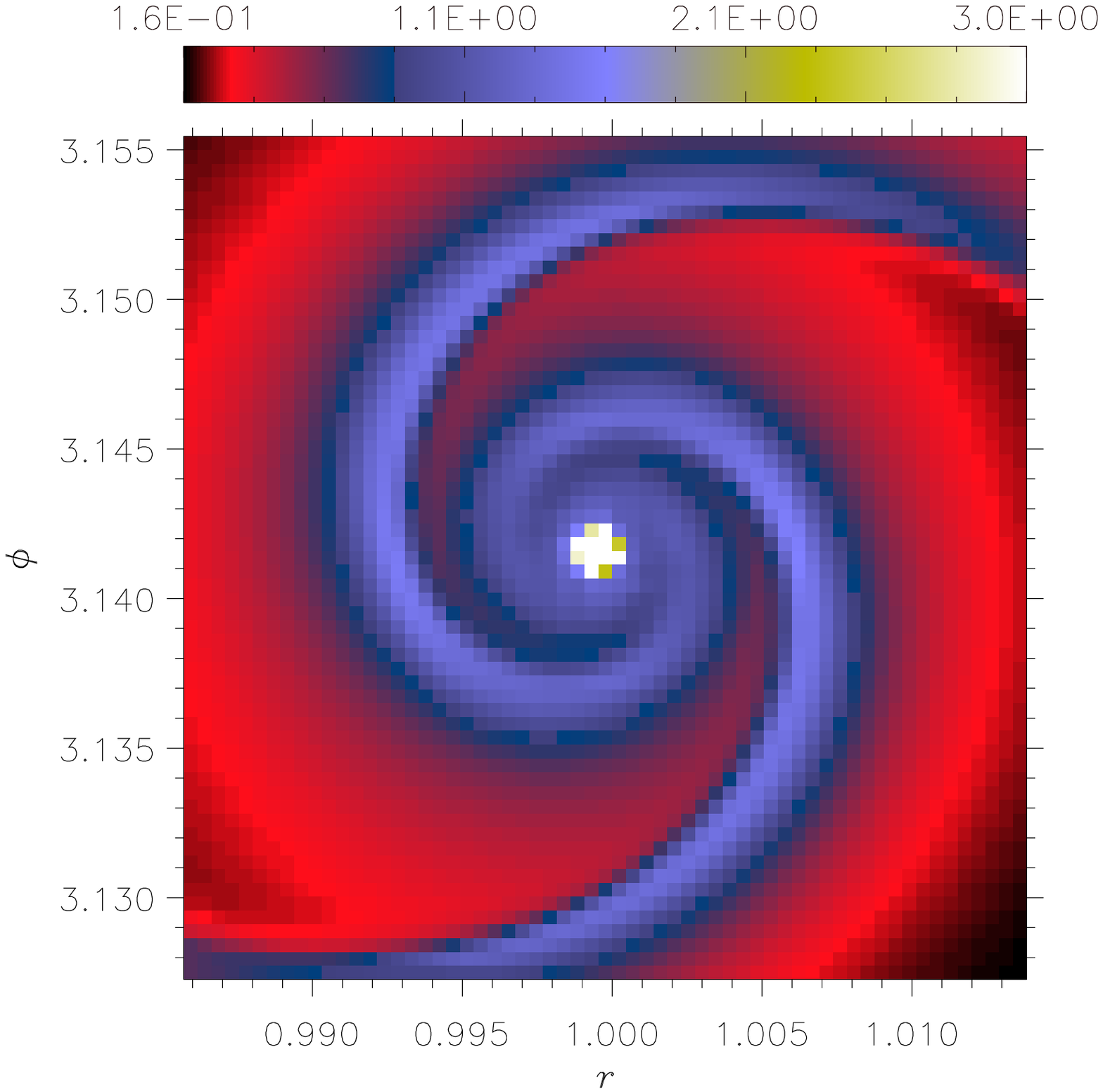}}
\end{center}
\caption{Overview of the surface density on the whole grid system
         for model \gino3 ($\Mp = 0.5$ \MJup), after 210 orbits.
         The over-plotted curve represents the Roche lobe of the 
         restricted three-body problem. Plus signs indicate the positions
         of L1 and L2 Lagrangian points, respectively on the left and 
         the right of the planet. The color scale is linear. 
         In order to enhance the contrast, in the two bottom panels 
         $\Sigma$ is cut at a value equal to 3 (saturated area).}
\label{img:overview}
\end{figure*}

Large-scale interactions (whose effects extend over
$2\,\pi$ in azimuth and cover a large radial extent)
of a Jupiter-mass planet with the surrounding environment,
have already been treated numerically in a number of papers 
(Artymowicz \cite{Artymowicz1992}; Kley \cite{kley1999}; 
Bryden et al. \cite{bryden1999}; Kley \cite{kley2001}).
An example of large-scale features can be seen in 
\Fig{img:overview}, where the comprehensive result
of a nested-grid computation is displayed. 

Planets with a lower value of $q$ should perturb the disk less
and have a weaker large-scale impact on it.
For these reasons we discuss only the medium
($|\vec{r}-\vec{r}_\mathrm{p}| \sim \Rhill$) 
and small
($|\vec{r}-\vec{r}_\mathrm{p}| \ll \Rhill$) 
scale effects of such interactions.

Nonetheless, it's worthwhile noticing that large-scale structures 
are clearly visible in our smallest mass models.
In \ciro3 ($\Mp=3.2$ \MEarth), a trailing density wave, 
emanating from the planet, spirals around the star for 
about $4\,\pi$, vanishing approximately at $r=2$.
In \pepp1 ($\Mp=1$ \MEarth), a similar feature spirals 
for almost $2\,\pi$, disappearing at $r=1.5$.
Although we did not investigate this issue any deeper, it may happen that
results from local simulations could be influenced by not taking into
account entirely such global features.

\begin{figure*}
\begin{center}
\resizebox{0.85\textwidth}{!}{%
\includegraphics[bb= 0 35 480 480, clip]{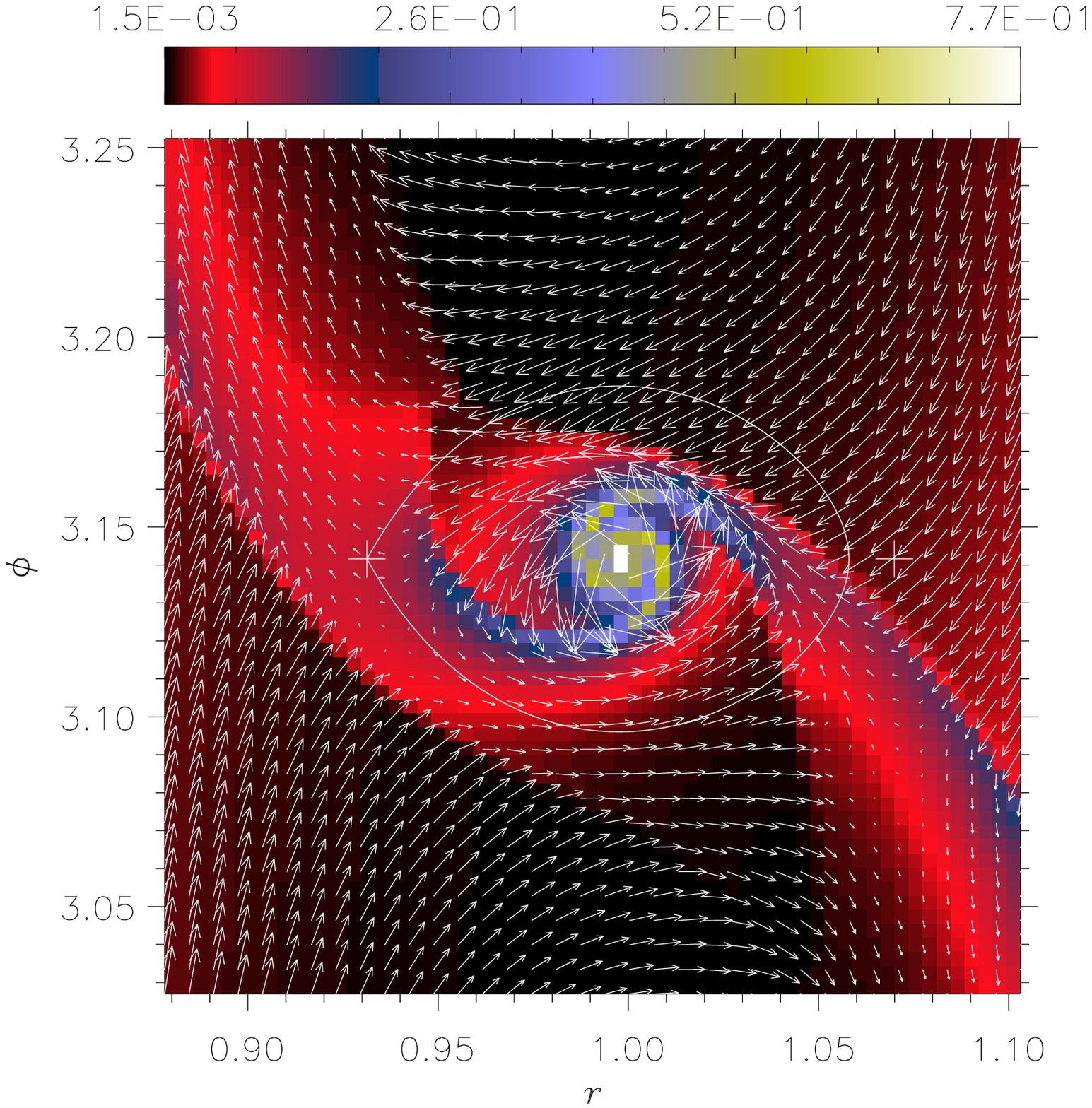}%
\includegraphics[bb=30 35 480 480, clip]{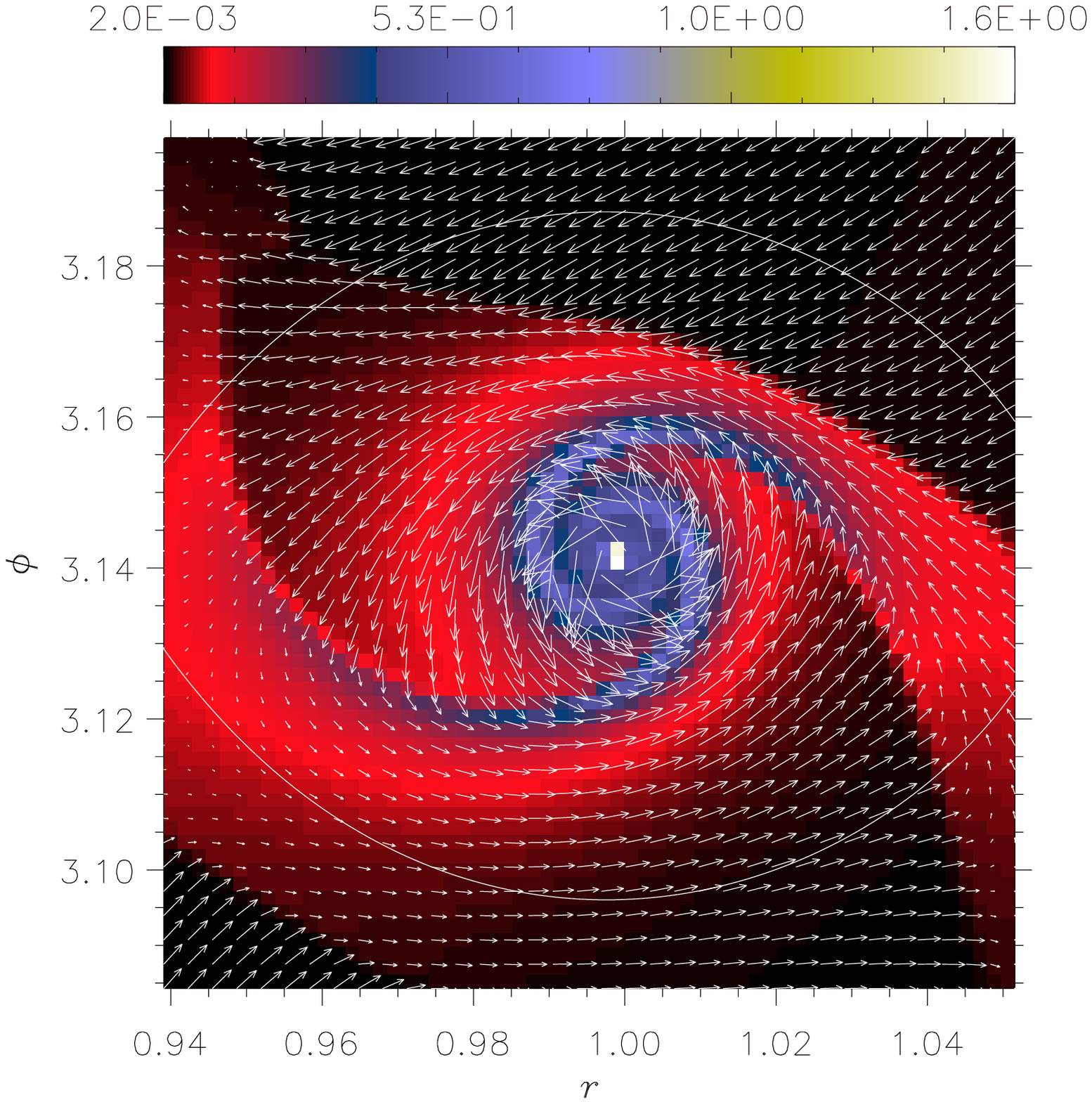}}
\resizebox{0.85\textwidth}{!}{%
\includegraphics[bb= 0 35 480 480, clip]{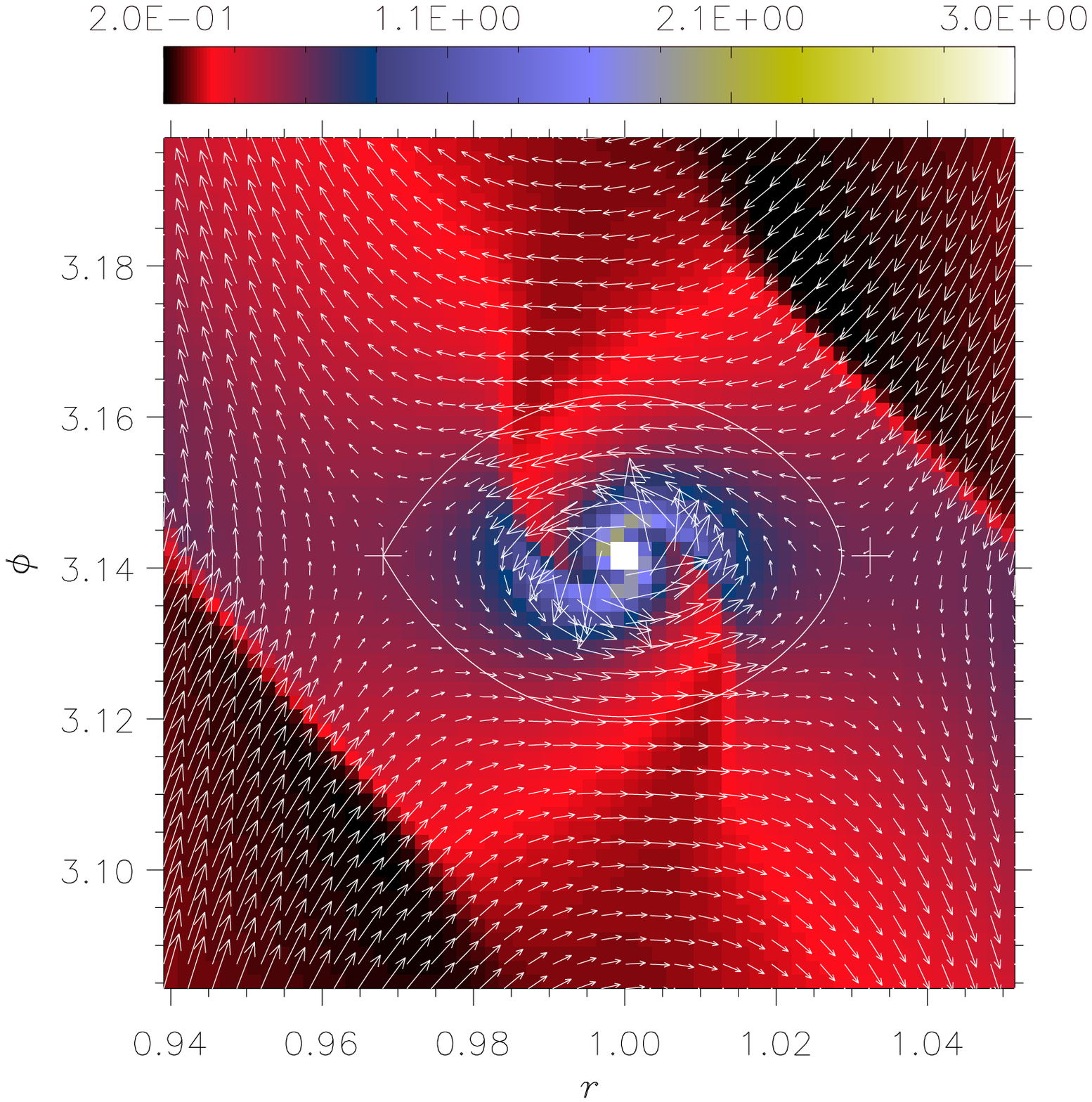}%
\includegraphics[bb=25 35 480 480, clip]{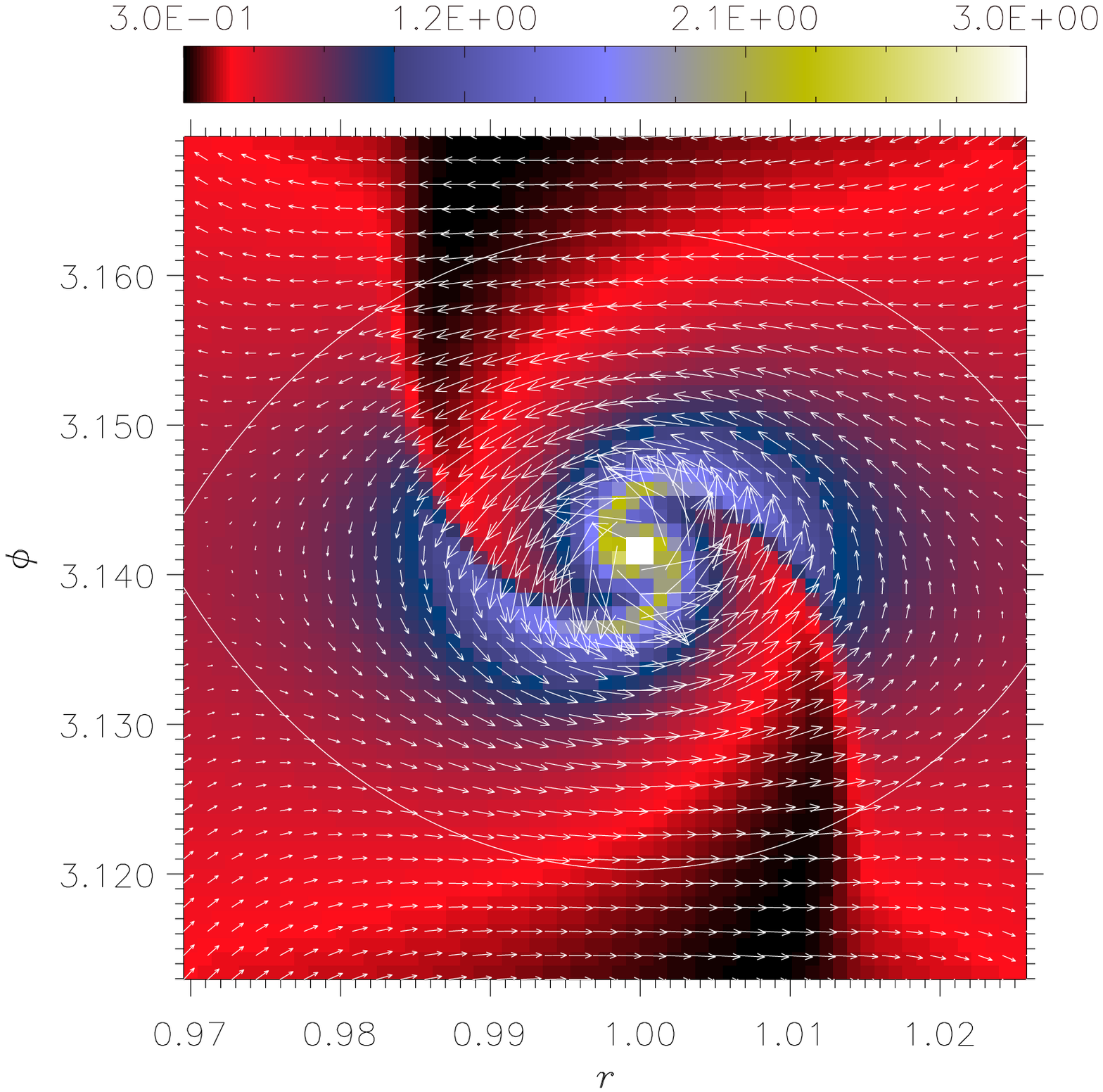}}
\resizebox{0.85\textwidth}{!}{%
\includegraphics[bb= 0  0 480 480, clip]{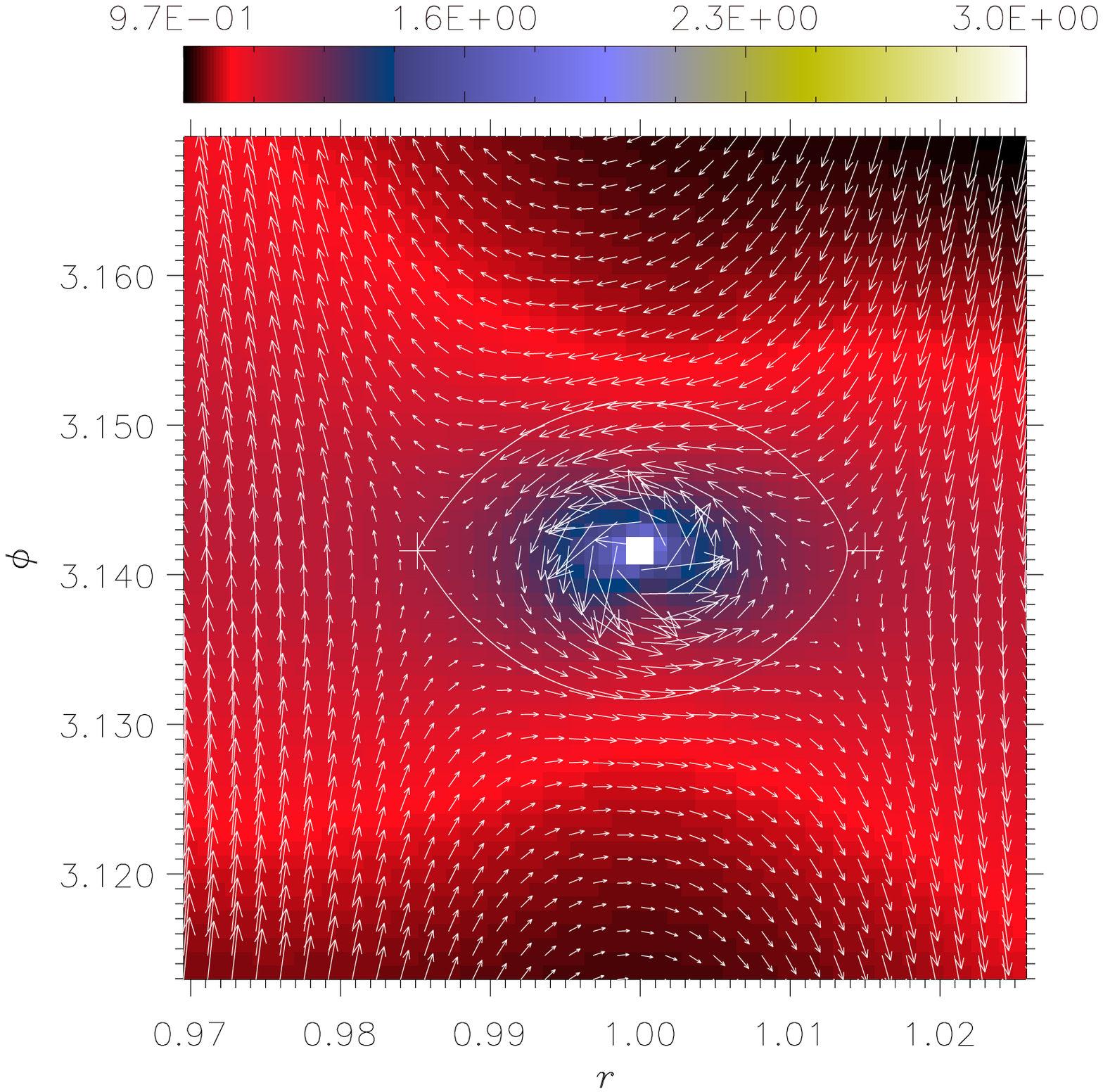}%
\includegraphics[bb=25  0 480 480, clip]{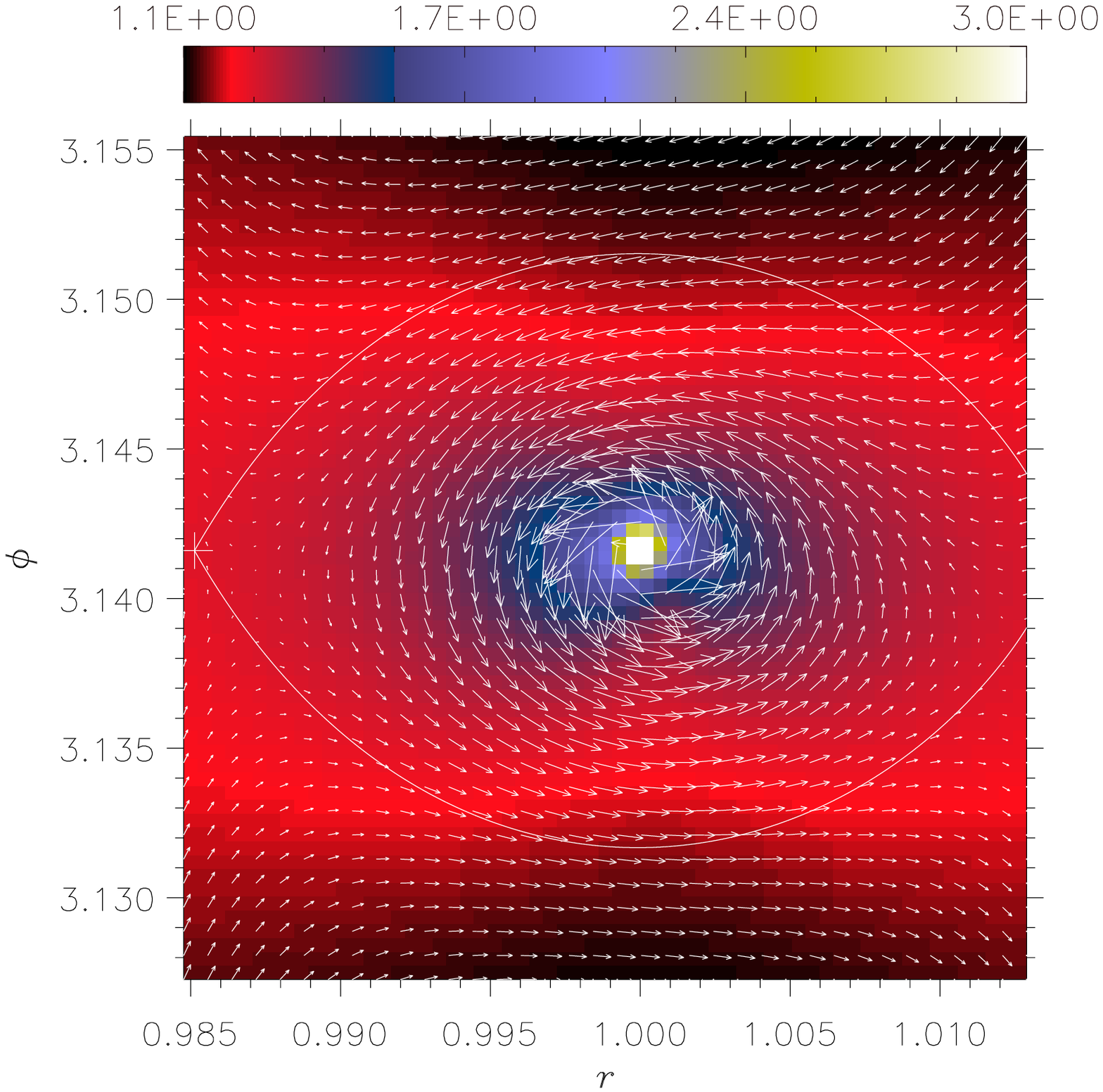}}
\end{center}
\caption{Surface density and velocity field for three selected models.
         \textbf{Top row}. \ciro1: grid level $l=3$ (left panel) and 
         $l=4$ (right panel) at $t=375$ orbits. 
         \textbf{Middle row}. \ciro2: grid level $l=4$ (left panel) and 
         $l=5$ (right panel) at $t=325$ orbits.
         \textbf{Bottom row}. \ciro3: grid level $l=5$ (left panel) and 
         $l=6$ (right panel) at $t=225$ orbits. 
         To avoid too much confusion, only $40$\% of the velocity field 
         vectors are drawn.
         The over-plotted curve and plus signs have the same meaning as
         in \Fig{img:overview}.}
\label{my_img}
\end{figure*}

In \Fig{my_img} the surface density is shown along with
the velocity field for the three \ciro-models.
As a reference, the Roche lobe 
(of the relative three-body problem) is over-plotted. 
From the upper row of this figure ($\Mp = 1$ \MJup) 
we can see the patterns of the two main spirals (left panel). 
They reach the Roche lobe, but do not enter it.
In fact, they are replaced by two streams of material 
which start from two points (located at $r=0.94$, $\varphi=3.12$
and $r=1.07$, $\varphi=3.07$, respectively), 
where the flow is nearly at rest with respect 
to the planet\footnote{They are designated
as ``X-points'' by Lubow et al. (\cite{lubow1999})}. 
Each of them enters the Roche lobe, encircles the planet
and hits the other one on the opposite side.
As a result of the collision, the material is shocked and the
flow is redirected towards the planet.
Hence, these streams assume the form of two spirals,
winding around the planet (right panel) for $2\,\pi$.
That such smaller scale spirals are detached from the main ones
can be inferred from the direction of the flow. 
Along the main spirals the material follows the disk rotation
around the star, moving away from the planet. 
Along the small ones the gas orbits the planet. 
In fact, they represent the outstanding features 
of a \textit{circumplanetary disk}.
A more detailed description of the flow regions around and inside
the Roche lobe, concerning a Jupiter-mass planet, can be found
in Lubow et al. (\cite{lubow1999}).

The case $\Mp = 32$ \MEarth\ (\Fig{my_img}, middle row)
has many analogies to the previous one.
This planet is able to open a gap in the disk, as a permanent
feature. However, it is neither so wide 
(the base width is $0.15$ against $0.4$) 
nor so deep ($40$\% against $0.5$\% of the maximum surface density) 
as it is for a Jupiter-mass planet.
The overall behavior of the matter entering the Roche lobe
is very similar (left panel).
The up-stream disk material, relative to the nearest X-point,
reaches it, inverts partially the direction of its motion and
flows into the Roche lobe.
The gas stream penetrating from the left X-point turns about
the planet, at $\varphi < \varphi\subscr{p}$, and collides
with the stream incoming from the other X-point, generating
the upper spiral arm (at  $\varphi > \varphi\subscr{p}$).
However, here the locations, from which these gas streams depart
($r=0.97$, $\varphi=3.14$
and $r=1.03$, $\varphi=3.135$, respectively), 
lie closer to the L1 and L2 points.
The circumplanetary spirals are less twisted around the planet
than before. They wrap around it for an angle $\pi$ (right panel). 

For the less massive planet, $\Mp = 3.2$ \MEarth, the situation
is somewhat different (bottom row of \Fig{my_img}). 
In fact, within the Roche lobe, the signs of spiral fronts are 
very feeble, though some traces can still be seen.  
They assume the shape of a bar-like structure which extends, 
for $0.3$ \Rhill,
from side to side of the planet at $\varphi\simeq \varphi\subscr{p}$. 
As indicated by the velocity field, the circumplanetary disk roughly 
occupies the entire Roche lobe (right panel).

Taking into account the other models as well, 
the following scenario can be proposed:
the lower the value of $q$, the shorter
and straighter the circumplanetary spirals become. 
For example, in model \pepp2, they track a tilde-like 
pattern, extending for a total length of $0.4$ \Rhill.
\subsection{Density in the planet's environment}
\begin{figure}
 \begin{center}
 \includegraphics[bb=0 18 470 470, clip, width=0.85\linewidth]{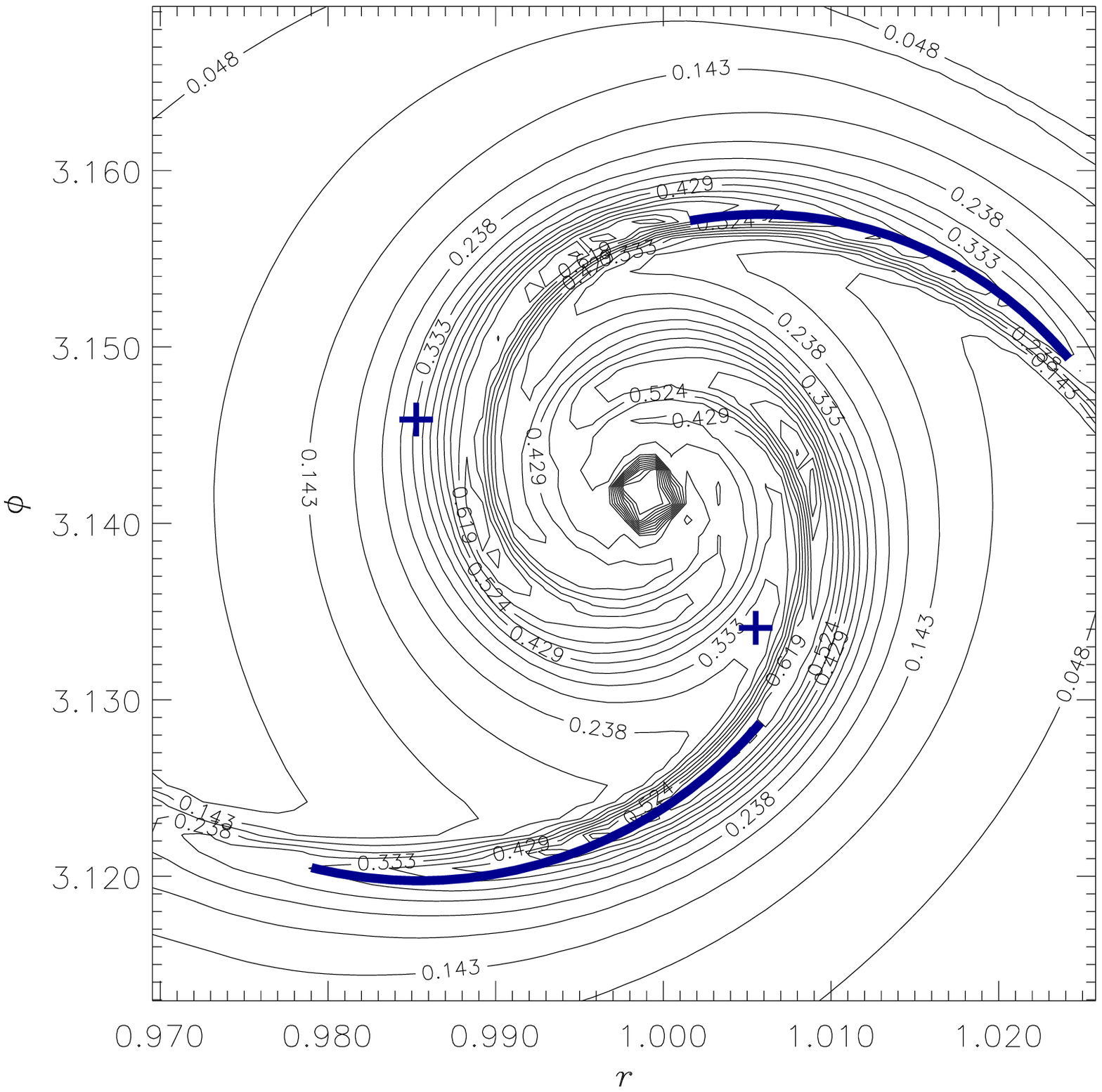}
 \includegraphics[bb=0 18 470 470, clip, width=0.85\linewidth]{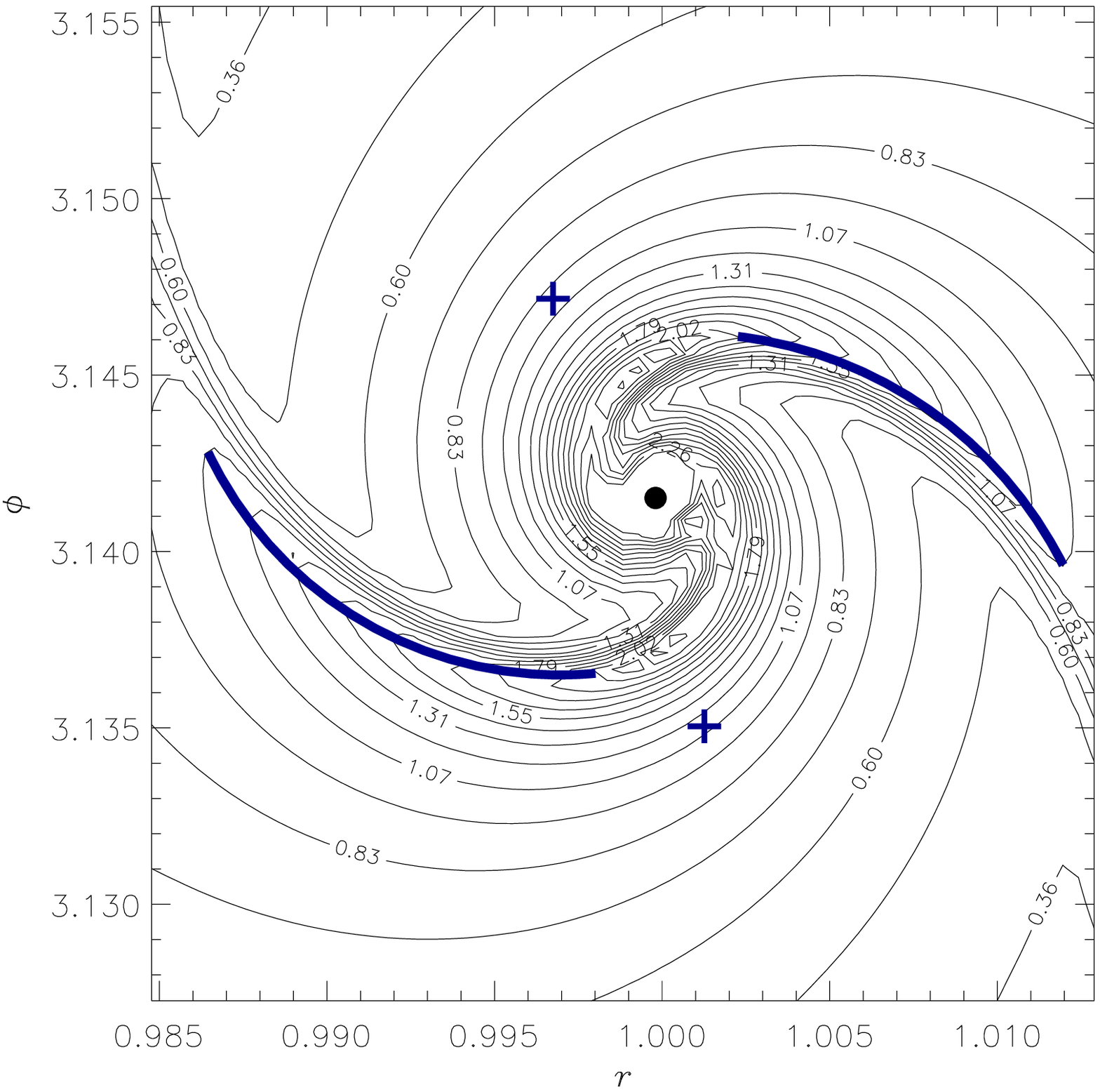}
 \includegraphics[width=0.85\linewidth]{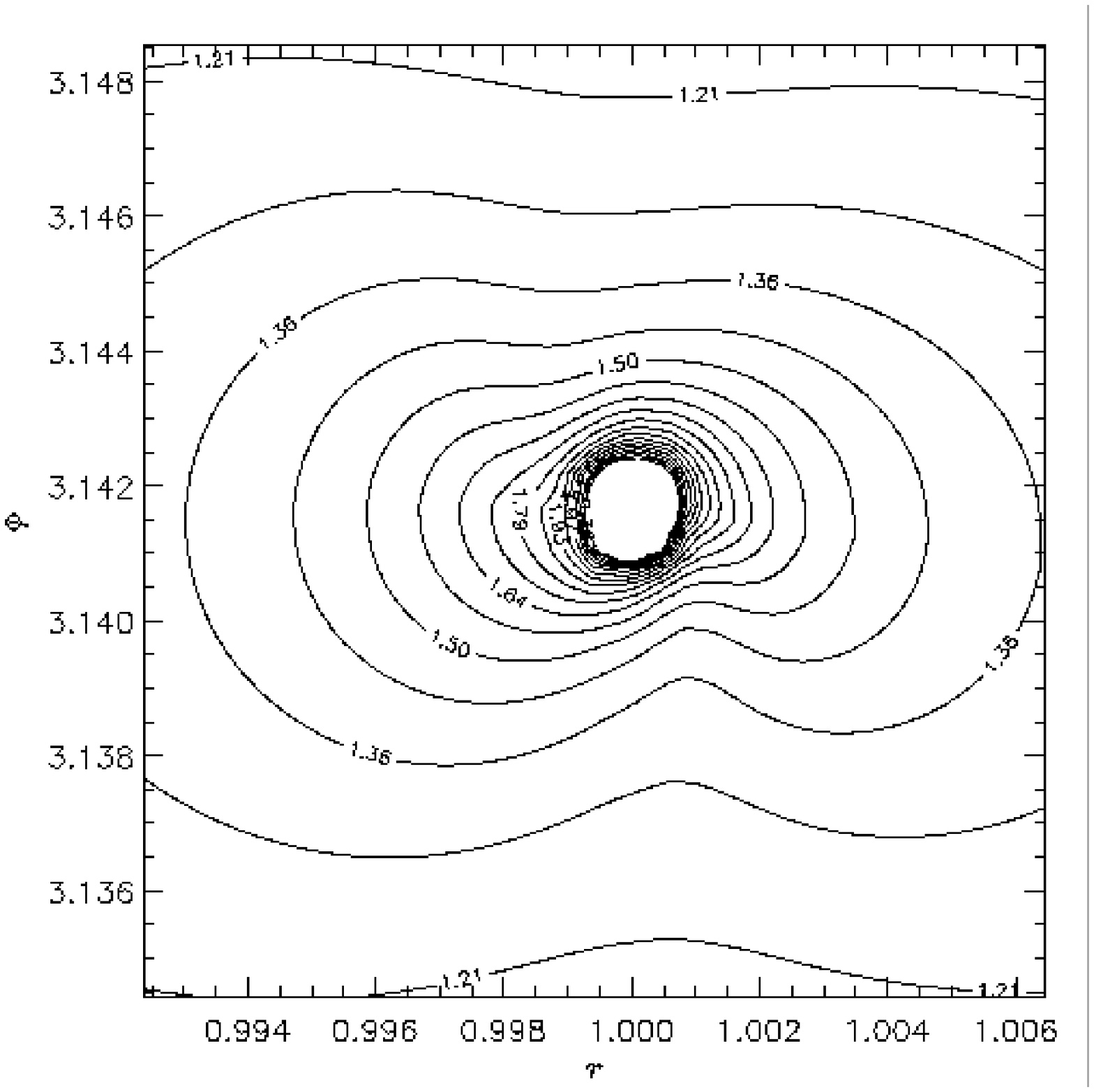}
 \end{center}
 \caption{Surface density contours on the finest grid level. 
          The evolutionary time is the same as in \Fig{my_img}.
          \textbf{Top}: \ciro1. \textbf{Middle}: \ciro2. 
          \textbf{Bottom}: \ciro3.
          See the text for an explanation
          of the thick bow-lines and the plus signs.}
\label{deep_cont}
\end{figure}

So far we have described qualitatively the main hydrodynamic 
structures which are present, near the planet,
on the length scale of a Hill radius.
We shall now discuss, more quantitatively, two aspects
of the surface density, observable on shorter length scales 
($\lesssim 0.5$ \Rhill). Both of them have repercussions on the
torque exerted on the planet by the close-by matter, as explained later.
First of all we should notice that $\Sigma(r,\varphi)$ is not
perfectly symmetric with respect to the planet, in none of
our reference models.
Such property can be checked by means of an accurate examination of 
\Fig{deep_cont}. In this figure
the contour lines of the surface density (i.e.\ lines on which
$\Sigma$ is constant) are plotted, for the finest grid levels $l=ng$.
In case of \ciro1 (top panel, \Fig{deep_cont}), one can 
see that the right-hand arm ($\varphi > \varphi\subscr{p}$) 
twists towards the planet more than the other one does. 
The curvature\footnote{The curvature of a circle of radius $R$
is $1/R$.} 
ratio of the more external spiral parts 
(indicated by thick lines in the figure) is $0.9$.
Furthermore, the right-hand arm lies closer to the planet
than the left-hand one, as can be evaluated from
the positions of the arc centers (marked with plus signs).
Contour line asymmetries are fainter in the case of 
\ciro2 (middle panel, \Fig{deep_cont}).
However, a quantitative analysis shows that the external ridges
of the spirals (thick lines) have slightly different curvatures. 
Moreover, their centers (plus signs) are not aligned 
with the planet position (represented by a big dot). 
In particular, if we consider the contours between $1.03$ and
$2.02$ (arbitrary units), the right-hand ridge is a little
straighter than the left-hand one. 
In fact, the ratio between their curvatures is $0.93$.
Also in this case, the arc centers indicate that external
parts of the right-hand arm are a little nearer to 
the planet than those of the left-hand arm.   
Indeed, surface density asymmetry is evident, 
in the case of \ciro3 (bottom panel, \Fig{deep_cont}).
At $r\simeq 1.001$ and  $\varphi < \varphi\subscr{p}$, 
the density is systematically lower than it is on the opposite side
($r\simeq 0.999$, $\varphi > \varphi\subscr{p}$). 
Following the shape of density contours, it is possible to track
what remains of the spiral arms.
The pre-shock material conveys the more convex form to these lines,
at $r\simeq 0.999$ and $r\simeq 1.001$ 
\subsubsection{The core}
\label{Subsec:core}
Finally, we would like to mention what happens, in our models,
at very short distances from the planet. We described, in 
\Sect{Subsec:accret}, how material is removed from the
neighborhood of the planet. 
This is usually a small fraction of the available matter,
during each integration time step of the main grid.
If $\Delta M\superscr{ev}$ is the evacuated amount of mass,
then typically
\[
\frac{\Delta M\subscr{ev}}{M}\approx
\frac{\Delta \Sigma}{\Sigma}\approx 10^{-2}
\]
(see the caption of \Fig{massacc}).
This choice accounts for the fact that a planet should not be able
to accept, very rapidly, all the material the surrounding environment 
can offer (Wuchterl \cite{wuchterl1993}).

\begin{figure}
 \begin{center}
 \includegraphics[width=\linewidth]{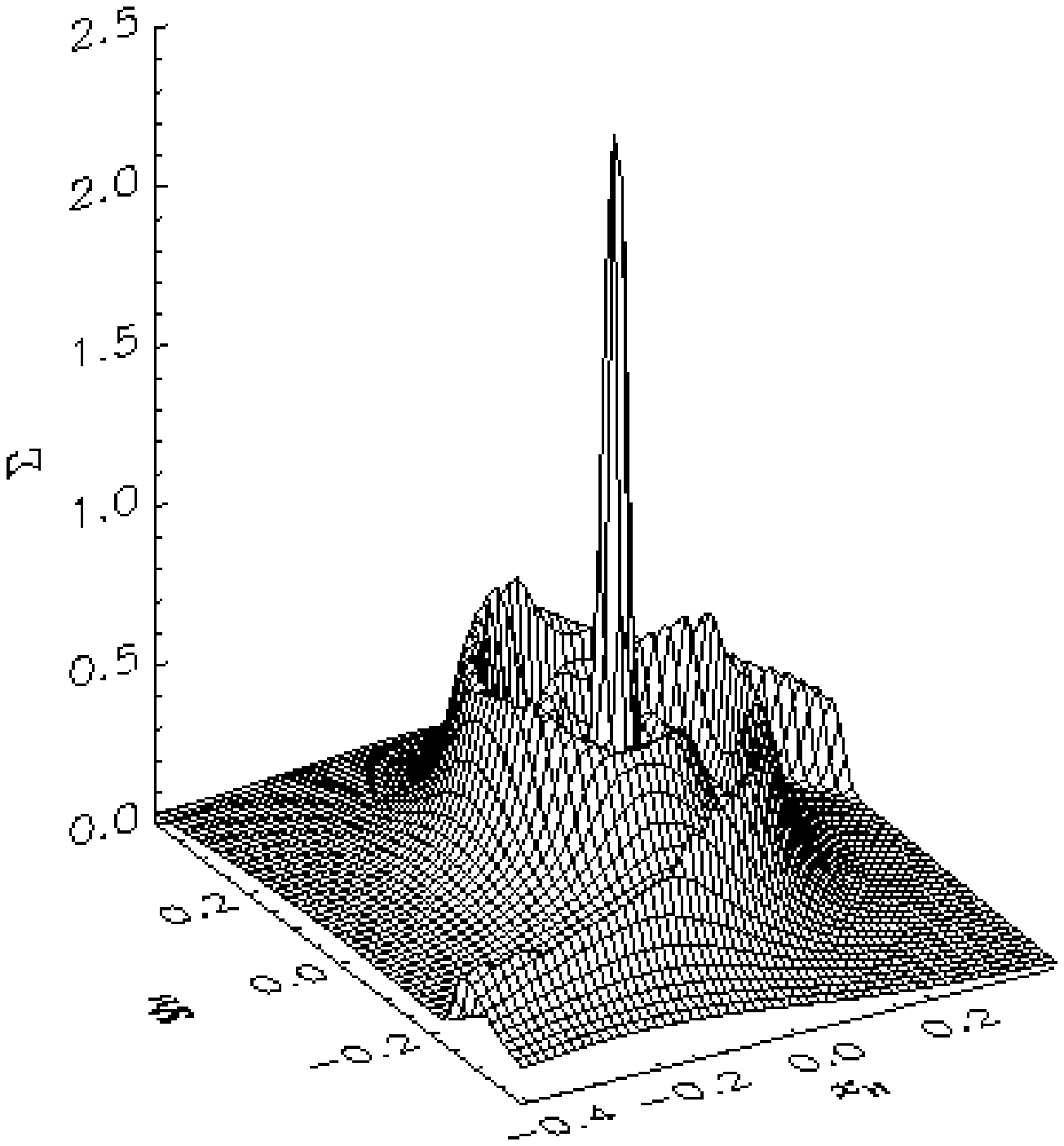}
 \includegraphics[width=\linewidth]{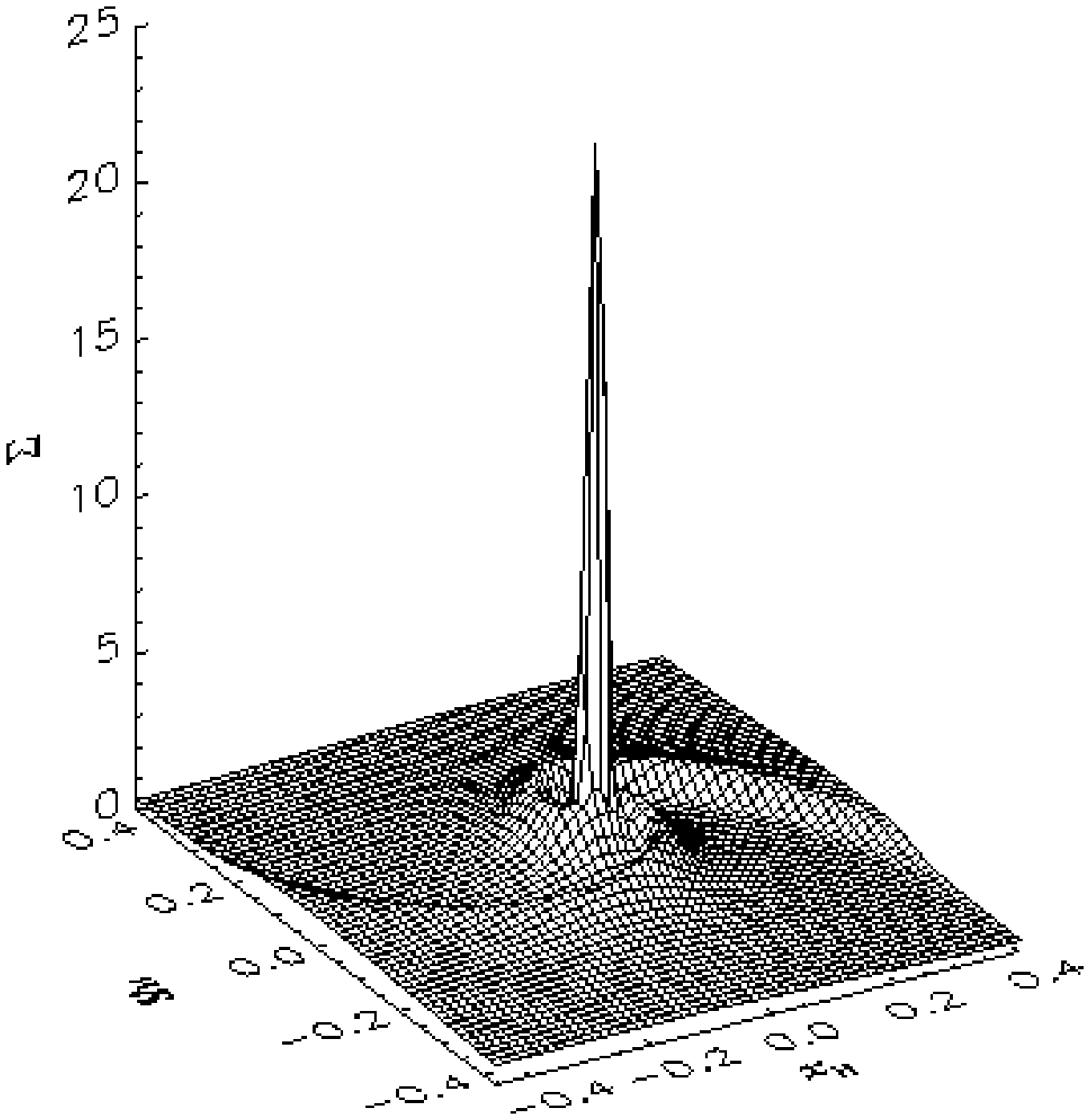}
 \end{center}
 \caption{Surface density plot of the highest grid level.
          \textbf{Top}: \ciro1 at $t=375$ orbits. 
          \textbf{Bottom}: \ciro2 at $t=325$ orbits.
          The Hill coordinates, $(x\subscr{H}, y\subscr{H})$, 
          are so defined:
          $x\subscr{H}= [r\subscr{p}\,\cos(\varphi\subscr{p}) - 
                         r\,\cos(\varphi)]/\Rhill$,
          $y\subscr{H}= [r\subscr{p}\,\sin(\varphi\subscr{p}) - 
                         r\,\sin(\varphi)]/\Rhill$.
          The star is situated along the negative $x\subscr{H}$-axis. 
          The azimuth $\varphi$ increases with $y\subscr{H}$.}
\label{core_look}
\end{figure}

Since not all of the matter is taken away, it should pile up at the 
location of the planet, eventually forming a very dense core. 
Indeed, this is what we find, as already visible
in \Fig{deep_cont}, where density contour lines crowd
around the planet at $(r\subscr{p},\varphi\subscr{p})$. 
Figure~\ref{core_look} displays better how this feature looks like
for models \ciro1 and \ciro2.
In order to make a comparison of the linear extension
of such cores with the Hill radius, we introduce the
\textit{Hill coordinates}. These are defined as
a Cartesian reference frame with origin on the planet and 
coordinates normalized to \Rhill\ (see the caption
of \Fig{core_look} for details).
In these two cases, the core width, $2\,\eta$ (exactly defined below),
can be estimated to be approximatively $0.1$ \Rhill.

One reason for the sharpness of these peaks is the very
small length scale we adopt to smooth the
potential of the planet (\Sect{Subsec:smooth}).
On the finest grid level the smoothing length $\delta$ is
equal to $\lambda(ng) \simeq 1.4\,\Delta r(ng)$ (see Eq.~\ref{lambda}).
From Table~\ref{models} one can deduce that $\delta \sim 10^{-2}$ \Rhill\
(though it changes a little for the  different values of $q$).
Despite that, the core width always equals at least $6\,\Delta r(ng)$.
Other two hints suggest that such features are not just a numeric 
product. According to models \ciro2 and \wpro1 (which differ only in
the number of grid levels), the core width
is very similar. If we cut the peak at $\Sigma=2$ on the finest 
level of each model, its extent, at $r=r\subscr{p}$, is $6$ and $11$ 
grid cells, respectively.
Models \ciro1 and \elen2 agree in a very similar way.
Furthermore, the structure of these peaks does not depend on
the exact placement of the 
planet within a grid cell, as models \ciro3 and \gino1 prove.

\begin{figure}
 \begin{center}
 \resizebox{0.9\linewidth}{!}{%
 \includegraphics{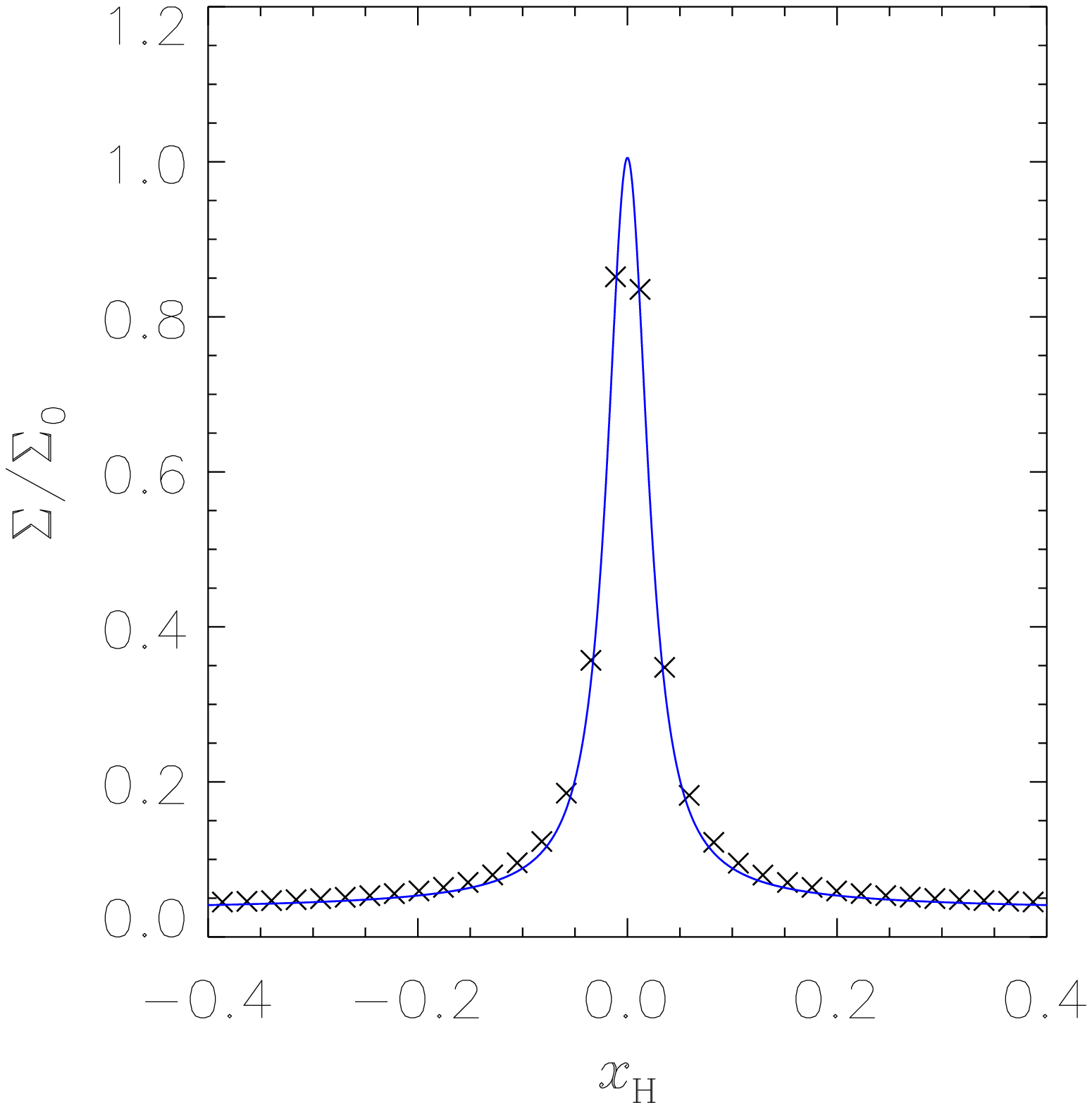}}
 \end{center}
\caption{The core width is generally much larger than that
         proper for an isothermal, hydrostatic, configuration.
         In fact the centrifugal force, due to axial rotation of 
         the gas around the planet, may play an important role in 
         supporting the structure.
         This is not the case for the core around an Earth-mass 
         planet (model \pepp1). Thus, the structure equation reads: 
         $d\,\Phi\subscr{p} = -c\subscr{s}^2\,d\,\Sigma/\Sigma$, 
         where $\Phi\subscr{p}$ is the smoothed potential of the 
         planet. The constancy of the sound speed 
         ($c\subscr{s} = h\, \sqrt{G\,\MStar/a}$) implies that
         the gas is isothermal. 
         In the figure, the surface density of \pepp1, 
         at $\varphi=\varphi\subscr{p}$ ($y\subscr{H}=0$), 
         is represented by ``$\times$'' signs. 
         The over-plotted curve is the solution of the previous
         structure equation.}
\label{core_match}
\end{figure}
However, in order to study the exact properties of such features
a more detailed physical treatment is required in proximity of 
the planet. This may be part of future computations
which include a more detailed treatment of the internal
constitution of the protoplanet.
Anyway, properties such as the local surface density profile 
and the velocity field indicate that the core structure could be 
approximated to that of a rotating and isothermal gas, in hydrostatic
equilibrium. 
In model \pepp1 (1 \MEarth), the rotation of the core is
very slow. Hence, its density profile can be well reproduced by that 
of an isothermal and hydrostatic gas, as demonstrated in
\Fig{core_match},
where we compare the numerical results with an analytic solution
for a hydrostatic isothermal core (solid line).

\begin{figure*}
 \begin{center}
 \hspace*{\fill}%
 \includegraphics[width=0.45\textwidth]{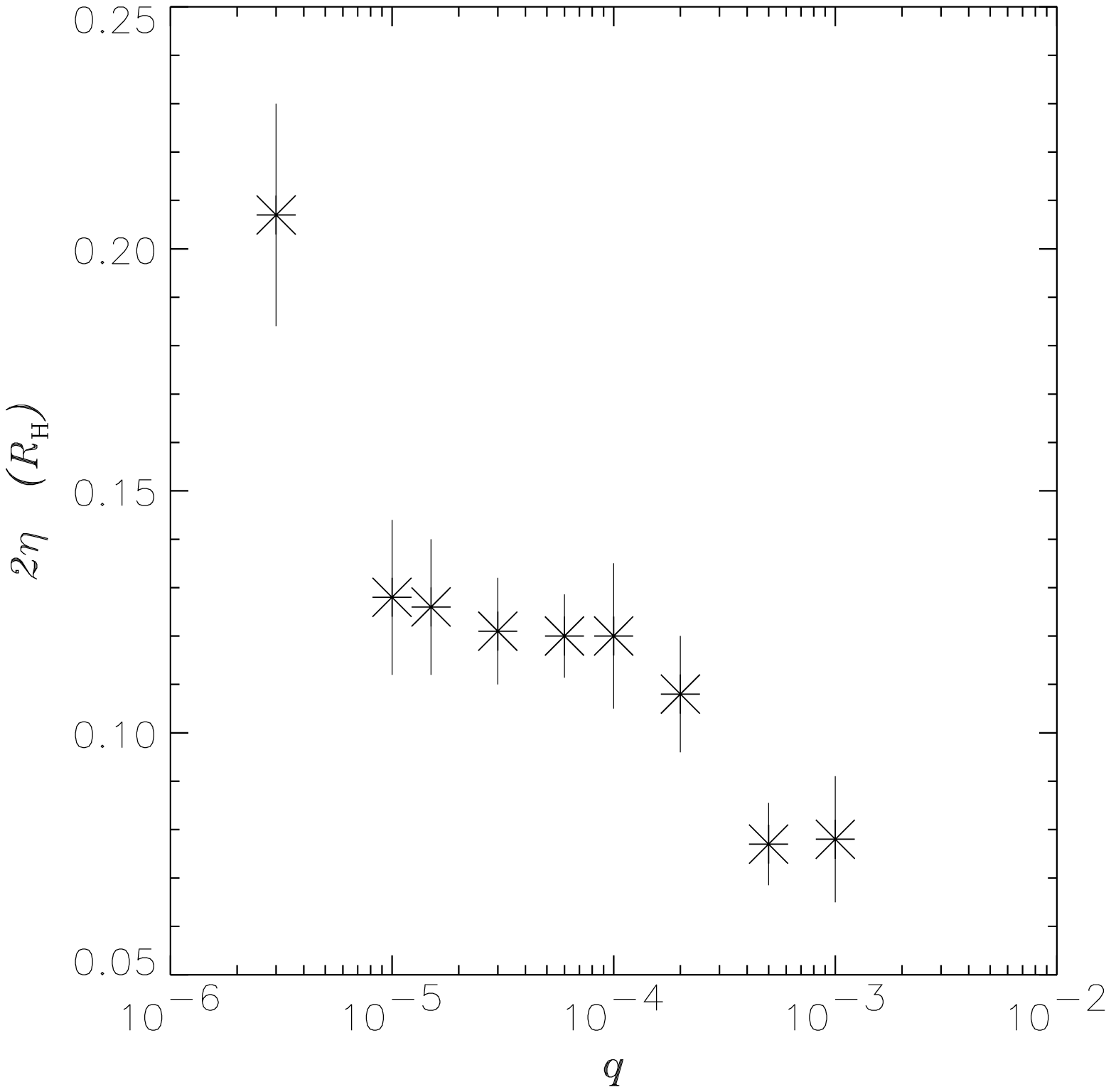}\hfill%
 \includegraphics[width=0.45\textwidth]{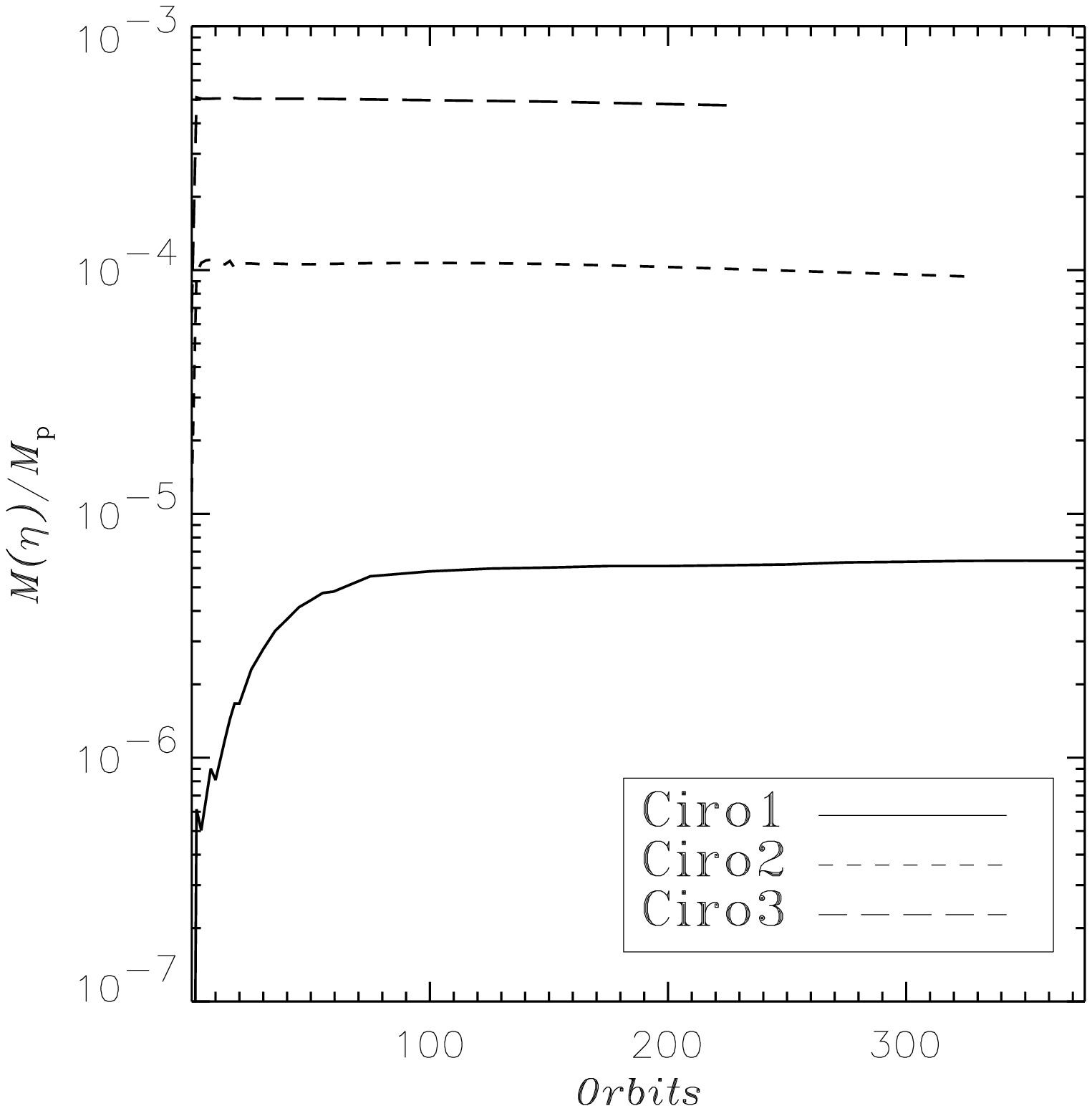}%
 \hspace*{\fill}
 \end{center}
\caption{\textbf{Left panel}. Core width, $2\,\eta$, in units of the
                 Hill radius as a function of the mass ratio $q$. 
                 \ciro\ and \pepp-models are considered along with
                 \gino2 and \gino3.
                 The measure errors are assumed to be equal
                 to $\xi$ (see Table~\ref{models}).
         \textbf{Righ panel}. Mass within a distance $\eta$ from
         the planet versus time. The core mass $M(\eta)$ is normalized to
         the mass of the planet.}
\label{core}
\end{figure*}

As pointed out,
the material within the core region is strongly coupled to the planet,
due to the small distances. In some way, 
it may be considered as part of the protoplanet itself, 
whose structure we may not resolve well enough in the present paper.
Whatever its nature, it is very likely that the angular momentum
transferred, by the core material to the planet, 
may influence the planet's spin angular momentum
rather than its orbital one.
As we are treating the planet as a point mass we cannot estimate its
spin. Therefore we decided to exclude this region from the torque
computation. To do that, we need a quantitative estimate of the 
core radius $\eta$, for every model in Table~\ref{models}.   
We adopt the following procedure: the average density
$\bar{\Sigma}$ is computed over the region
$0.2\ \Rhill \leq |\vec{r}-\vec{r}_\mathrm{p}| \leq 0.5\ \Rhill$.
Then we define the core width ($2\,\eta$) to be that where
$\Sigma = 2\,\bar{\Sigma}$. 
In \Fig{core} (left panel) the dependence of $2\,\eta$
versus $q$ is displayed. The ratio $\eta/\Rhill$ decreases 
for increasing values of $q$. However, between $3.2$ and $32$ \MEarth,
it seems to vary very little. 
Our measure of the core sizes is performed at a particular 
time (for \ciro-models, they are indicated in \Fig{my_img}).
Anyway, such estimates are not affected much by this choice 
because the cores reach a steady state,
early during the system evolution. As an example,
\Fig{core} (right panel) shows how the core
mass $M(\eta)$ assumes a static value very soon.

As already mentioned, the spiral features vanish for low 
values of $q$, and for Earth-mass planets the core becomes
the most prominent feature within the Roche lobe.
\subsection{Torque exerted on the planet} 
\label{Subsec:Teop}
Any protoplanet embedded in a protostellar environment 
suffers gravitational torques, exerted by the surrounding disk material.
If the density distribution were symmetric with respect to the planet,
or with respect to the line connecting the star with the planet,
the resultant total torque would be zero.
However, we have just seen that this is not the case,
and even around a planet as small as $3.2$ \MEarth,
the matter distribution is not axially symmetric. 
Thus, we expect a non-zero total torque from the disk,
acting on the planet. 
In response to it, because of the conservation of 
the orbital angular momentum, the planet has to adjust 
its semi-major axis, which leads to a migration phenomenon.

In the present computations, we evaluate the torque exerted
on the planet in a straightforward way. First the gravitational
force acting on the planet $\vec{f_g}(i,j)$, 
due to each grid cell $(i,j)$, is calculated. 
The torque, with respect to the star, is then
\begin{equation}
\vec{t}(i,j)= \vec{r}\subscr{p} \times \vec{f_g}(i,j).
\label{t_ij}
\end{equation}
Since we are interested in the $z$-component of this vector, 
we perform the scalar product 
$\hat{\vec{z}}\cdot\vec{t}(i,j) = t_z(i,j)$, where
$\hat{\vec{z}}$ is the unit vector of the vertical direction.
Finally, by summing over all $i$ and $j$, we obtain 
the total disk torque $\mathcal{T}\subscr{D}$,
whereas the simple contraction over the azimuthal index $j$ gives
the radial distribution of the torque.

The quantity $t_z(i,j)$ is computed on each grid level. 
Where the computational domain is covered by more than one level,
the torque $t_z(i,j)$ on the finest grid is considered for 
the evaluation of $\mathcal{T}\subscr{D}$. 

In order to avoid the region dominated by the core, 
we exclude a certain area from the computation of 
$\mathcal{T}\subscr{D}$. 
Because of the way we
defined the core radius $\eta$, some core material may still lie
outside $|\vec{r}-\vec{r}_\mathrm{p}|=\eta$.
Therefore, for safety reasons, we choose not to take into account 
the planet neighborhood  defined by 
$|\vec{r}-\vec{r}_\mathrm{p}|\leq \beta = 2\,\eta$. 
The only level, upon which this operation is relevant, is the
highest since it provides the gravitational torque form the 
regions closest to the planet. 
On coarser levels such operation is meaningless since 
inner torques are taken from elsewhere. But it can be useful
to confer a more regular look to the radial torque profile.
We generally adopt the value $\beta = 5\,\delta(l)$ for $l<ng$.

However, as this might be somewhat arbitrary, we discuss in 
a separate section how the choice of $\beta$, on the finest 
level, affects the value of the total torque $\mathcal{T}\subscr{D}$.

\begin{figure*}
 \begin{center}
 \resizebox{0.9\textwidth}{!}{%
 \includegraphics[bb=0 30 470 470, clip]{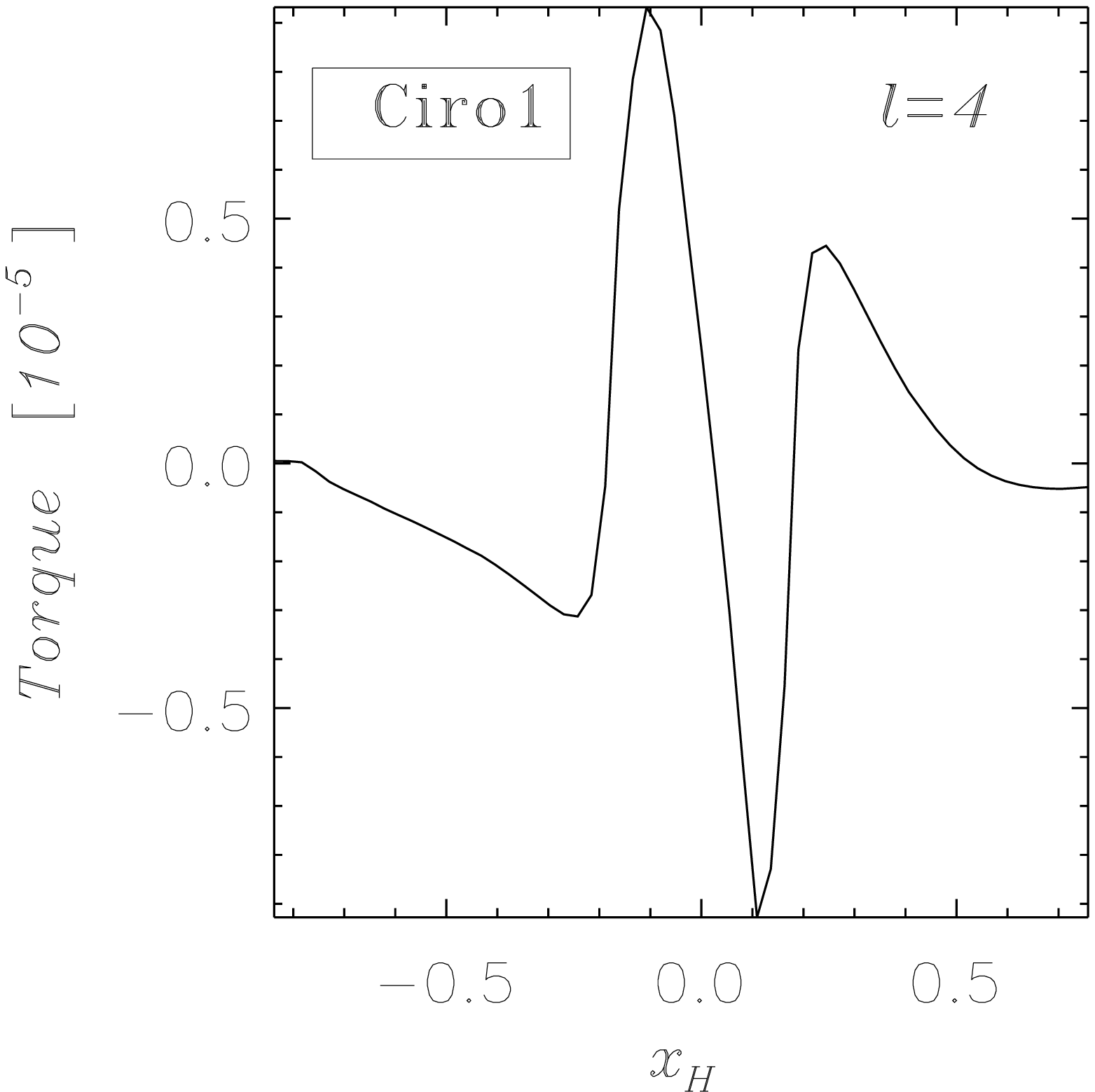}%
 \includegraphics[bb=0 30 470 470, clip]{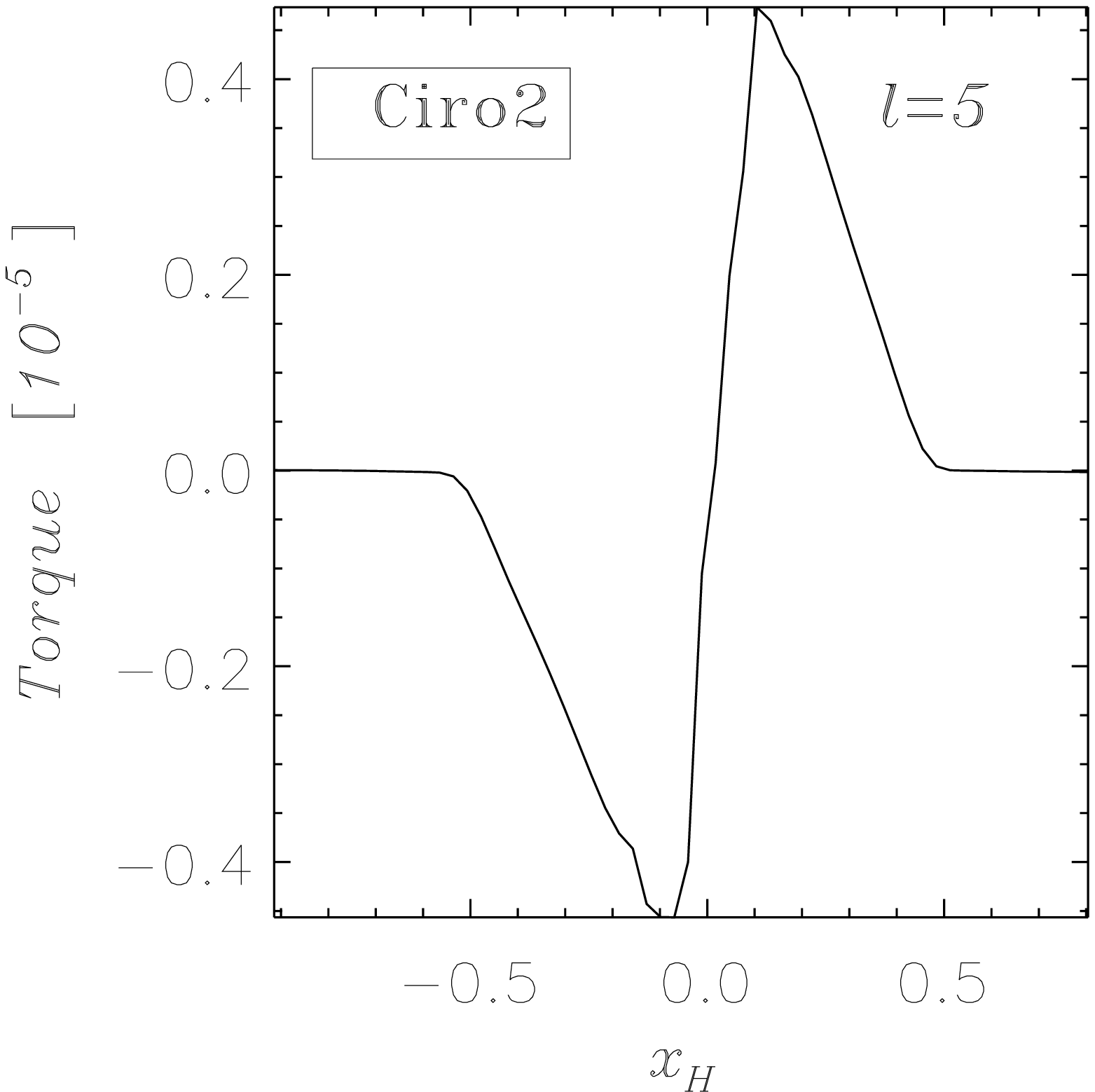}%
 \includegraphics[bb=0 30 470 470, clip]{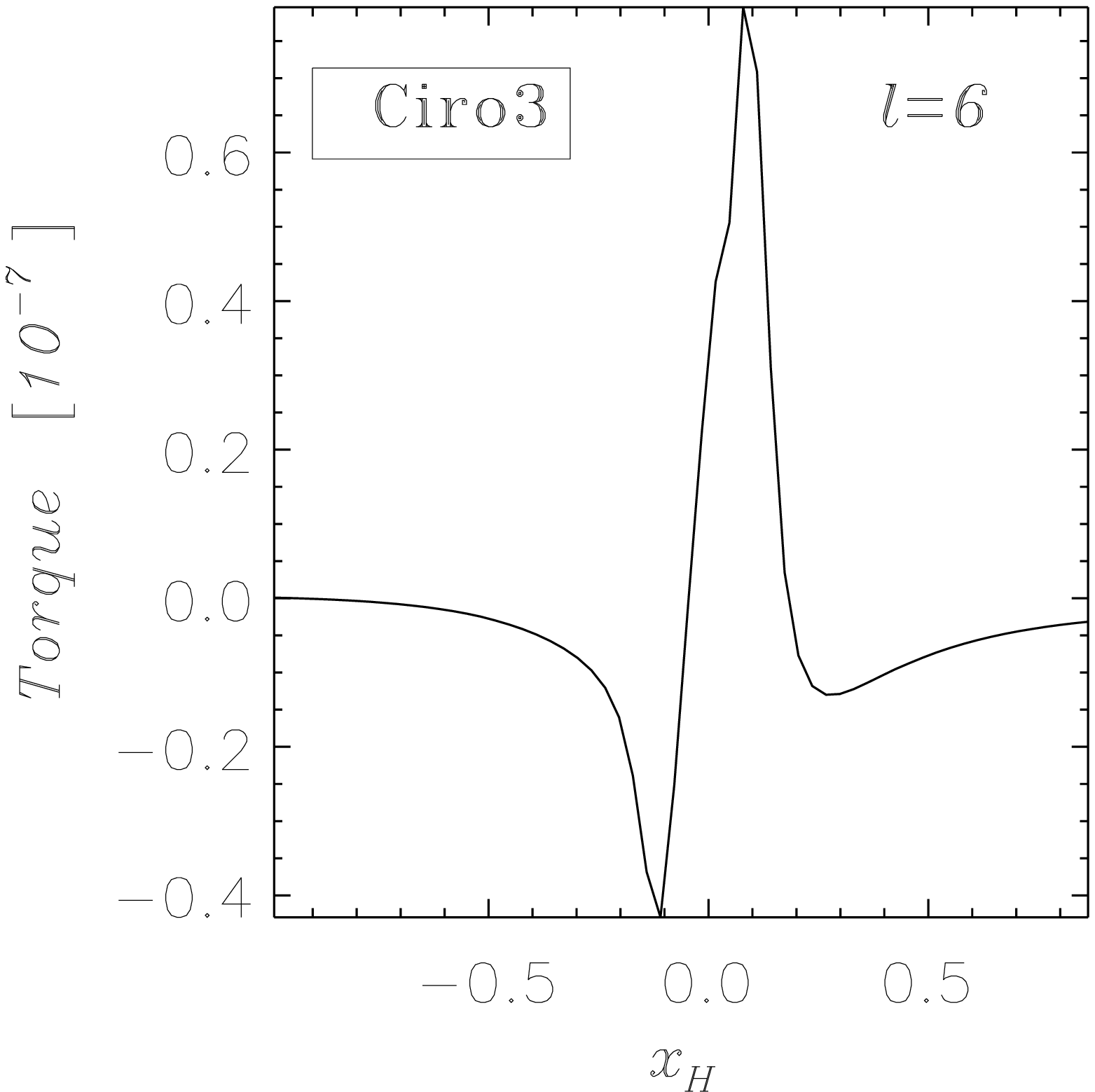}}
 \resizebox{0.9\textwidth}{!}{%
 \includegraphics[bb=0 30 470 470, clip]{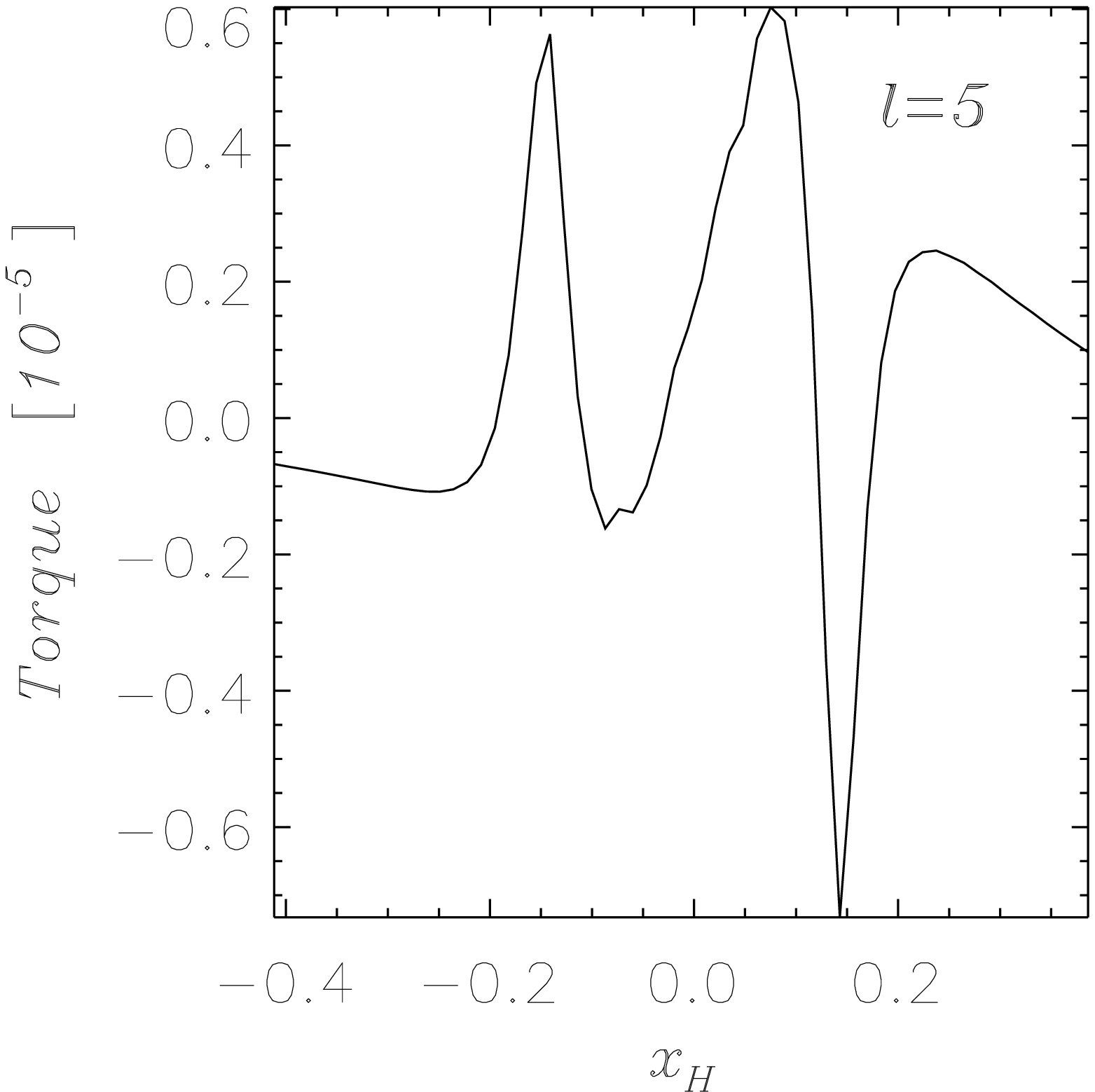}%
 \includegraphics[bb=0 30 470 470, clip]{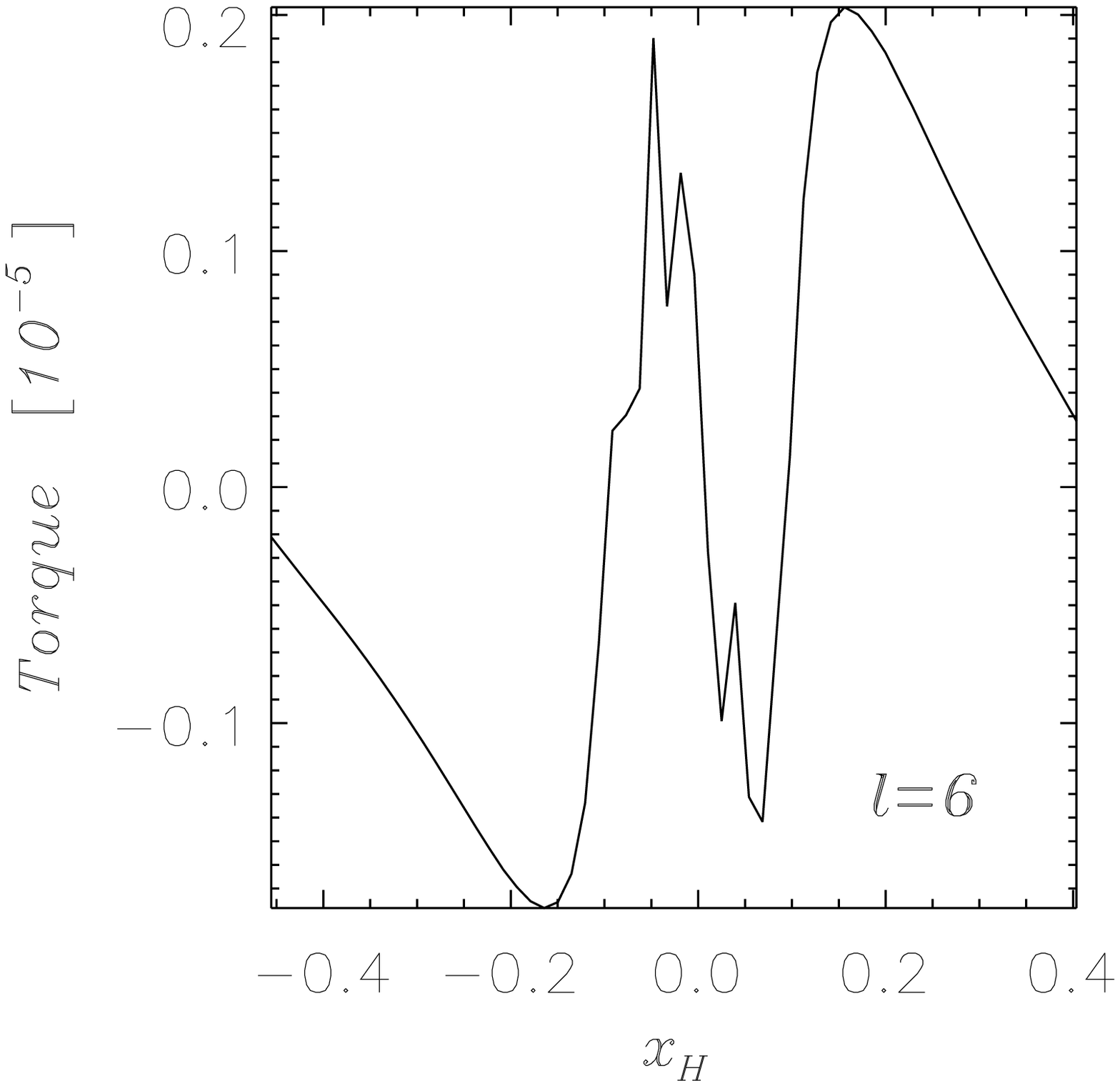}%
 \includegraphics[bb=0 30 470 470, clip]{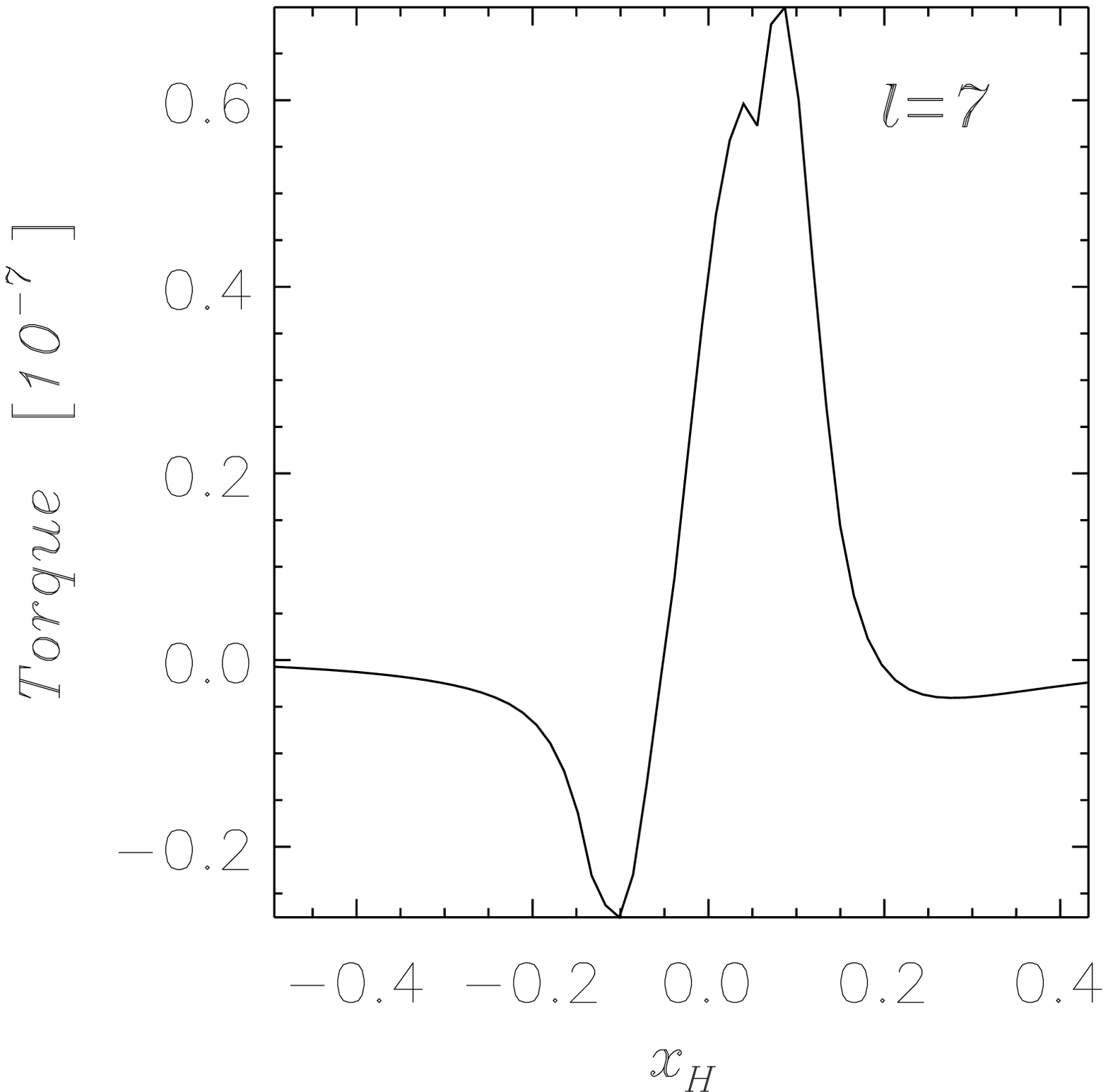}}
 \resizebox{0.9\textwidth}{!}{%
 \includegraphics[bb=0 0 470 470, clip]{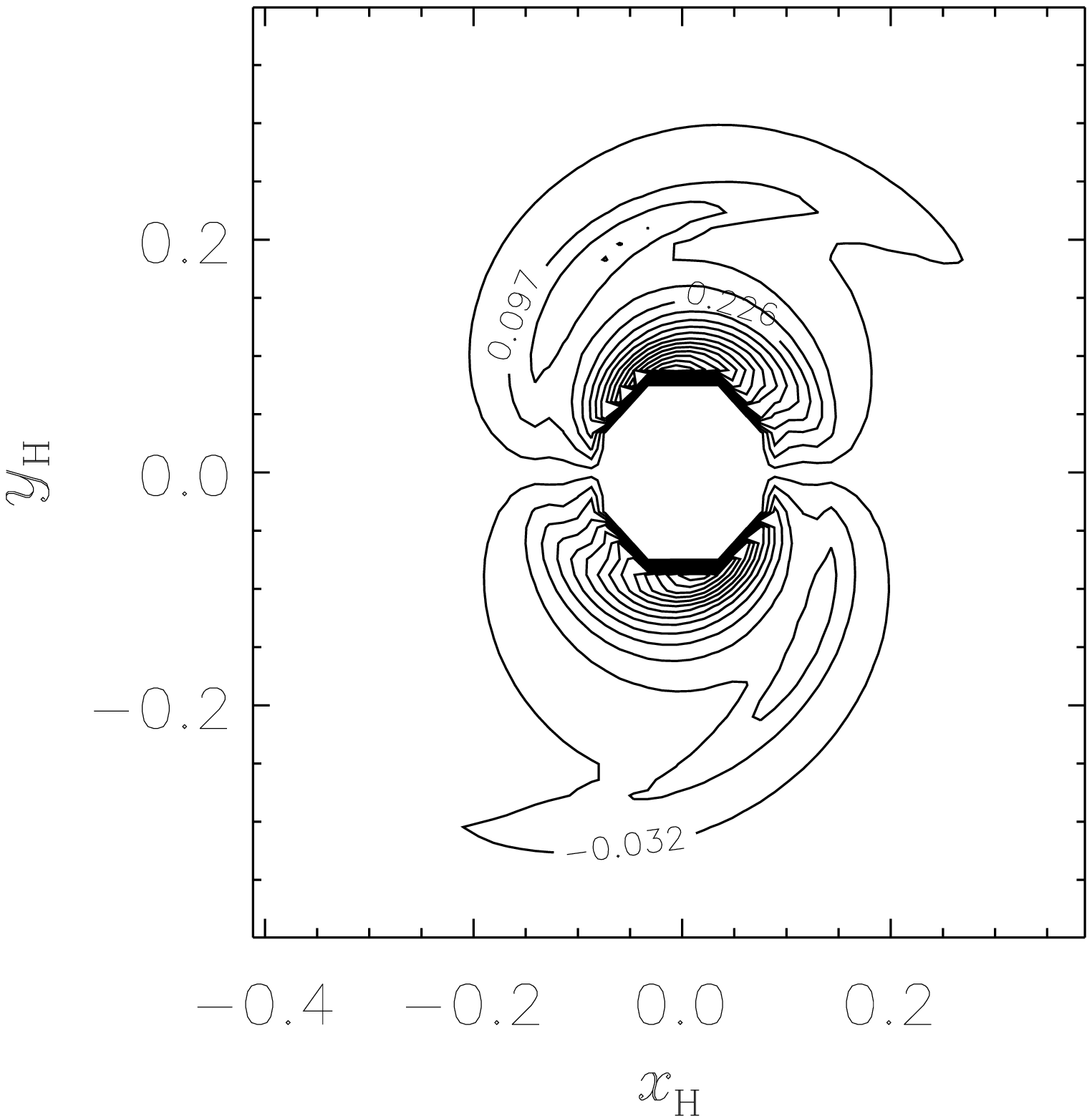}%
 \includegraphics[bb=0 0 470 470, clip]{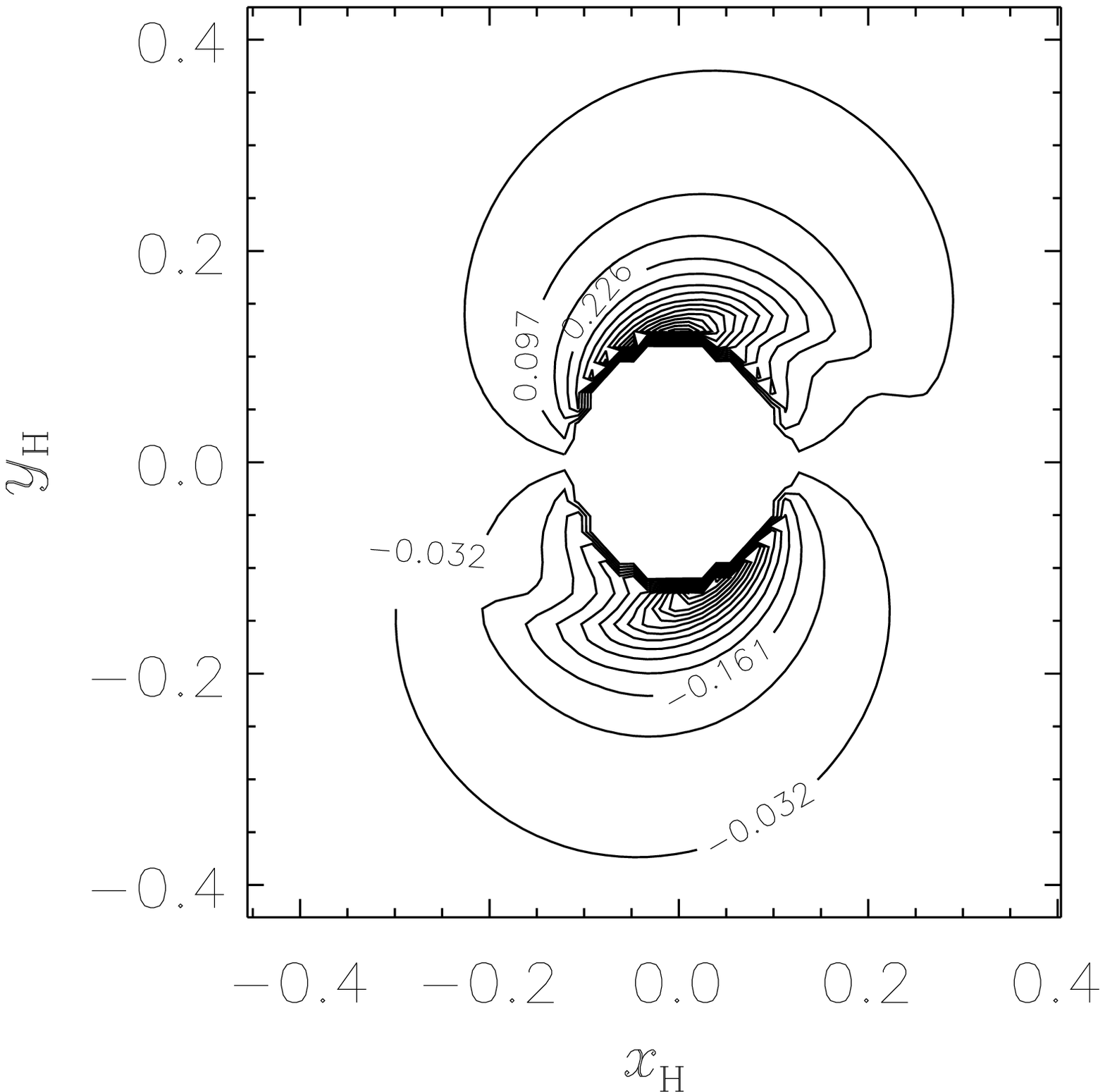}%
 \includegraphics[bb=0 0 470 470, clip]{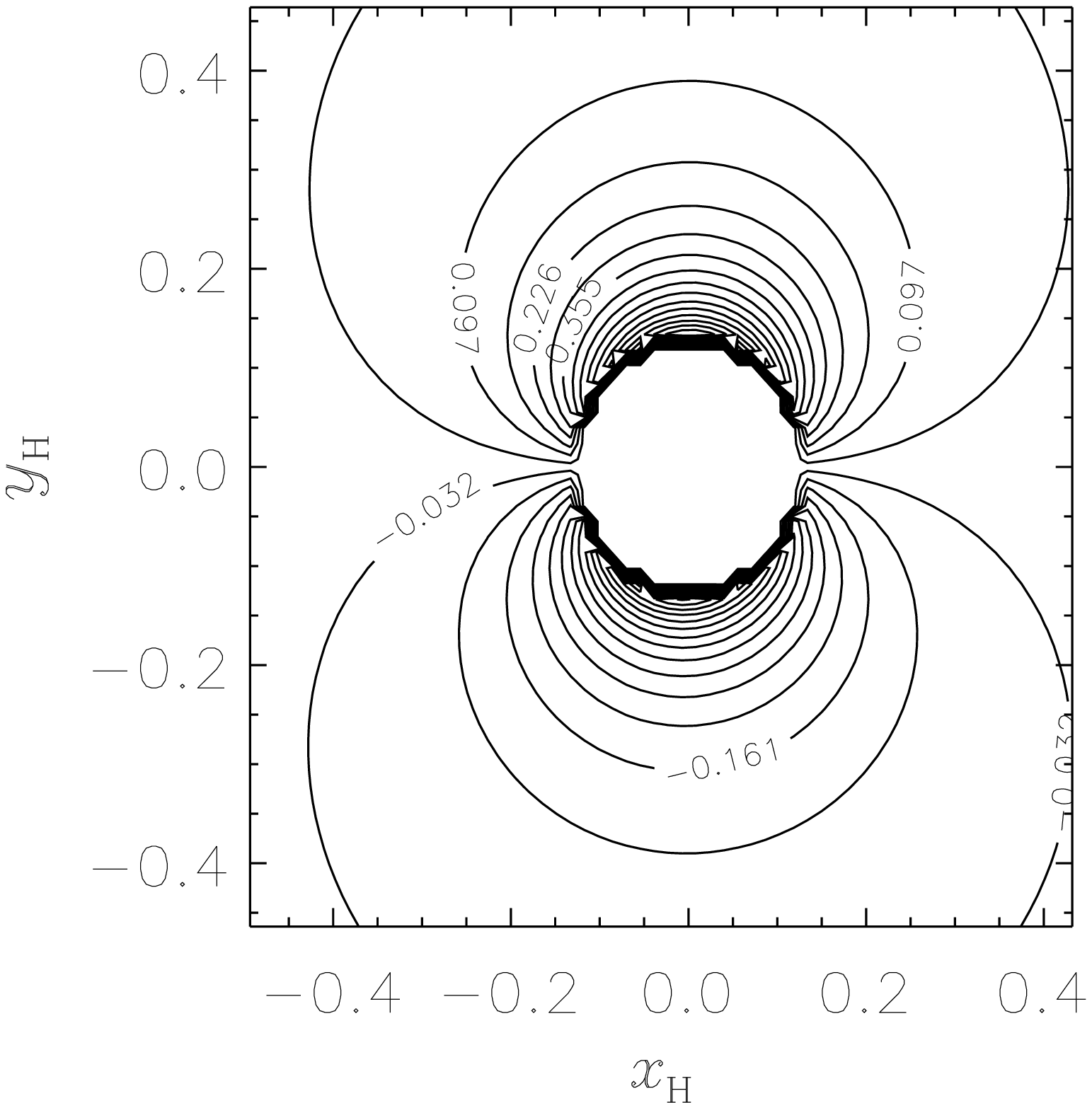}}
 \end{center}
 \caption{Radial distribution and two-dimensional contour map of the
          gravitational torque exerted by the disk on the planet.
          \textbf{Left panels}. \ciro1: levels $l=4$ and $l=5$.
          \textbf{Center panels}. \ciro2: levels $l=5$ and $l=6$. 
          \textbf{Right panels}. \ciro3: levels $l=6$ and $l=7$. 
          The two-dimensional torque distribution (bottom panels)
          is reported for the finest grid level and is normalized 
          to its maximum, absolute, value.}
\label{rad_tor}
\end{figure*}
Because of the global angular momentum transfer, the disk material 
(orbiting the star), at $r>r\subscr{p}$, exerts a negative 
torque on the planet whereas the inside-orbit gas tends to increase 
its angular momentum.
This tendency changes, as material closer to the planet is 
accounted for, and it may reverse eventually, 
once in the circumplanetary disk (for $\Mp=1$ \MJup, 
such behavior was also found by Lubow et al. \cite{lubow1999}). 
The radial torque distribution, from \ciro-models, is shown in
\Fig{rad_tor}.
The two sets of profiles belong to the grid levels $l=ng-1$
(top row) and $l=ng$ (middle row).
In the case of \ciro1 (top row, left panel), the sign reversal of 
the torque is not completed yet. However some negative torques are exerted 
from regions inside the planet orbit and some positive torques are
exerted from the opposite side.
On the domain covered by this grid (fourth level), the torque 
contribution coming from $x\subscr{H} < 0$ is positive while the one 
coming from $x\subscr{H} > 0$  is negative.
The resulting net torque is positive, as the magnitude of
the latter contribution is $2.4$ times smaller than that of the former.
The torque behavior gets more complex if we restrict to a region
closer to the planet (middle row, left panel). Though not evident
at a first glance,
the signs are reversed if compared to the preceding grid level. 
The torque exerted by the region $x\subscr{H} > 0$ is definitely 
positive and, in magnitude, almost $30$ times as large as that 
arising from the region $x\subscr{H} < 0$. 
Thus, this region exerts a strong, positive, net torque.
Indeed, the phenomenon of the torque sign reversal is clear in the
case of \ciro2 and \ciro3 (\Fig{rad_tor}, center and right panels). 
For both grid levels, inside-orbit material lowers the
angular momentum of the planet while outside-orbit material
acts in the opposite direction. 
On the finest level of \ciro2, the ratio of the negative
to the positive torque contribution is just $0.96$ (in absolute value),
whereas it is $0.3$ for \ciro3. 

\begin{figure}
 \begin{center}
 \resizebox{0.9\linewidth}{!}{%
 \includegraphics{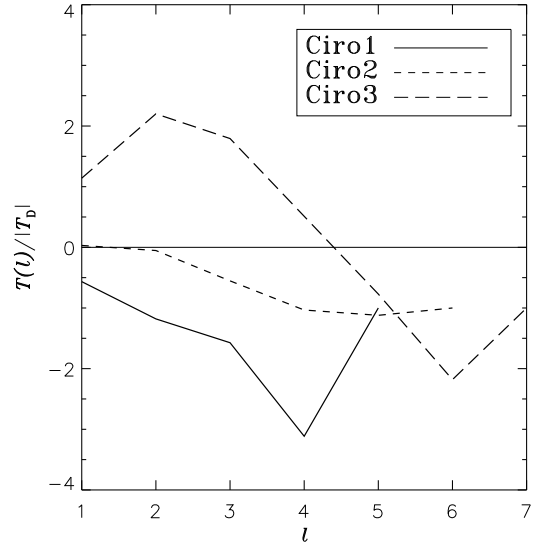}}
 \end{center}
\caption{Partial torque $\mathcal{T}(l)$,
         normalized to  $|\mathcal{T}\subscr{D}|$,
         plotted against the grid level $l$ for:
         \ciro1, \ciro2 and \ciro3. 
         $\mathcal{T}(ng)/ |\mathcal{T}\subscr{D}| = -1$ 
         because, by definition, 
         $\mathcal{T}(ng)=\mathcal{T}\subscr{D}$ and
         $\mathcal{T}\subscr{D}$ is negative for these models.}
\label{tor_vs_l}
\end{figure}
In order to check which is the overwhelming contribution,
between positive and negative torques, on the various grid
levels, we can define the partial torque $\mathcal{T}(l)$.
This represents the total torque computed over the entire domain 
but the part covered by the grid level $l+1$ and such that 
$\mathcal{T}(ng)=\mathcal{T}\subscr{D}$.
\Fig{tor_vs_l} illustrates the sign of the partial torque
and its relative strength,
on each grid level, for \ciro-models. We can see that the total
torque is negative in all of the three models. In \ciro1 and \ciro2
all levels, but the highest, contribute to lower the planet 
angular momentum. On the contrary, the matter inside the finest level
raises the overall torque (of a fair amount in case of \ciro1). 
In \ciro3, a positive torque is exerted by the
material outside a region, around the planet,
with a linear extension $\sim 2\,h\,r\subscr{p}$.
Instead, levels $3$, $4$, $5$ and $6$ provide negative torques,
which are then weakened considerably by the positive torque 
coming from the finest level.

\begin{table}
 \caption{Gravitational torque exerted on the planet arising
          from different disk regions, for the relevant models
          of Table~\ref{models}. The entire domain is divided
          into three region: outside the Hill circle; inside the 
          circle of radius $\Rhill/2$ and the zone in between.
          Then they are divided further in order to distinguish
          between inside-orbit (\textit{in}) and 
          outside-orbit (\textit{out}) contribution.
          The torque is computed over each of these regions,
          by employing $\beta=2\,\eta$ and  
          according to the value of the normalized distance
          from the planet
          $d=|\vec{r}-\vec{r}_\mathrm{p}|/\Rhill$.
          Torques are normalized to $\mathcal{T}\subscr{D}$.}
 \label{tor_tab}
 \begin{center}
 \begin{tabular}{lrrrrrr}
 \hline
 \hline
         &%
 \multicolumn{2}{c}{$d \leq 0.5$}&%
 \multicolumn{2}{c}{$0.5 < d < 1$}&%
 \multicolumn{2}{c}{$d \geq 1$}\\ 
 \cline{2-7}
 \raisebox{1.5ex}[-1.5ex]{Model}&%
  \multicolumn{1}{c}{\textit{in}} &  \multicolumn{1}{c}{\textit{out}} &%
  \multicolumn{1}{c}{\textit{in}} &  \multicolumn{1}{c}{\textit{out}} &%
  \multicolumn{1}{c}{\textit{in}} &  \multicolumn{1}{c}{\textit{out}} \\
 \hline
 \ciro1  &  $ -0.03$  & $ 2.08$  &  $-1.18$  &  $-0.43$  &  $0.53$  & $-1.97$ \\
 \ciro2  &  $ -3.07$  & $ 3.23$  &  $-1.06$  &  $ 0.94$  &  $2.04$  & $-3.08$ \\
 \ciro3  &  $ -1.93$  & $ 3.13$  &  $-0.49$  &  $-0.97$  &  $7.26$  & $-8.00$ \\
 \pepp1  &  $  0.33$  & $-0.40$  &  $ 0.33$  &  $-0.36$  &  $4.22$  & $-5.12$ \\
 \pepp2  &  $ -2.32$  & $ 2.05$  &  $-0.18$  &  $-0.18$  &  $1.91$  & $-2.28$ \\
 \pepp3  &  $ -2.61$  & $ 2.07$  &  $-0.16$  &  $-0.01$  &  $0.85$  & $-1.14$ \\
 \pepp4  &  $ -3.01$  & $ 2.41$  &  $-0.29$  &  $ 0.25$  &  $0.84$  & $-1.20$ \\
 \gino2  &  $  2.47$  & $-1.65$  &  $-1.44$  &  $ 1.02$  &  $2.56$  & $-3.96$ \\
 \gino3  &  $ -1.54$  & $ 5.12$  &  $-2.06$  &  $-0.34$  &  $1.90$  & $-4.08$ \\
 \hline
 \end{tabular}
 \end{center}
\end{table}
%
An overview of the torques exerted by different
portions of the disk, for all the relevant models, 
is given in Table~\ref{tor_tab}.  
First of all we notice that, within $0.5$ \Rhill, 
the phenomenon of the sign reversal is observed in all 
of the models, but \pepp1 and \gino2. 
Anyway, in the latter case, the torque inside this 
region is positive.

In \ciro-models and in model \gino3, the reversal of the 
sign is such to produce a positive torque over the domain 
$|\vec{r}-\vec{r}_\mathrm{p}| < 0.5$ \Rhill.
This is because the surface density of the leading material 
is slightly higher than that of the trailing matter
(\Sect{Subsec:ofs} and \Fig{img:overview}).
In contrast, the torque is negative in case of \pepp2, 
\pepp3 and \pepp4.
We note, however, that the positive torques
exerted by the outside-orbit material, within
this region, strongly attenuate the magnitude of
the negative net torque exerted by the rest of the disk.

Clearly, since neighboring material may tend to reduce
the magnitude of negative torques acting on the planet, 
it could either slow down its inward migration 
or reverse the direction of its motion.  

As anticipated,
the sign reversal of the radial torque distribution 
is not observed in model \pepp1 ($\Mp = 1$ \MEarth).
Due to the very low mass of the perturber,
the only structure present inside
the Roche lobe is the density core. Whatever level
is considered, the inside-orbit gas always provides
positive torques while negative ones come from
the outside (see also Table~\ref{tor_tab}). 
Negative torques are somewhat stronger, on any grid level. 
Almost the 50\% of the total torque is generated 
between $\sim 0.5\,h\,r\subscr{p}$ and $\sim h\,r\subscr{p}$,
at the starting positions the disk spirals.
\subsubsection{2D-torque distribution}
\label{Sect:2d-torque}
The detailed balance of the torques, arising from 
very close material, depends on the medium and small-scale density
structures around the planet, such as the shape of spirals. 
Therefore, referring to what was stated in the previous section, 
we can deduce that the radial torque asymmetry is a 
direct consequence of the asymmetric distribution of the gas
with respect to the planet.

Thus, a more comprehensive description of the torque behavior
requires its full two-dimensional distribution. 
In the bottom panels of \Fig{rad_tor}, the contour lines 
of the two-dimensional torque, $t_z$, are shown for each reference model.
The interesting point to notice here is that 
the largest magnitude torques arise
from the corotation locations, i.e.\ where $r=r\subscr{p}$
($x\subscr{H} \simeq 0$). Here in fact, $\vec{f_g}(i,j)$ is perpendicular 
to $\vec{r}\subscr{p}$ and the cross-product in \Eq{t_ij}
achieves its maximum (minimum).
The material leading the planet 
(at $\varphi>\varphi\subscr{p}$ or $y\subscr{H}>0$), 
pulls it ahead and makes it gain angular momentum. 
The trailing material brakes the planet making it lose
angular momentum. 

Let's consider two fluid elements
at $(0,\pm\,y\subscr{H})$ and write their mass density as $\Sigma^\pm$. 
Then we can write 
$|t_z^\pm| \propto r\subscr{p}\,\Sigma^\pm/y\subscr{H}^2$,
which yields:
\begin{equation}
|t_z^+| - |t_z^-| \propto r\subscr{p}\,\frac{\Delta \Sigma}{y\subscr{H}^2},
\label{tmis_n}
\end{equation}
where $\Delta \Sigma = \Sigma^+ - \Sigma^- $. 
Any mismatch of the surface density, $\Delta \Sigma$, 
causes a torque mismatch amplified by an amount equal 
to $y\subscr{H}^{-2}$. 
It's worth noticing that, on larger distances  
$|\vec{r}-\vec{r}_\mathrm{p}| \sim r\subscr{p}$,
the torque mismatch is amplified less, in fact:
\begin{equation}
|t_z^+| - |t_z^-| \propto \frac{\Delta \Sigma}{r\subscr{p}}.
\label{tmis_f}
\end{equation}
This is the reason why surface density asymmetries
near the planet have a very strong impact on $\mathcal{T}\subscr{D}$, 
and they can easily prevail against the more distant ones.

The region responsible for the maxima and minima of the radial 
torque distributions in the middle row of \Fig{rad_tor} 
can be identified by means of the 2D-torque maps.
For example, in the case of \ciro1, we see that 
the maximum at $x\subscr{H}=-0.15$ and the minimum at 
$x\subscr{H}\simeq 0.15$ are produced at  
$y\subscr{H}\simeq 0.1$ and $y\subscr{H}\simeq -0.1$, respectively.
In the other two cases, radial distribution extremes rise from 
regions where the torque function $t_z$ is steeper than it is
on the opposite side of the planet. 
\subsection{Planet migration}
\label{Sect:migration}
\begin{figure}
 \begin{center}
 \resizebox{\linewidth}{!}{%
 \includegraphics{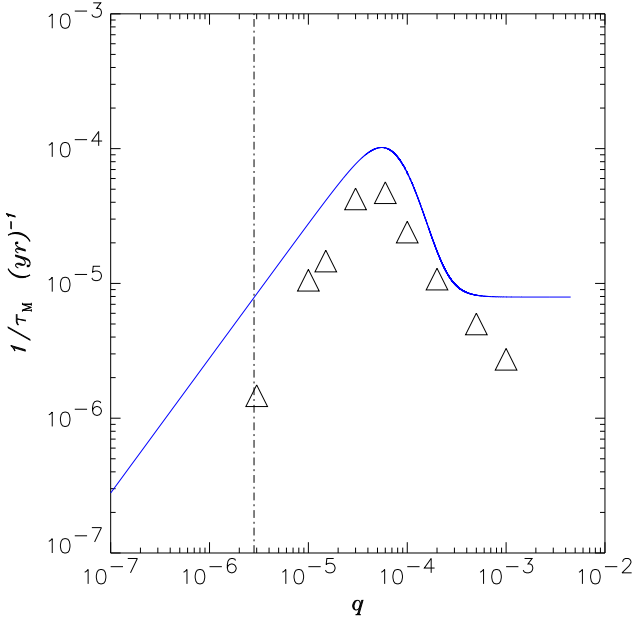}}
 \end{center}
\caption{Migration time scale $\tau\subscr{M}$ versus the
         mass ratio $q$. Open triangles represent the results 
         from models \ciro, \pepp, \gino2 and \gino3. 
         The total disk torque $\mathcal{T}\subscr{D}$ is
         computed assuming $\beta=2\,\eta$, for each model. 
         This means that the region lying inside 
         $|\vec{r}-\vec{r}_\mathrm{p}|= 2\,\eta$ is not taken into
         account. 
         In order to express  $\tau\subscr{M}$ into physical units, 
         we suppose that the planet orbits at $a=5.12$ \AU\ 
         in a disk with mass $\Md = 3.5 \times 10^{-3}$ \MStar.
         The curve over-plotted represents predictions of the 
         analytical theory as formulated by Ward (\cite{ward1997}), 
         for the case of ``strong'' viscosity ($\alpha=4\times10^{-3}$
         in our case) and accounting \emph{only} for Lindblad torques. 
         It is derived assuming an unperturbed constant surface density 
         and a disk temperature dropping as $1/r$. 
         The behavior of this curve reduces to \Eq{typei} 
         and \Eq{typeii} letting $q \rightarrow 0$
         and $q \rightarrow \infty$, respectively. 
         The vertical line marks the value given by \Eq{q_fm}.}
\label{tau_vs_q}
\end{figure}
If we consider a planet moving on a circular orbit, we find
that the rate of change of its semi-major axis $a$, caused by an
external torque $\mathcal{T}\subscr{D}$, is
\begin{equation}
\frac{da}{dt} = \frac{2\,\mathcal{T}\subscr{D}}{\Mp\,a\,\Omega\subscr{p}}.
\label{adot}
\end{equation}

Analytical estimates of $\mathcal{T}\subscr{D}$ show that 
two limiting cases exist, depending on whether the planet 
is massive enough to generate a gap or not.
In the first case of small planetary masses,
we have the so-called \textit{type I} migration.
Ward (\cite{ward1997}) derives the following expression:
\begin{equation}
\left(\frac{da}{dt}\right)\subscr{I} \simeq -\frac{1}{2}\, 
q\,\,h^{-3}\,a\,\Omega\subscr{p}\left(\frac{\pi\,a^2\,\Sigma}{\MStar}\right).
\label{typei}
\end{equation}
The direction of the migration is inwards because of the 
dominating role of the outer Lindblad resonances\footnote{%
This is true as long as the temperature gradient within
the disk is negative (Ward \cite{ward1986}).} 
(Ward \cite{ward1986,ward1997}).
In the second case the planet is more massive, opens up a gap,
and the evolution is \textit{locked} to that of the disk.
As a general trend, the disk material drifts inwards on the viscous 
time scale, and so does the planet. It follows that 
\begin{equation}
\left(\frac{da}{dt}\right)\subscr{II} = -\frac{3\,\nu}{2\,a}
         = -\frac{3}{2}\,\alpha \, h^2 \, a\, \,\Omega\subscr{p},
\label{typeii}
\end{equation}
which is known as \textit{type II} migration.
Comparing \Eq{typei} with \Eq{typeii}, it turns out that
type I drift is faster than type II (i.e.\ faster than viscous
diffusivity) whenever
\begin{equation}
q \gtrsim 3\,\alpha \, h^5 \, \left(\frac{\MStar}{\pi\,a^2\,\Sigma}\right).
\label{q_fm}
\end{equation}
The parameter values employed here\footnote{If we take into account
the dependence of the unperturbed surface density upon the radial
distance $r$, $\pi\,a^2\,\Sigma=0.38\,\Md$.} 
yield a right-hand side equal
to $2.8\times10^{-6}$, which is just a bit smaller 
than an Earth mass ($0.93$ \MEarth).
Fast type I migration should continue till the planet grows enough
to impose a gap on the disk. By that time, however, it could have
already impacted the parent star.
Once entered this fast drifting regime, it seems that 
the planet may survive only if the growth time scale 
$\tau\subscr{G}\equiv \Mp/\dot{M}\subscr{p}$
is much smaller than the migration time scale
$\tau\subscr{M}\equiv a/|\dot{a}|$.

In \Fig{tau_vs_q} the migration time scale
$\tau\subscr{M}$ is shown for the main models listed in 
Table~\ref{models}. 
The drifting motion, is directed inwards, in all cases.
The lowest migration velocity belongs to the Earth-mass
planet (\pepp1). The second lowest drifting velocity is that
of the Jupiter-mass planet (\ciro1).
The most rapidly migrating planet is the one having
$\Mp = 20$ \MEarth\ (\pepp4).

In agreement with predictions of analytical linear theories, 
the drift velocity $|\dot{a}|$ increases for increasing planet mass,
just as prescribed by type I migration (Eq.~\ref{typei}).
The fast speed branch has a turning point around 
$q\simeq 6\times 10^{-5}$, after which migration 
slows down considerably.
Past this point, $|\dot{a}|$ drops as the planet mass increases.
According to the linear theory, this property announces 
the transition to type II migration (Eq.~\ref{typeii}).
As a comparison, the complete theoretical behavior of
$\tau\subscr{M}$ is also reported in  \Fig{tau_vs_q}. 
It was derived by Ward (\cite{ward1997}) for viscous disks with
$\alpha\gtrsim 10^{-4}$. 
Eqs.~(\ref{typei}) and (\ref{typeii})
represent the asymptotic branches of this curve for
very light and very heavy planets, respectively. 
The vertical line, in the figure, indicates the 
mass value above which migration is faster than 
inward viscous diffusion.

In all the models under examination, numerical
simulations predict a slower drift than analytic 
linear theory does.
Roughly, migration is two times as slow for models
\gino2, \gino3, \pepp3 and \pepp4; 
three times as slow for \ciro1, \ciro2, \ciro3 and \pepp2.
Planets as massive as $3.2$ and $64$ Earth-masses migrate
on the viscous time scale of the disk (Eq.~\ref{typeii}).  
$\tau\subscr{M}$ is six times as large in case of the 
Earth-mass planet, when compared to analytic predictions. 
Such discrepancies are likely to arise also because the theoretical
curve in \Fig{tau_vs_q} does not include corotation torques. 
Yet, we have seen that just these torques slow down inward migration,
preventing the total torque $\mathcal{T}\subscr{D}$ from being
very negative.

It's worthy to note here that the relevance of coorbital torques,
to the orbital evolution of a protoplanet, was already pointed
out by Ward (\cite{ward1993}).
\subsubsection{$\beta$-dependence}
\label{Sect:beta-dep}
\begin{figure}
 \begin{center}
 \includegraphics[bb=30 40 475 460,clip,width=0.8\linewidth]{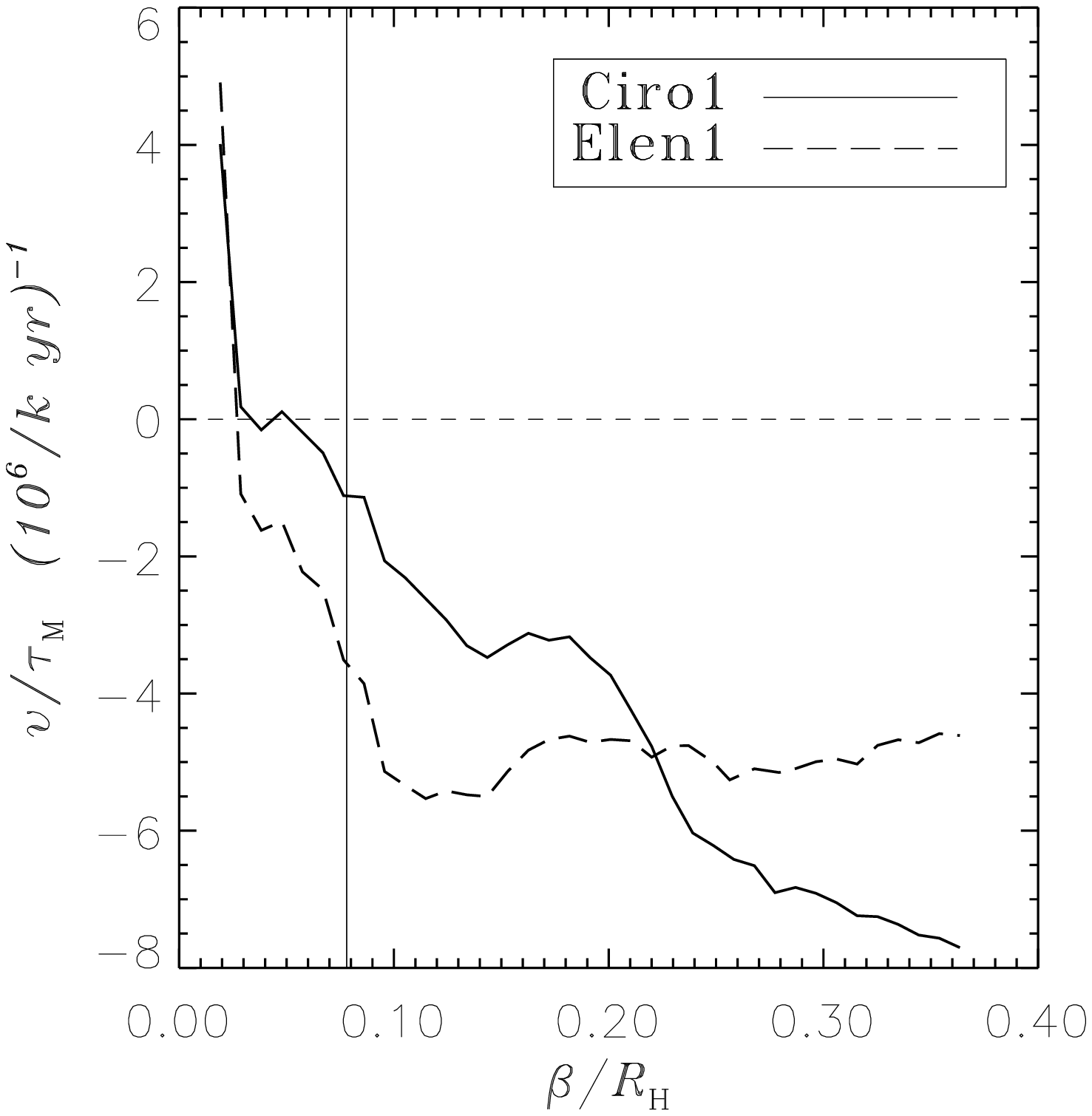}
 \includegraphics[bb=30 40 475 460,clip,width=0.8\linewidth]{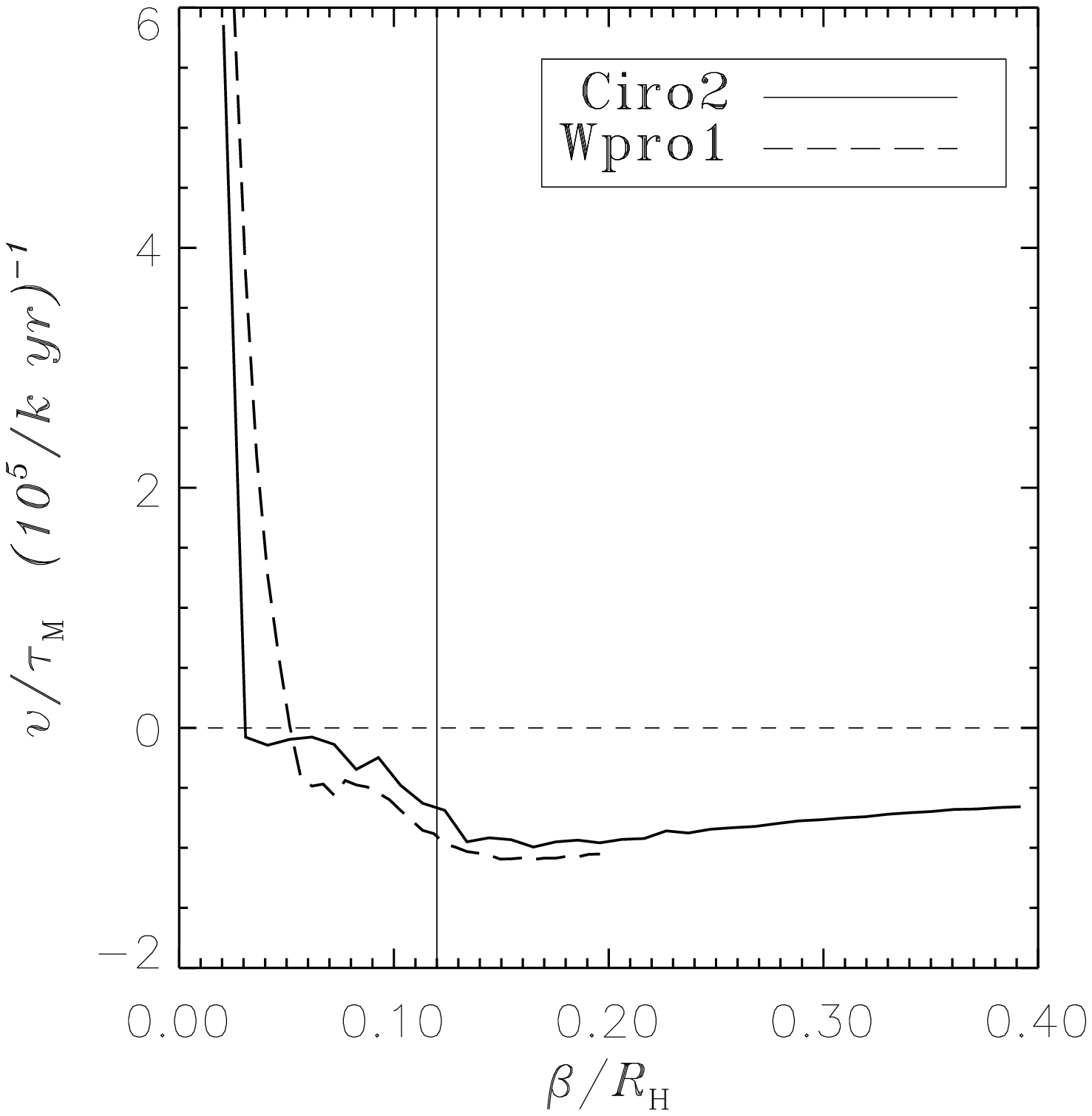}
 \includegraphics[bb=30 10 475 460,clip,width=0.8\linewidth]{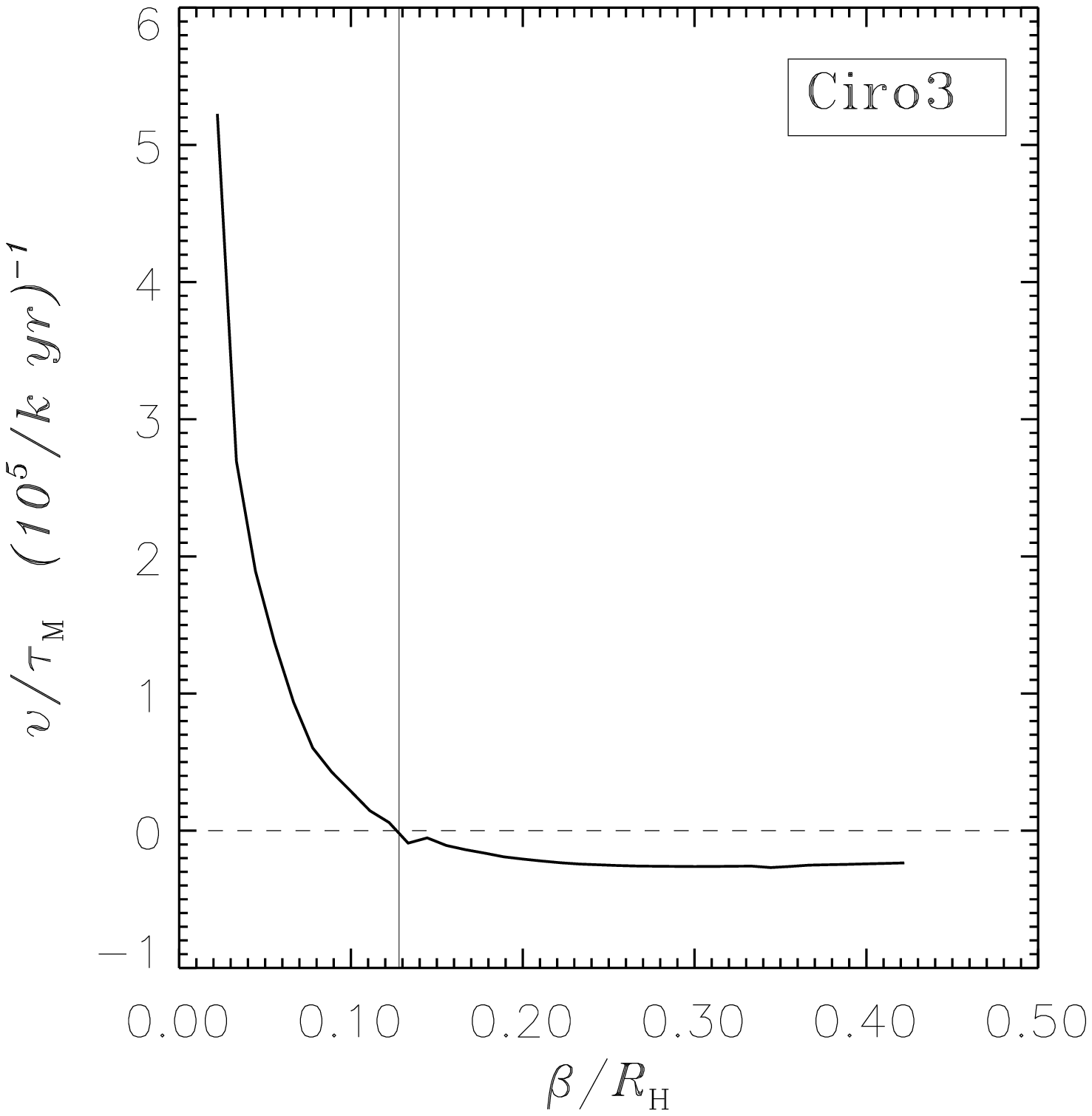}
 \end{center}
\caption{Migration time $\tau\subscr{M}$ as a function of $\beta$,
the radius of the region excluded from the computation of the torque.  
The ratio $v=\dot{a}/|\dot{a}|$ indicates the direction of the planet 
drift: $v<0$ for inward migration.
The semi-major axis $a$ has the same value as in \Fig{tau_vs_q}.  
The disk mass is cast into the form $\Md = k \times 10^{-3}$ \MStar.
Since in this paper we are using the value $k=3.5$, the factors 
in round brackets become $2.9 \times 10^5$ and $2.9 \times 10^4$ years, 
respectively.
The solid, vertical, line indicates the length $\beta = 2\,\eta$
(see \Fig{core}).}
\label{tau_vs_b}
\end{figure}
In the above discussion, as explained in \Sect{Subsec:Teop}, 
we do not account for the torque exerted by matter closer than 
$\beta=2\,\eta$ to the planet. 
Now we would like to relax this assumption and consider how
material, lying even closer, affects the overall torque  
$\mathcal{T}\subscr{D}$.
We mentioned already that the tendency of the nearby gas is 
to increase the angular momentum of the planet. 
As we enter the core dominated zone, such tendency may grow 
stronger and stronger, reducing more and more the magnitude 
of the negative torque exerted by the rest of the disk. 

An overview of the effects, due to nearby matter, on the
migration time scale $\tau\subscr{M}$, is given in \Fig{tau_vs_b}.
Actually, these plots show the dependence of  
$v/\tau\subscr{M} \propto \mathcal{T}\subscr{D}$ upon
$\beta$, where $v=\dot{a}/|\dot{a}|$ indicates the 
direction of the planet's migration. 
The distance where $\beta=2\,\eta$ is marked with a vertical, 
solid line. 
$\mathcal{T}\subscr{D}$ is directly proportional to
the mass of the disk \Md.
Therefore, to remove this potential restriction, in these plots 
we let it as a free parameter and write \Md\ as
$k \times 10^{-3}$ \MStar, although we use $k=3.5$ 
for our estimates. 
In \Fig{tau_vs_b}, 
$\mathcal{T}\subscr{D}=\mathcal{T}\subscr{D}(\beta)$
is presented for all of the three reference models. 

In \ciro1 (top panel), $\mathcal{T}\subscr{D}$ becomes larger
as $\beta$ gets smaller. 
The sign of the total torque changes around $\beta=\eta$, 
the threshold of the density core.
As a comparison, the behavior of $\mathcal{T}\subscr{D}$,
versus $\beta$, is reported also for the model \elen1.
In this case, because of the closed inner radial border,
the amount of matter inside the orbit of the planet is five 
times as large as that of \ciro1.
Outside the Hill circle, the torque exerted by the 
inner-disk, in case of \elen1, is twice as large as that 
measured in \ciro1. Instead, torques arising from the 
outer-disk nearly coincide. Inside the Hill circle, 
in \elen1, the contribution to $\mathcal{T}\subscr{D}$ 
is relatively small down to $\sim 0.1$ \Rhill\ whereas, 
in \ciro1, it never appears to be negligible.

\ciro2 (middle panel) behaves somewhat differently from \ciro1. 
The total torque attains a minimum around $\beta= 0.15$ \Rhill,
where $\tau\subscr{M}\simeq 3 \times 10^4$ years.
Then the positive torques, exerted by close matter, increase 
the total torque, though it remains negative all the way down to
$\beta=0.03$ \Rhill.
Below such value, $\mathcal{T}\subscr{D}$ diverges positively.
Results from the higher resolution model, \wpro1, do not differ
significantly (dash-line in \Fig{tau_vs_b}).

$\mathcal{T}\subscr{D}$ varies smoothly, as a function of $\beta$,
in case of \ciro3. The torques arising from the region enclosed
between $\beta\approx 0.2$ \Rhill\ and $\beta\approx 0.4$ \Rhill\
almost cancel out, so that the total torque appears nearly constant
($\tau\subscr{M}\approx 10^5$ years).
At shorter distances, positive torques prevail over the negative 
ones and $\mathcal{T}\subscr{D}$ starts to increase.
The sign of the total torque reverses at $\beta\simeq 0.12$ \Rhill.
For example, at $\beta=\eta$, its value is quite positive, imposing 
an outward migration rate $\tau\subscr{M}\simeq 3 \times 10^4$ years. 

Some comments should be devoted to how the smoothing length 
affects the total torque. We did not try to reduce 
further its value, however we ran some models, identical to 
\ciro1, but multiplying $\lambda$ (see Eq.~\ref{lambda})
by some integer number. 
A larger smoothing length tends to smear out more the surface 
density nearby the planet. Besides, it also causes the material 
to be distributed more symmetrically around it.
Both tendencies contribute to reduce the magnitude of the net 
torque arising from a region with radius $\approx \lambda$.
\subsection{Circumplanetary disk: gas flow} 
\label{Sect:CD}
\begin{figure*}
 \begin{center}
 \resizebox{\textwidth}{!}{%
 \includegraphics[bb=20 40 470 470, clip]{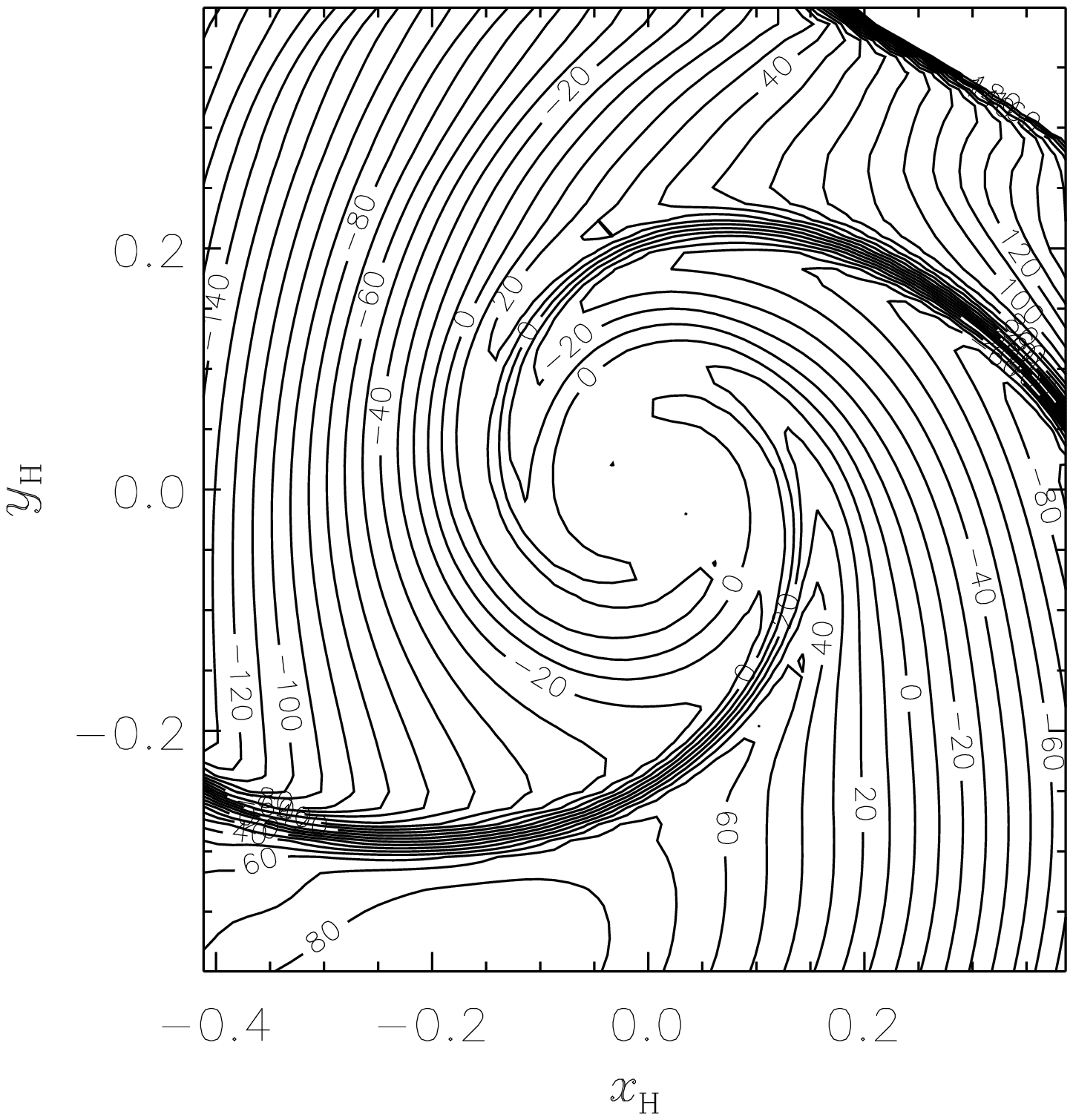}%
 \includegraphics[bb=55 40 470 470, clip]{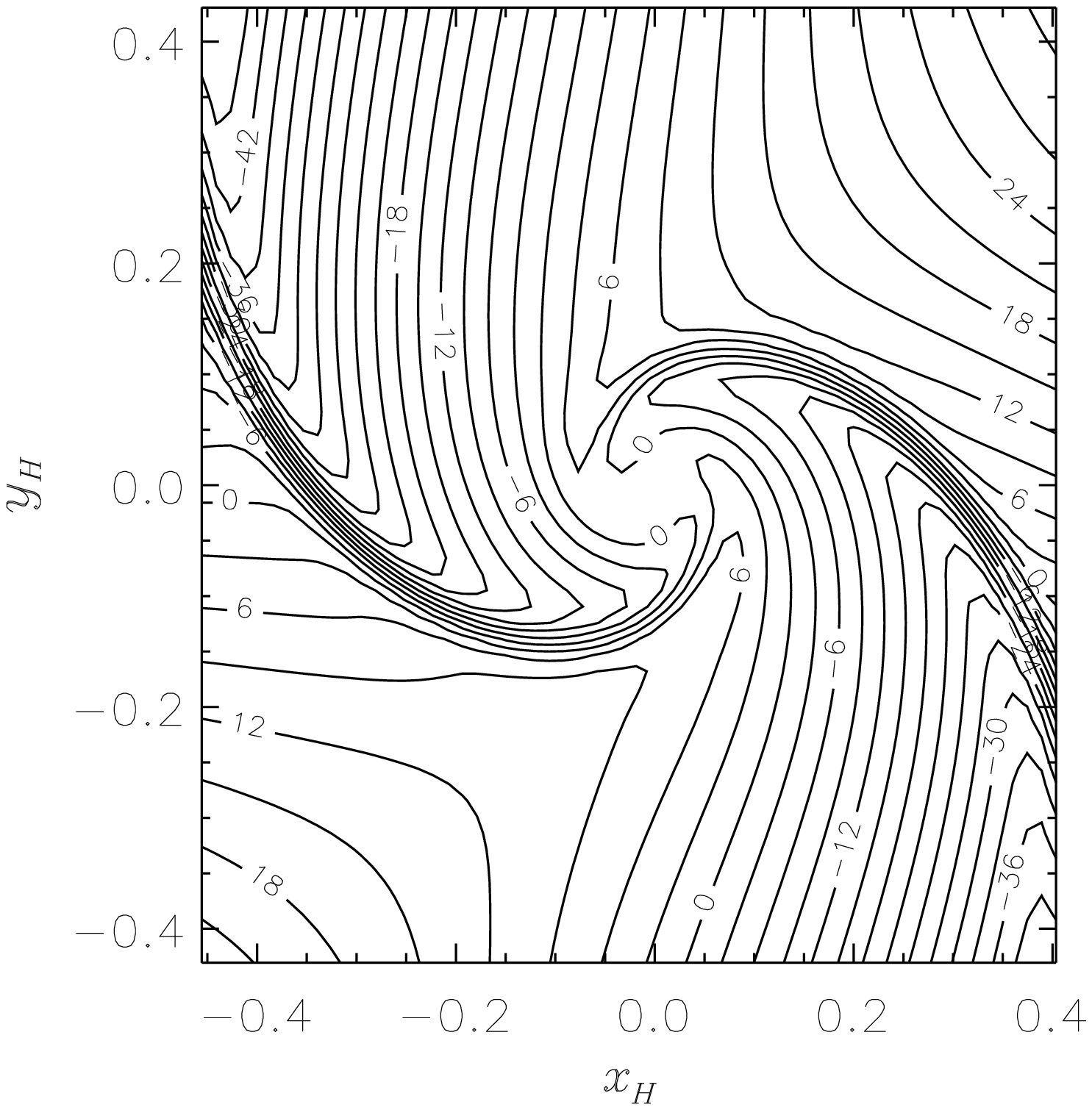}%
 \includegraphics[bb=55 40 470 470, clip]{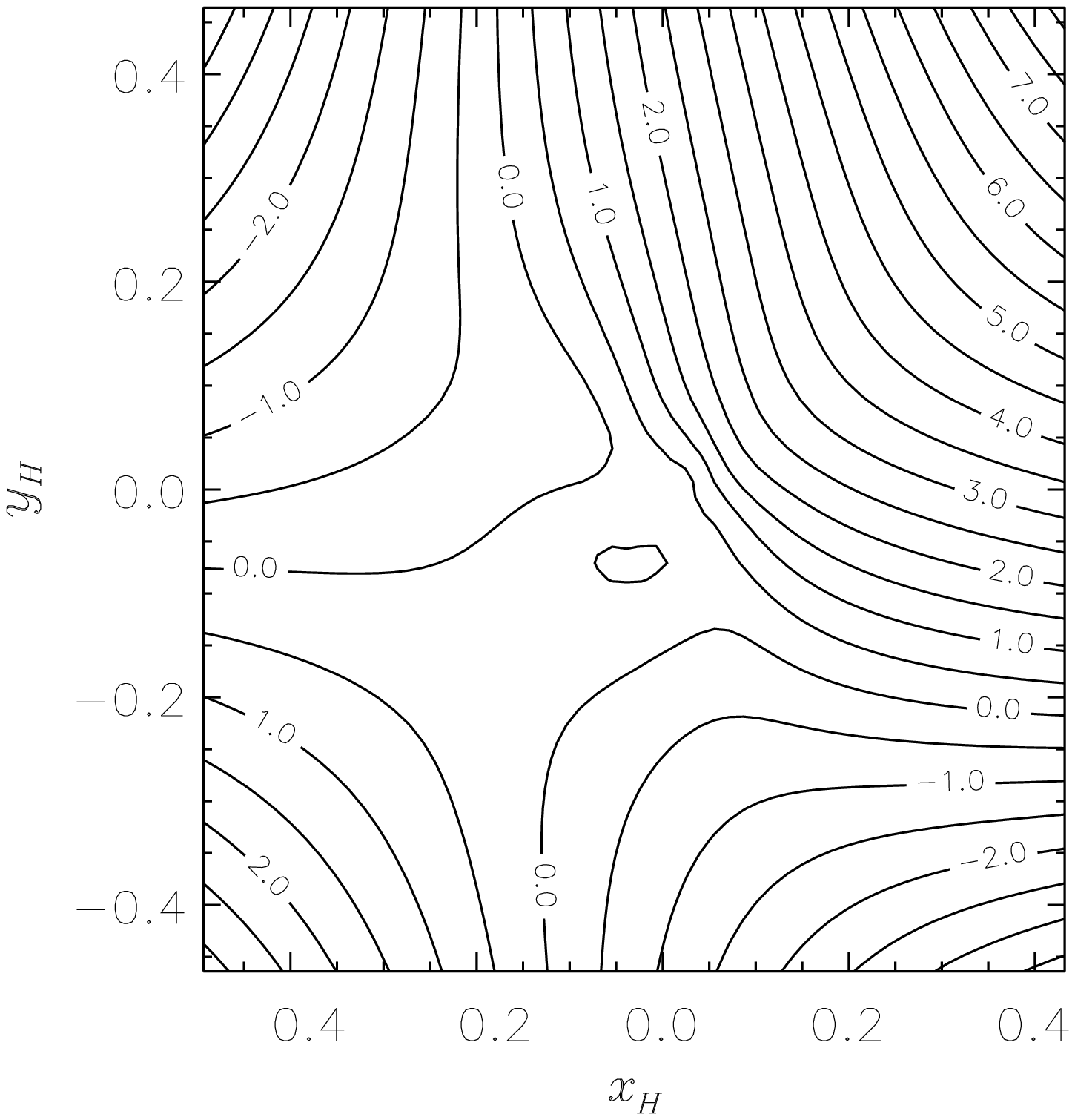}}
 \resizebox{\textwidth}{!}{%
 \includegraphics[bb=20 10 470 470, clip]{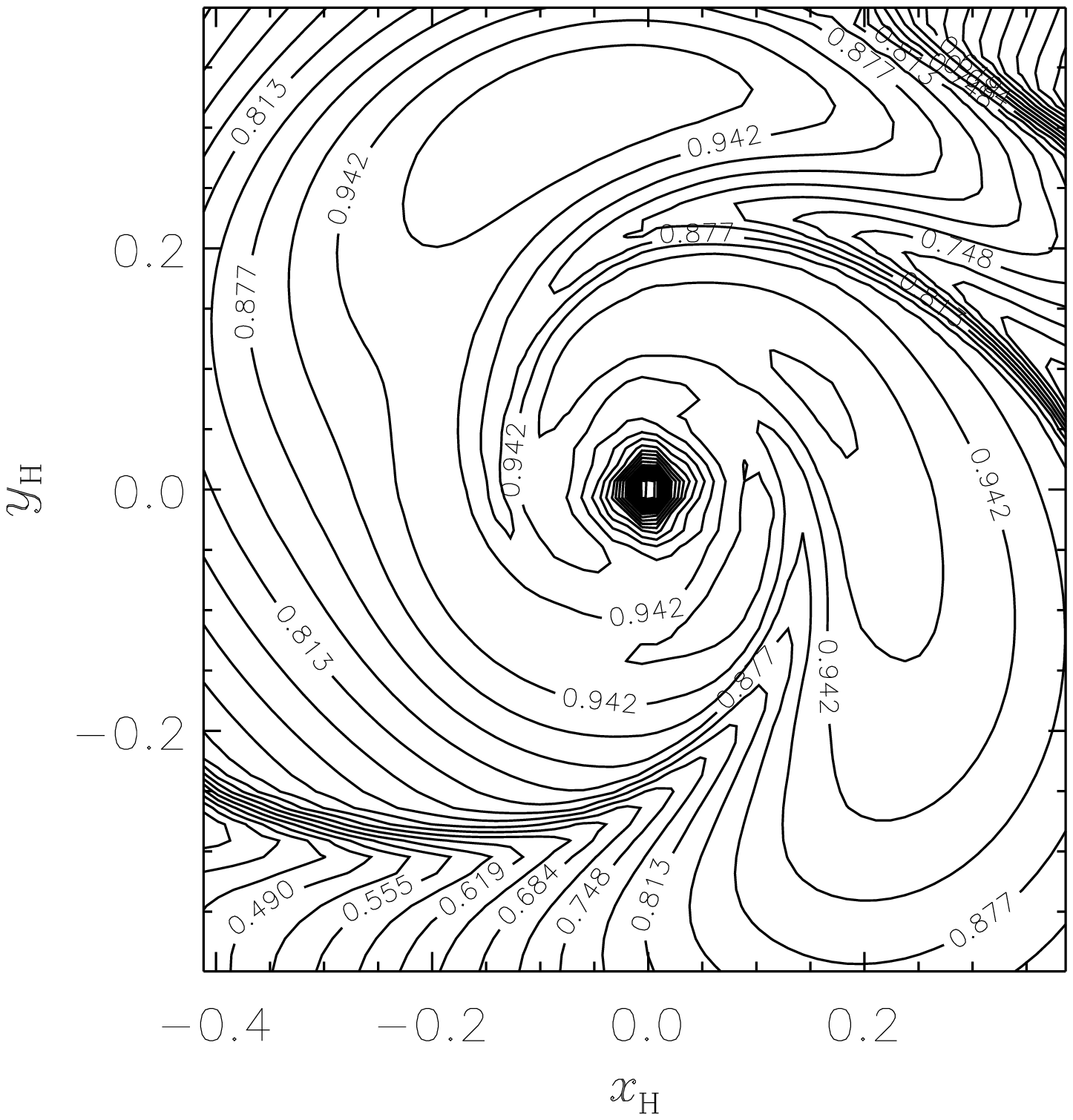}%
 \includegraphics[bb=55 10 470 470, clip]{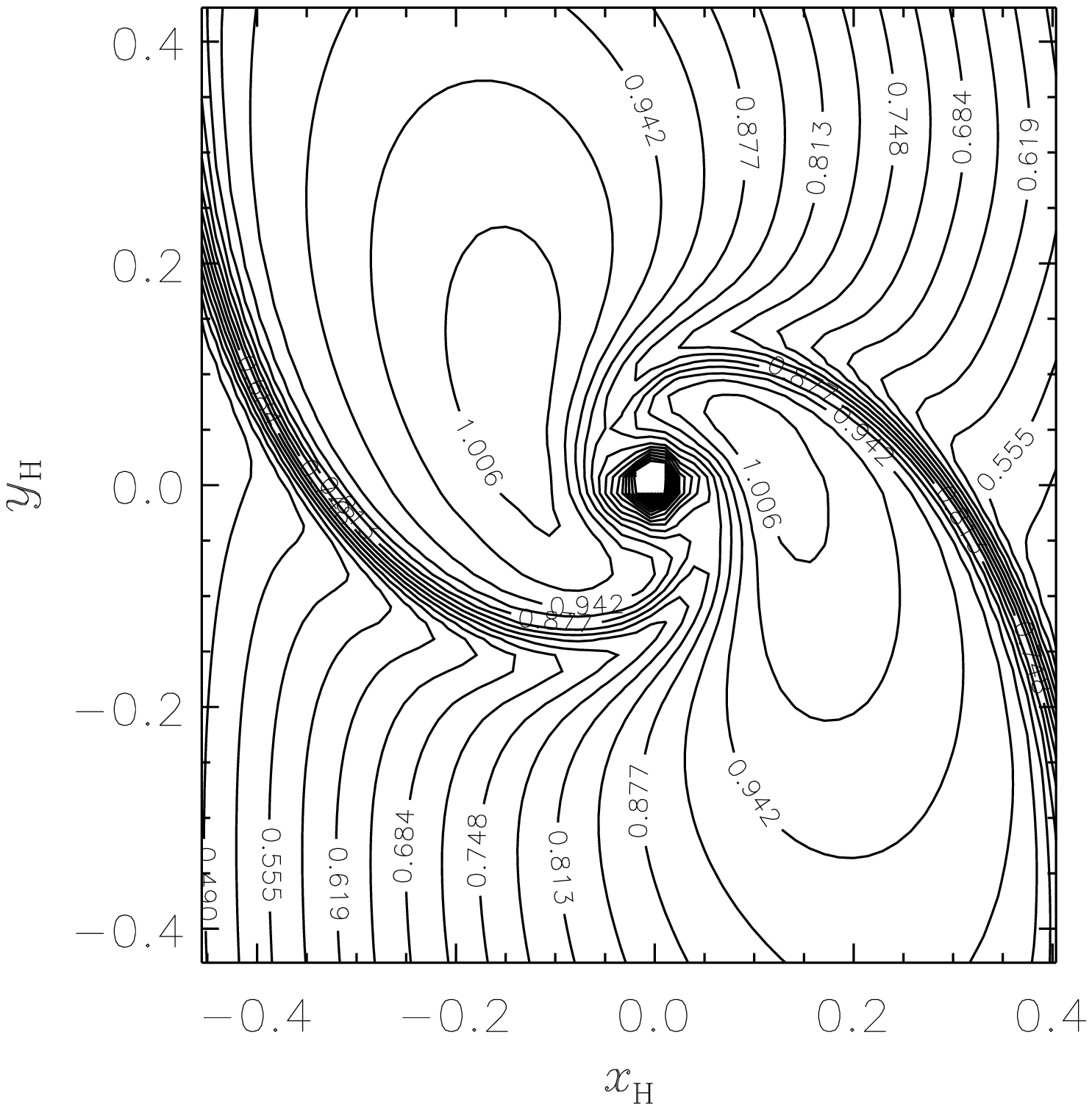}%
 \includegraphics[bb=55 10 470 470, clip]{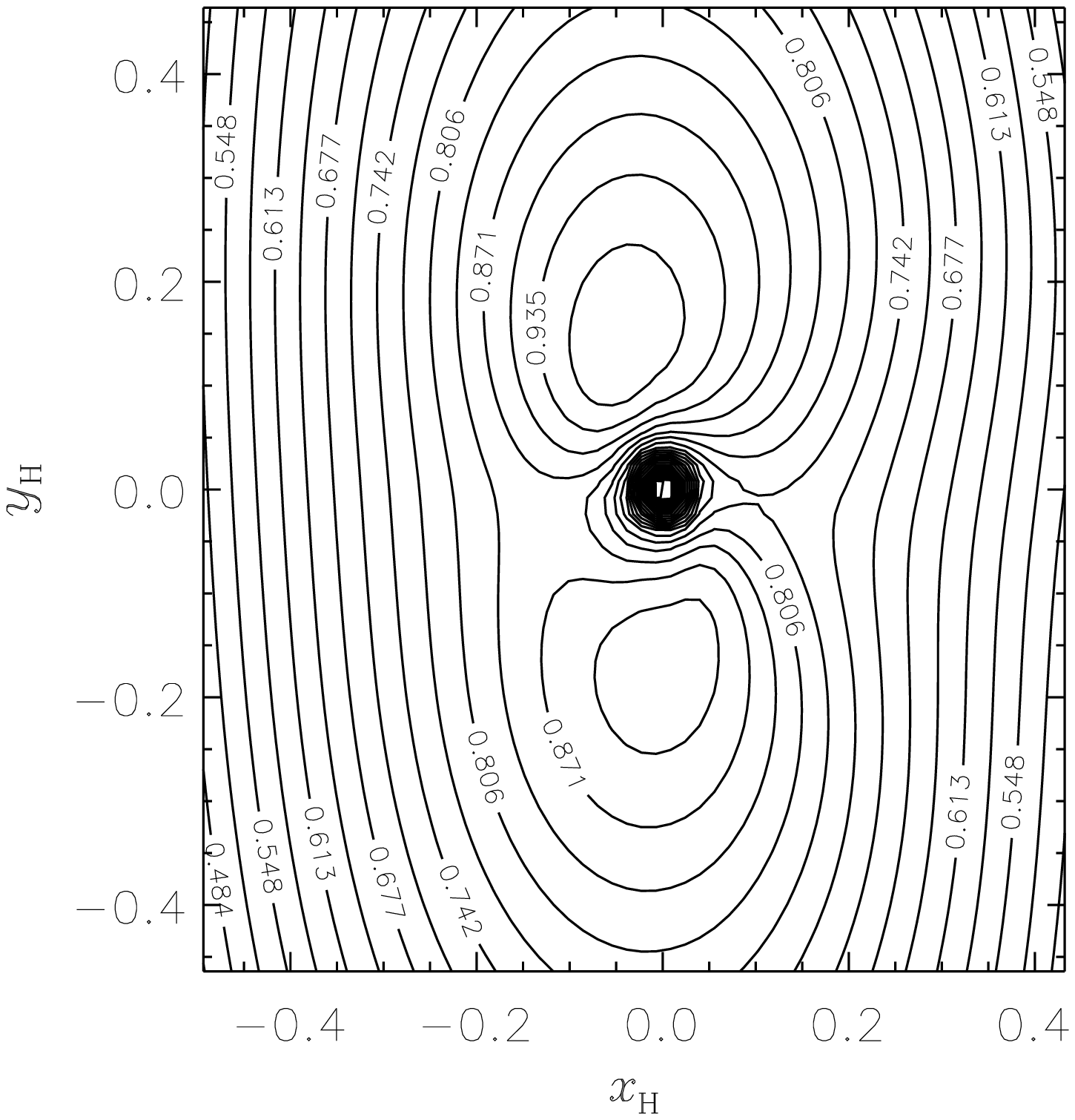}}
 \end{center}
 \caption{Contour lines of velocity ratios 
          $w\subscr{in}/w\subscr{in}\superscr{D}$ (\textbf{top panels}) and
          $w\subscr{rot}/w\subscr{rot}\superscr{K}$ (\textbf{bottom panels}).
          From left to right, we report the above quantities, as
          computed on the highest level, of models: \ciro1, \ciro2 and \ciro3. 
          $w\subscr{in}>0$ contours indicate places where the flow approaches 
          the planet. At locations where the in-fall velocity is 
          zero the flow performs a pure rotation around the planet.}
\label{nearflow}
\end{figure*}
In \Sect{Sect:results} we have described in detail 
how gas flows into the Roche lobe of a planet, for three 
particular values of its mass.

If a planet is massive enough, say $\Mp \gtrsim 10$ \MEarth, 
the streams of matter, entering the Roche lobe, produce strong 
shock waves which then rule the gas flow inside this region.
Material passing through the shock fronts is deflected towards 
the planet, tightening its orbit on it.
Less massive planets are not able to cause strong perturbations 
inside the Roche lobe. As a consequence, the flow pattern appears 
more uniform around the planet.

Now we would like to investigate quantitatively 
the rotational regime of the gas inside the circumplanetary disks. 
In particular, we would like to estimate how much it resembles a 
Keplerian one.
In order to address this issue we decompose the local velocity
field $\vec{u}$, in two components, representing
the in-fall velocity $w\subscr{in}$ and the rotational velocity 
$w\subscr{rot}$ of the fluid \textit{relative to the planet}.
The first component is defined as:
\[
w\subscr{in} = - \vec{u}\cdot \frac{\vec{r}-\vec{r}_\mathrm{p}}%
                                 {|\vec{r}-\vec{r}_\mathrm{p}|},
\]
which is positive if a fluid element moves towards the planet.
The quantity $w\subscr{rot}$ is the projection of $\vec{u}$ along the 
direction orthogonal to $\vec{r}-\vec{r}_\mathrm{p}$ and
such that 
$(\vec{w}\subscr{rot} \times \vec{w}\subscr{in}) \cdot \hat{\vec{z}}$
is positive. With this choice $w\subscr{rot}$ is positive
for a counter-clockwise rotation. 

If circumplanetary disks were regular accretion disks, 
we should expect them to be in a ``Keplerian'' regime.
This is characterized by the rotational velocity
\begin{equation}
w\subscr{rot}\superscr{K}= 
 \sqrt{\frac{G\,\Mp}{|\vec{r}-\vec{r}_\mathrm{p}|}},
\label{wkrot}
\end{equation}
and the inward viscous diffusion
\begin{equation}
w\subscr{in}\superscr{D}= 
\frac{3\,\nu}{2\,|\vec{r}-\vec{r}_\mathrm{p}|}.
\label{wdin}
\end{equation}
Equations (\ref{wkrot}) and (\ref{wdin}) don't include 
the smoothing length, $\delta$, because its effects were
checked to be unimportant.
Comparing $w\subscr{rot}$ and $w\subscr{in}$ with \Eq{wkrot} 
and \Eq{wdin}, respectively, we can estimate how much the 
circumplanetary flow is close to be Keplerian, i.e.\ close that 
of an unperturbed viscous disk.

Figure~\ref{nearflow} shows, for \ciro-models, the contour 
lines of $w\subscr{in}$ normalized to $w\subscr{in}\superscr{D}$ 
(top panels) and $w\subscr{rot}$ normalized to 
$w\subscr{rot}\superscr{K}$ (bottom panels).
 
As first remark we note that, if we compare \Fig{nearflow} 
to \Fig{deep_cont}, lines of equal surface density perturbation 
are also lines of equal velocity perturbation, as spiral wave 
theory predicts. 

From the top panels of \Fig{nearflow},
we can see that material approaches the planet along well
defined patterns.  
Contours $w\subscr{in}=0$ mark locations where the
flow rotates around the planet without altering its distance
from it. 
They also separate regions in which material proceeds towards
the planet from those where it moves away. 
One of these contours runs along the spiral ridges. 
Across it, the in-fall velocity changes abruptly its sign.

The ratio $w\subscr{in}/w\subscr{in}\superscr{D}$ becomes
smaller as the gas comes closer to the planet.
Since the viscous diffusion $w\subscr{in}\superscr{D}$ is
not related to \Mp, it's possible to compare the magnitude
of $w\subscr{in}$ for the different cases. Contour level
values indicate that it gradually reduces as \Mp\ gets smaller.

\begin{figure}
 \begin{center}
 \resizebox{\linewidth}{!}{%
 \includegraphics{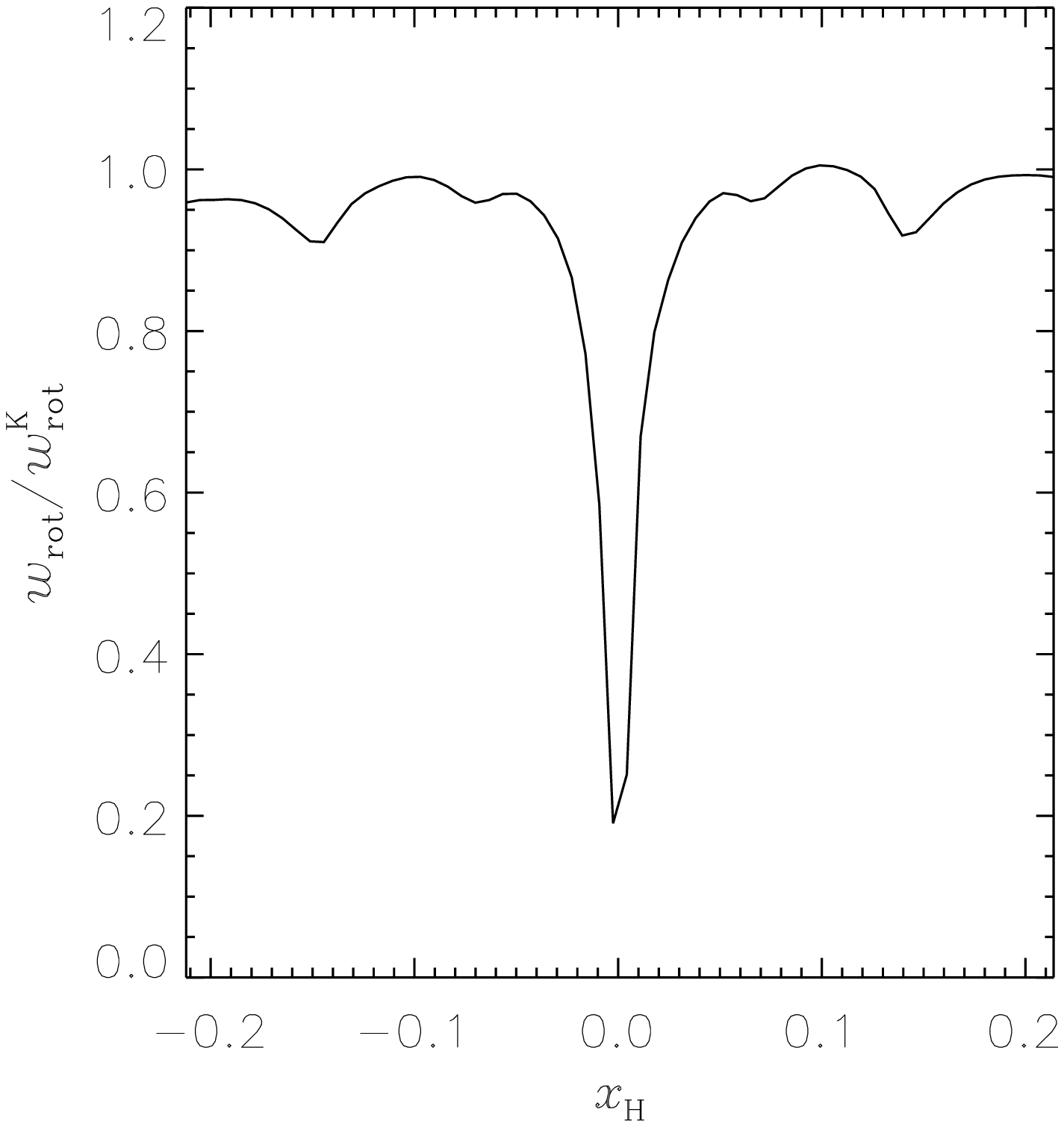}}
 \end{center}
\caption{The plot shows the ratio $w\subscr{rot}/w\subscr{rot}\superscr{K}$
at $\varphi=\varphi\subscr{p}$ ($y\subscr{H}=0$), 
as computed on the grid level $l=6$ of 
the model \elen{2} ($\Mp = 1$ \MJup). 
The core area, in the Jupiter-mass case, extends for 
$[0.08 \times 0.08]$  $\Rhill^2$ (see \Fig{core}, left panel). 
The resolution, in \elen2, is such that this is covered by 
$\sim 12\times12$ grid cells.}
\label{wrot_willy}
\end{figure}

As regards the rotational component of the velocity field 
$w\subscr{rot}$ (\Fig{nearflow}, bottom panels), we can see 
that it is generally slightly below $w\subscr{rot}\superscr{K}$. 
For the Jupiter-mass case, this can be seen also in \Fig{wrot_willy}.
Centrifugal over-balance regions are not present around
the smallest planet ($\Mp =3.2$ \MEarth). 
Instead, they are observed in \ciro1, 
at $(x\subscr{H},y\subscr{H}) \approx (-0.1,0.3)$ and
$(0.15,0.1)$, and in \ciro2, where they are labeled.
In both cases, anyway, the ratio 
$w\subscr{rot}/w\subscr{rot}\superscr{K}$
is very close to unity.
Centrifugal under-balance is mainly established along
spiral ridges. This is in agreement with the idea that 
spiral arms are zones of compression, hence pressure
plays a more active role in supporting the gas. 

In all of the three cases shown in \Fig{nearflow}, $w\subscr{rot}$ 
reveals somewhat a circular symmetry only within a distance $\sim \eta$
from the planet.
Yet, we have to notice that this coincides nearly with the region 
from which matter is removed to simulate the gas accretion. 
Figure~\ref{wrot_willy} also indicates that the core material 
(\Sect{Subsec:core}) rotates slower, when it approaches its center. 
\subsection{Accretion onto the planet} 
\label{Sect:AOP}
\begin{figure}
 \begin{center}
 \resizebox{\linewidth}{!}{%
 \includegraphics{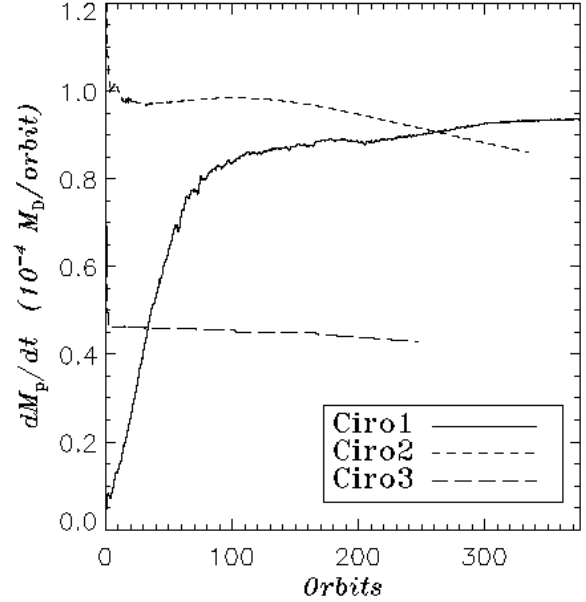}}
 \end{center}
\caption{Mass accretion rate onto the planet versus time, in units of
       $10^{-4}$ disk masses per orbital period of the planet. 
       For $M\subscr{D}=3.5 \times 10^{-3}$ \MStar\ and $a = 5.2$ \AU,
       one dimensionless unit corresponds to 
       $2.95 \times 10^{-5}$ $\MJup\,\mathrm{yr}\superscr{-1}$.
       The initial accretion rate is very small
       for model \ciro1 because of the imposed initial gap.}
\label{mdot_vs_t}
\end{figure}
Gas matter closely orbiting the planet is eligible to be accreted
once its distance, $|\vec{r}-\vec{r}_\mathrm{p}|$, is less than 
$\kappa\subscr{ac}\approx 9 \times10^{-2}$ \Rhill. The details
of this process are described in \Sect{Subsec:accret} 
(see Fig.~\ref{massacc}).  
In general, the mass accretion rate of a planet, $\dot{M}\subscr{p}$,
becomes relatively constant after the gap (if any) has evolved to
a quasi-stationary state. 

For a Jupiter-mass planet (\ciro1) this happens around 100 
orbital periods, as indicated by the solid line in \Fig{mdot_vs_t}.
The theoretical gap imposed at the beginning of the evolution
(see \Fig{sigma0}) is deeper and wider, at
$\varphi=\varphi\subscr{p}$, than it is later on. 
For this reason $\dot{M}\subscr{p}$
is negligibly small at early evolutionary times. 
The partial replenishment is related to the formation
of the circumplanetary disk which supplies matter 
for the accretion process.

Smaller planets dig narrower and shallower gaps so this 
quasi-steady regime is reached even earlier. 
For both \ciro2 (\Fig{mdot_vs_t}, short-dash line) and
\ciro3 (long-dash line), $\dot{M}\subscr{p}$ reduces a little 
as the evolution proceeds. This is likely due to the depletion 
of the inside-orbit disk. In case of \ciro1, most of the
inner-orbit material is cleared out during the transitional
phase (75\% after 100 orbits). Therefore, it does not contribute
much during the quasi-steady phase.

\begin{figure*}
 \begin{center}
 \hspace*{\fill}%
 \includegraphics[width=0.48\textwidth]{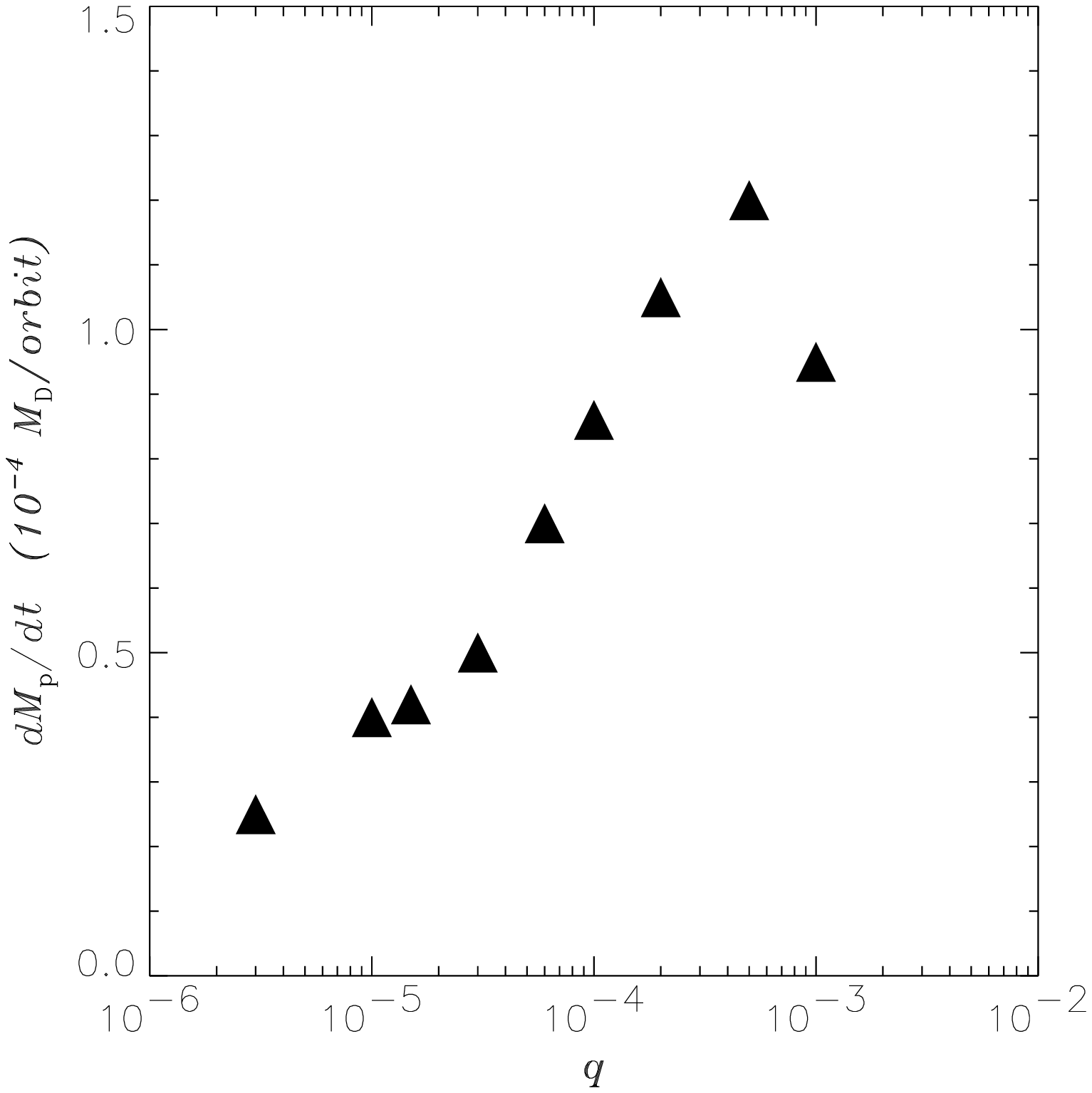}\hfill%
 \includegraphics[width=0.48\textwidth]{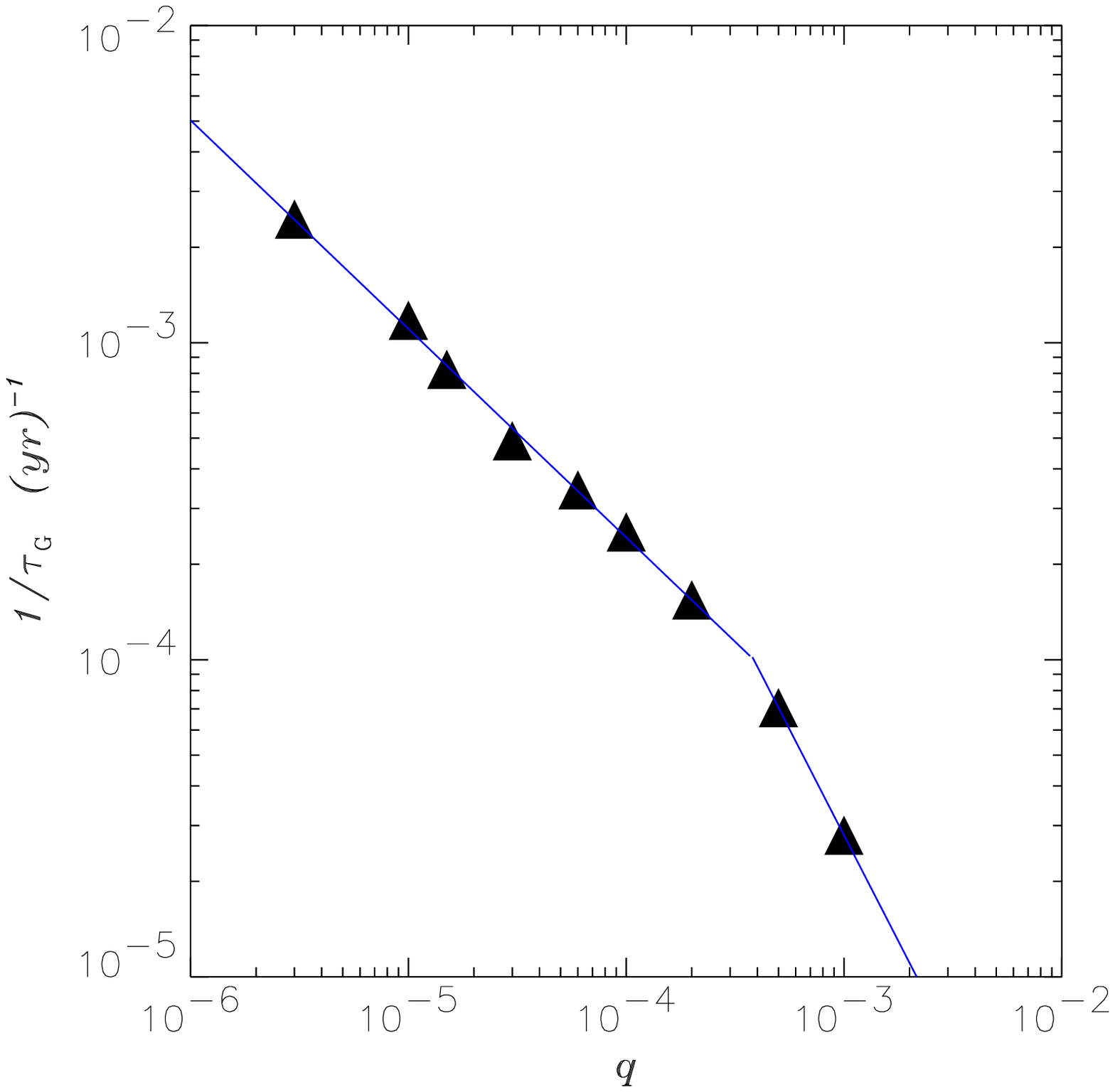}%
 \hspace*{\fill}
 \end{center}
\caption{\textbf{Left panel}. Mass accretion rate 
         $\dot{M}\subscr{p}$ as function of the mass ratio 
         $q$. Quantities are normalized to the disk mass \Md.
         \textbf{Right panel}. Dependence of the growth time
         scale $\tau\subscr{G} \equiv \Mp/\dot{M}\subscr{p}$
         on the planet mass (filled triangles).
         $\Md = 3.5 \times 10^{-3}$ \MStar\ is employed to
         express $\tau\subscr{G}$ into years. 
         The two-branch solid line represents the first-order 
         polynomial fitting the data (see Eqs.~\ref{tau_G-qa} 
         and \ref{tau_G-qb}).}
\label{mdot_vs_q}
\end{figure*}

In \Fig{mdot_vs_q}, the accretion rate is plotted against the 
mass ratio (left panel), in dimensionless units as in \Fig{mdot_vs_t}.
$\dot{M}\subscr{p}$ increases as the planet mass increases and
reaches a maximum around $0.5$ \MJup\ (model \gino3).
Beyond this point the curve starts to drop. In fact, the accretion 
rate of \ciro1 lies between those of
\ciro2 ($\Mp = 32$ \MEarth) and \gino2 ($\Mp = 64$ \MEarth).
The accretion rates for planets with masses above about $10^{-4}$ \MStar\
($\sim 30$ \MEarth) are consistent with those obtained
by recent models studying the evolution of giant proto-planets
(Tajima \& Nakagawa \cite{tajima1997}).
For smaller masses our accretion rates are substantially higher.
However, all
the more detailed studies of protoplanetary evolution are
spherically symmetric (see the review by Wuchterl et al. \cite{wuchterl2000}),
and accretion via a planetary accretion disk
may allow for higher accretion rates.

We did not perform computations involving planets heavier than 
one Jupiter-mass so we cannot follow the trend of the curve for 
larger values of $q$. 
However, Lubow et al. (\cite{lubow1999}) found that $\dot{M}\subscr{p}$ 
decreases in the mass range from 1 to 6 \MJup.  

In the right panel of \Fig{mdot_vs_q} the growth time scale
$\tau\subscr{G}\equiv \Mp/\dot{M}\subscr{p}$ is plotted versus $q$. 
Logarithmic scaling of the axes shows that the curve decreases 
almost linearly with respect to the mass of the planet.
If we perform a linear least-square fit of the values for 
which $q \ge 2\times 10^{-4}$, corresponding to $\Mp=64$ \MEarth, 
we get:
\[
\frac{d\log(1/\tau\subscr{G})}{d\log(q)} = -0.66.
\]
This equation yields:
\begin{equation}
\tau\subscr{G} \propto q^{0.66} \simeq q^{2/3}.
\label{tau_G-qa}
\end{equation}
At higher values of $q$, the curve steepens, decreasing more rapidly.
Taking into account the growth time scales relative to the
two most massive planets, one finds 
\begin{equation}
\tau\subscr{G} \propto q^{1.34} \simeq q^{4/3}.
\label{tau_G-qb}
\end{equation}
Therefore, as a planet becomes more massive, the growth time scale 
grows with an increasing power of its mass. 
Thus, we can argue that very high mass planets should 
be extremely rare.
\subsubsection{Influence of numerics} 
Finally, we comment on the influence of some numerical 
parameters upon the mass accretion rate. 
Table~\ref{mdot_tab} lists some results, after the same
number of orbits, for various models which can be useful 
to the goal.
The value of $\dot{M}\subscr{p}$ may depend on either
the radius $\kappa\subscr{ac}$ of the accretion region and
the mass evacuation rate $\kappa\subscr{ev}$. 
By comparing models \ciro2 and \wpro2, it turns out that
doubling $\kappa\subscr{ev}$, the accretion rate is only
11\% higher. 
In order to estimate how relevant the extension of the
accretion region may be, we ran a model having an accretion
length scale ($\kappa\subscr{ac}$)
$0.6$ times as small as that of \gino3  
(the rest of the model set-up being identical). 
We obtain an accretion rate 7\% smaller.

Further, $\dot{M}\subscr{p}$ could depend on the numerical 
resolution as well. Indeed, this dependency turns out to be 
very weak, as indicated by two sets of models.
\elen2 and \wpro2 have, in the accretion region, a resolution 
two times as high as that of \ciro1 and \ciro2,
respectively (see Table~\ref{models}). Despite this fact,
in the first case, the accretion rates differ by just 2\%, 
whereas the difference amounts to 4\% in the second. 

As mentioned in \Sect{Sect:MS}, the planet in not symmetrically
centered within a grid cell. 
Model \gino1 was designed to accomplish a fully symmetric 
configuration (as already explained in \Sect{Sect:MS}).
However, the planet accretion rate is not affected by
this position shift, as can be seen by comparing
the values in Table~\ref{mdot_tab}, belonging to \gino1
and \ciro3.
\begin{table}
 \caption{Mass accretion rate onto the planet for different numerical
          parameters. This quantity is reported at the same evolutionary
          time for similar models. 
          $\dot{M}\subscr{p}$ is given in units of disk masses per orbit.}
 \label{mdot_tab}
 \begin{center}
 \begin{tabular}{lccc}
 \hline
 \hline
 Model   &  $q$                  & $\dot{M}\subscr{p}$   & Orbits \\
 \hline 
 \ciro1  &  $1.0 \times 10^{-3}$ & $8.7 \times 10^{-5} $ & 200 \\ 
 \elen2  &  $1.0 \times 10^{-3}$ & $8.9 \times 10^{-5} $ & 200 \\ 
 \ciro2  &  $1.0 \times 10^{-4}$ & $9.0 \times 10^{-5} $ & 280 \\ 
 \wpro1  &  $1.0 \times 10^{-4}$ & $8.6 \times 10^{-5} $ & 280 \\
 \wpro2  &  $1.0 \times 10^{-4}$ & $1.0 \times 10^{-4} $ & 280 \\
 \ciro3  &  $1.0 \times 10^{-5}$ & $4.5 \times 10^{-5} $ & 120 \\ 
 \gino1  &  $1.0 \times 10^{-5}$ & $4.5 \times 10^{-5} $ & 120 \\ 
 \hline
 \end{tabular}
 \end{center}
\end{table}
\section{Conclusions}
\label{Sect:conclusions}
A number of numerical simulations concerning disk-planet interactions 
have been performed to get new insights into the scenario of the joint
evolution of protoplanets and their environment.
They have confirmed analytical theories for gap formation and planet 
migration.

However, many open questions still remain. The most important
unsolved issue is the influence of the ambient gas on the
dynamical evolution of a planet. 
Another one is the way disk-planet interaction changes when small 
planets, in the mass range of Neptune and Earth, are considered.

We began to investigate in both directions by means of a nested-grid 
technique, which is particularly suitable for treating these problems.
The main asset of this numerical scheme is the possibility of achieving,
locally, a very high spatial and temporal resolution.
With such a method we are able to resolve very accurately both,
the inner parts of the Hill sphere of the planet \emph{and}
the global structure of the disk, because
we treat the whole azimuthal extent of the disk.

Concerning the issues mentioned above, with the present
paper we tackled some outstanding problems concerning the
growth and migration of protoplanets, covering a range 
from one Jupiter-mass down to one Earth-mass.
Thus, even though we do not include the detailed energetic balance
of the planetary structure, which is still beyond present day
computer facilities, this study represents a definite improvement
in the determination of the torque balance on protoplanets.

Our main achievements can be summarized as follows.
\begin{enumerate}
\item  Inside the Hill sphere of the planet a circumplanetary disk
       forms. Within it strong spiral density waves develop, if the 
       planet mass is larger than $\sim 5$ \MEarth. 
       These waves assume the shape of a two-arm pattern. The two spiral 
       arms are slightly asymmetric with respect to the planet.
       For decreasing planet masses, they stretch and shorten. 
       Matter is observed to pile up at the location of the planet,
       generating a very high density zone (we named as density 
       ``core''), 
       which might represent its primordial gaseous envelope.
\item  Nearby material exerts positive torques on the planet,
       slowing down, considerably in some cases, its inward 
       migration.
       Most of these torques arise from corotation regions,
       i.e. from gas lying on the planet's orbit.
       Analytical models about migration do not account for them. 
       This can be one of the reasons why our estimates
       of the migration time scales give somewhat higher values than
       those predicted by such theories.
\item  Within a distance of $\sim 0.1$ \Rhill\ from the planet,
       the point-mass approximation becomes too restrictive
       and maybe not appropriate.
       Therefore, the structure of the planet should be also 
       taken into account over such a length scale. 
       This is absolutely necessary if one wants to evaluate how much 
       of the angular momentum, transferred by closely orbiting matter, 
       is conveyed to the spin of the planet rather than to its orbital 
       angular momentum.
\item  The Keplerian rotational regime of circumplanetary disks
       is affected by spiral perturbations. Just as for the mass 
       density, the more massive the planet is, the stronger such 
       perturbations are.
       Gas material, passing through the spiral fronts, is
       deflected towards the planet. Instead, in the inter-arm 
       regions it moves away from it. 
       This is in analogy with the spiral wave theory in galaxies.
       Around Earth-mass planets, the rotation of the gas is
       very slow if compared to the Keplerian rotation. 
       In fact, in this particular case, the density core has 
       nearly a hydrostatic structure.
\item  The mass accretion rate, as a function of the mass of the 
       planet, has a maximum around $\Mp = 0.5$ \MJup. 
       As long as $\Mp \lesssim 0.2$ \MJup, the growth time scale 
       of a planet increases, approximatively, as $\Mp^{2/3}$. 
       For more massive planets, it increases roughly
       as $\Mp^{4/3}$. 
       Such a dependence may contribute to limit the size of
       a massive planet.
\end{enumerate}
Since we just started to explore these new grounds,
each of the items above may deserve a more specific and
dedicated study.

Next efforts should be devoted to refine the physical model,
especially in the vicinity of the planet. 
Henceforth, some of the future developments could be:
\begin{itemize}
\item including an energy equation, by implementing an approximate 
      treatment of radiative transfer and viscous dissipation;
\item improving the equation of state by using an alternative form 
      which accounts for the planet's structure;
\item evaluating possible effects due to the two-dimensional
      approximation of the disk, via three-dimensional simulations.
\end{itemize}
\begin{acknowledgements}
We would like to thank Udo Ziegler for making the FORTRAN 
Version of his code \textsc{Nirvana}\footnote{%
\texttt{http://www.aip.de/~ziegler/}}
available to us.
We would like to thank also Dr. Ewald M\"uller for stimulating 
discussions on nested-grid calculations.
This work was supported by the German Science Foundation (DFG) 
under grant KL 650/1-1.
The numerical computations have been carried out at the Computer 
Center of the University of Jena.
\end{acknowledgements}
%

\end{document}